\documentclass[11pt]{article}

\usepackage{amssymb,amsmath,amsfonts,eurosym,geometry,ulem,graphicx,caption,color,setspace,comment,footmisc,caption,natbib,pdflscape,array,hyperref,enumitem,threeparttable,booktabs,tabularx,bbm,pifont,float,graphicx,subcaption,rotating,makecell,afterpage,dsfont,adjustbox,newpxtext,newpxmath}

\hypersetup{colorlinks,linkcolor={blue},citecolor={blue},urlcolor={red}}

\DeclareMathOperator*{\argmax}{arg\,max}

\newcommand{\R}{\mathbb{R}}

\newcommand{\E}{\mathbb{E}}

\newcolumntype{L}[1]{>{\raggedright\let\newline\\arraybackslash\hspace{0pt}}m{#1}}
\newcolumntype{C}[1]{>{\centering\let\newline\\arraybackslash\hspace{0pt}}m{#1}}
\newcolumntype{R}[1]{>{\raggedleft\let\newline\\arraybackslash\hspace{0pt}}m{#1}}

\begin{document}

\begin{titlepage}
\title{
To Each Their Own: Heterogeneity in Worker \\ 
Preferences for and Responses to Peer Information\thanks{I thank especially Mark Dean for invaluable guidance and support. I thank Sandra Black, Alessandra Casella, Lorenz Goette, Shizuka Inoue, Judd Kessler, Robert Metcalfe, Suresh Naidu, Michael Woodford, Hanyao Zhang, as well as workshop and conference participants at Columbia University, the 2025 SWEET Conference, and the 2025 North American ESA Meeting for helpful feedback and suggestions. Claire Zhao provided excellent research assistance. This work is supported by the Columbia Experimental Laboratory for Social Sciences (CELSS), the Program for Economic Research (PER) at Columbia University, and the NUS Development Grant. This study was approved by the Columbia University Institutional Review Board (IRB-AAAU7355) and pre-registered on AsPredicted.org (\#210555). All errors remain my own.}
}
\author{Zhi Hao Lim\thanks{Department of Economics, Columbia University. Email: zl2969@columbia.edu.} 
\\[2mm]
\textit{\small Columbia University}
\\[6mm]
}

\date{\today}
\maketitle

\begin{abstract}
\singlespacing
Information about peers’ performance is pervasive in workplaces, yet its effects on worker behavior are mixed. We show that a key reason is that workers differ in how they value such information. In a real-effort experiment with 793 workers, we elicit willingness-to-pay for peer information delivered either before or after the task. We document substantial heterogeneity in demand for peer information: some workers are indifferent, some prefer to avoid it before the task, and others value it more as their relative performance increases. These differences strongly predict effort responses to peer information. Notably, 15\% of workers would pay to avoid information ex ante due to stress and exhibit no productivity gains from it. We further show that uniform feedback policies can impose welfare losses on such workers, while tailoring the timing of peer information increases welfare by up to 48\%. Our results highlight the importance of accounting for heterogeneous information preferences when designing workplace feedback policies.
\bigskip \\[4mm] 
\end{abstract}

\setcounter{page}{0}
\thispagestyle{empty}
\end{titlepage}
\pagebreak \newpage

\onehalfspacing

\section{Introduction} \label{sec:introduction}

Peer performance information is a common feature of modern organizations. Firms use dashboards, rankings, and performance reviews to inform workers how they compare with their peers, often with the goal of improving effort and productivity. Yet the evidence on the effects of such feedback is mixed: in some settings, peer information increases productivity, while in others it has little effect or even backfires \citep{eriksson2009feedback, barankay2011rankings, barankay2012rank, kuhnen2012feedback, charness2014dark, huet2015social, gill2019first, wu2021social, ashraf2025performance}. One possible reason is that workers differ in how they value and respond to peer information. If so, the key managerial challenge is not simply whether peer information works on average, but which workers are likely to benefit from it and when it should be delivered.

Workers may seek or avoid peer information for different reasons. Some may not care about it as it does not affect their own performance. Others may find it motivating because it provides a salient benchmark for comparison. Some may dislike receiving it because it creates pressure or distraction. Still others may value it instrumentally, using it to assess whether they are underperforming and should adjust their strategy. These motives imply not only different willingness to receive peer information, but also different responses when information is provided. Understanding this heterogeneity is therefore central to the design of effective workplace feedback policies. 

This paper studies heterogeneity in worker preferences for peer information and examines whether such differences help explain how workers respond to it. We measure workers' willingness-to-pay (WTP) for receiving peer information and use these elicited preferences to identify distinct patterns of information demand. In particular, we ask how heterogeneous workers’ preferences for peer information are, what these differences imply for effort responses and worker welfare, and whether an incentivized measure of information demand can help identify when peer information is beneficial and when it may be counterproductive.

We develop a conceptual framework that links workers' demand for peer information to subsequent effort responses. The framework considers several motives for seeking or avoiding peer information: indifference, social comparison, stress avoidance, and instrumental learning. These motives generate distinct predictions for how workers value peer information when it is received before performing the task (ex ante) versus after completing it (ex post), and for how their effort should respond once information is delivered. We use these predictions as a disciplined way to organize heterogeneity in the data and to guide our empirical analysis.

We test these predictions in a preregistered real-effort experiment with 793 workers on Prolific. Workers complete the same task over two work periods under piece-rate incentives, with the first period serving as a baseline. Before the second period, we elicit each worker's WTP for learning the average performance of 100 prior workers who completed the same task, under two timing scenarios: ex ante or ex post. WTP is elicited for each possible realization of first-period performance using an incentive-compatible Becker-DeGroot-Marschak (BDM) mechanism \citep{becker1964measuring, butera2022measuring}. To preserve random assignment, WTP responses determine information receipt with a 10\% probability. Otherwise, workers are randomly assigned to a control group that receives no peer information, an ex ante information group that receives peer information before the second work period, or an ex post information group that receives peer information only after the second work period. Finally, we ask workers to provide open-ended explanations for their information choices in the ex ante and ex post scenarios.

We present four main sets of results. First, we document substantial heterogeneity in workers' preferences for peer information. Using WTP measures, we identify four distinct types of information demand. \textbf{Type 1 (indifferent)} workers (32\%) are indifferent to peer information, whether provided ex ante or ex post. \textbf{Type 2 (stress-avoidant)} workers (15\%) strictly avoid receiving information before the task but are indifferent to receiving it afterward. \textbf{Type 3 (competitive)} workers (23\%) value peer information more as relative performance increases, consistent with competitive motives. \textbf{Type 4 (learning-oriented/residual)} workers (30\%) exhibit positive WTP across performance levels and timing scenarios, suggestive of instrumental and other motives.

Second, this heterogeneity strongly predicts how workers respond to peer information. Average treatment effects of peer information mask substantial variation across workers. Type 1 (indifferent) and Type 2 (stress-avoidant) workers exhibit no change in effort from receiving information, whether ex ante or ex post. Type 3 (competitive) workers increase effort significantly by 8.7\% when information is provided ex ante, with a smaller, insignificant increase when it is provided ex post. Type 4 (learning-oriented/residual) workers also respond more strongly to ex ante than to ex post information. These patterns are broadly consistent with the predictions of the conceptual framework.

Third, we provide supporting evidence using workers' stated motives from free-form responses. Type 1 (indifferent) workers often express apathy toward peer information, while Type 2 (stress-avoidant) workers cite stress or distraction as reasons for avoiding information ex ante. Type 3 (competitive) workers frequently mention social comparisons or using the peer average as a target to beat. Type 4 (learning-oriented/residual) workers reference both learning and curiosity as key motivations for seeking information. Complementing this, a data-driven text analysis identifies two distinct clusters: one with a flat WTP profile and no effort response, and another with an upward-sloping WTP profile and strong effort responses. Together, these findings suggest that workers’ heterogeneous information preferences are not random noise, but reflect meaningful differences in how they interpret and use peer information.

Fourth, using WTP as a measure of worker welfare, our results imply that feedback should not be timed uniformly across workers. While peer information raises welfare on average, a uniform policy can impose negative payoffs on a sizable subset of workers who are willing to pay to avoid information before the task. A simple policy that provides peer information ex ante to most workers, but ex post for this Type 2 (stress-avoidant) group, yields welfare gains of up to 48\% relative to uniform provision.

Overall, this paper provides the first evidence that workers differ substantially in how they value peer information, and that these differences strongly predict how they respond to it. This highlights the importance of accounting for heterogeneity in information preferences, as average effects may reflect offsetting responses across worker types and lead to misleading conclusions about the effectiveness of peer information. Notably, we identify a sizeable subset of workers (15\% in our sample) who would pay to avoid peer information ex ante, cite stress or distraction as a reason for doing so, and show no productivity gains from receiving it. These findings underscore an important challenge for the design of performance feedback policies in organizations: information that motivates some workers may impose psychological costs on others. Beyond shaping effort in the short run, peer information may also affect worker retention in the longer run. As a result, uniform feedback policies can be suboptimal, and tailoring information provision to worker preferences can improve aggregate welfare. More broadly, we provide a framework for measuring heterogeneous information preferences and linking them to subsequent effort outcomes. Such tools can help organizations better predict how workers respond to peer information and design more effective feedback systems to improve worker outcomes.

The rest of the paper proceeds as follows. Section 2 discusses our contributions to the literature. Section 3 presents the theoretical framework. Section 4 describes the experimental design. Section 5 details the results. Section 6 concludes.

\section{Related Literature} \label{sec:literature}

First, this paper advances the literature on performance feedback and peer effects (for a recent review, see \citealp{villeval2020performance}). We speak to a large body of work documenting the impact of peer information (typically in the form of relative performance feedback) on worker effort and productivity \citep{hannan2008effects, eriksson2009feedback, blanes2011tournaments, barankay2011rankings, barankay2012rank, kuhnen2012feedback, charness2014dark, huet2015social, azmat2016provision, gjedrem2018relative, gill2019first, wu2021social, ashraf2025performance}. While many studies find modest positive effects on average, others report null or even negative effects. Our results provide a unifying explanation for these mixed findings by showing that workers differ systematically in whether they value peer information and, in turn, how they respond to it. We directly measure workers' demand for peer information and link it to subsequent behavior, providing new evidence on how information preferences shape effort responses. Without accounting for this heterogeneity, estimated average effects can mask offsetting effects across workers, leading to misguided conclusions about the effectiveness of peer information.

Closest in spirit to our paper is \citet{senn2023leveraging}, who study workers' preferences for social comparisons by allowing them to choose whether and whom to compare themselves to.\footnote{In their main treatment (ENDO), workers could choose whether to observe a reference worker of high, average, or low productivity, or not at all during a real-effort task.} They find that endogenous peer choice improves productivity as much as targeted peer assignment, but with lower stress. While we share their focus on endogenous choice, our study differs in both scope and method. We focus on whether workers have a significant demand for peer information and develop a theoretical framework to identify distinct types of information preferences based on WTP across two timing scenarios. Our framework is complementary to the network literature on peer effects, where own and peer actions are jointly determined in equilibrium (e.g., \citealp{boucher2024toward}). We instead adopt a partial equilibrium approach and take peer performance as given so as to isolate the informational channel through which peer information affects worker behavior.

Our design also advances the experimental paradigm used in the performance feedback literature in two key respects. First, we employ a minimal form of peer information, providing only average performance rather than full rankings, and show that even simple comparative feedback can influence worker behavior. Second, we explicitly vary the timing of information by comparing feedback delivered before versus after task completion. With the exception of \citet{kuhnen2012feedback}, most studies examine feedback provided only before the task (i.e., ex ante in our terminology).\footnote{In \citet{kuhnen2012feedback}, subjects are told at the start of the first round whether they will receive feedback about their ranking afterward, which corresponds to the ex post treatment in our terminology. This allows comparison of ex post feedback vs. no information, but only in the first round. In subsequent rounds, feedback precedes the task, so the estimated effects correspond to receiving information ex ante. Therefore, the estimated ex ante and ex post effects may not be directly comparable due to potential round effects.} Our design therefore allows a direct comparison of ex ante and ex post effects and shows that effort responses are stronger when peer information is provided before the task.

Next, our paper contributes to a growing literature on the welfare effects of behavioral interventions by eliciting individuals' WTP to capture non-monetary costs and benefits \citep{allcott2019welfare, butera2022measuring, andor2023differences}. We build on \citet{butera2022measuring}, who introduce the ``strategy-method'' approach to elicit WTP for public recognition and study its welfare effects. Adapting this approach to a workplace setting, we elicit WTP for peer information and use these measures in a theory-driven way to document distinct patterns of information demand. This allows us to quantify heterogeneous welfare effects and to examine how tailoring information provision across workers can improve aggregate welfare relative to uniform provision.

Finally, our paper contributes to an emerging literature that uses open-ended survey questions to study mechanisms underlying economic behavior (for a recent review, see \citealp{haaland2025understanding}). Recent work uses survey responses to understand what shapes productivity at work \citep{abeler2023incentive, senn2023leveraging, kaur2025financial}.\footnote{See in particular \cite{senn2023leveraging}, who analyze workers' free-form responses to examine their motives for choosing which peer to compare themselves to. They document heterogeneity in preferences, with most reporting a desire to motivate themselves by observing a reference worker, and a smaller fraction citing stress or distraction as reasons to avoid observing one.} First, we show that workers' open-ended explanations strongly predict their information preferences and effort responses, highlighting the value of qualitative data. Second, we apply a fully data-driven approach to analyze the text data, embedding responses using a pre-trained language model (BERT) and applying clustering techniques to uncover latent groups of workers with similar information preferences \citep{devlin2019bert, subakti2022performance}.\footnote{This contrasts with recent studies that leverage large language models, such as GPT-4, to annotate text data \citep{arrieta2024procedural, bordalo2023people, bursztyn2023product}. While more efficient than manual coding, such approaches typically require researchers to specify a coding scheme and provide example responses (i.e., few-shot prompting), which may introduce bias through researcher discretion over category definitions.} This approach provides further support for the heterogeneity patterns and demonstrates how qualitative data can help uncover differences in behavioral responses.

\section{Theoretical Framework} \label{sec:theory}

We develop a unified framework to distinguish between two broad classes of mechanisms through which peer information may affect worker behavior. In the first, peer information directly enters utility by activating nonstandard preferences (e.g., social preferences or psychological costs). In the second, peer information has instrumental value by changing beliefs through learning. Each model delivers testable predictions for workers' WTP for peer information and their effort responses.

In our setting, the worker chooses an effort level $e \in \R_{+}$ to complete a real-effort task and is paid a piece rate $w>0$ per unit of effort. Utility is assumed to be separable in the consumption utility from earnings and the disutility from effort, taking the form of:
\begin{align}
    U(e) = m(we) - c(e), 
\end{align}
where $m(we)$ denotes the consumption utility from earning $we$, and $c(e)$ denotes the cost of effort that is strictly increasing and convex: $c'(e)>0$ and $c^{''}(e)>0$ for all $e>0$. 

For tractability, we impose two functional form assumptions throughout. First, consumption utility is linear in earnings, so $m(we) = we$. Second, the cost of effort takes a quadratic form, $c(e) = \frac{c}{2} e^2$, for some cost parameter $c>0$, following \citet{butera2022measuring}.

In each model below, we consider two scenarios that differ in the timing of information provision. In the ex ante scenario, the worker receives peer information about the average effort level $\bar{e}$ before choosing effort. In the ex post scenario, the worker is informed that they will receive the information only after the task, so they must choose effort before learning $\bar{e}$. We treat $\bar{e}$ as exogenous to the worker's decision, focusing on partial equilibrium responses to peer information.\footnote{In the experiment, $\bar{e}$ is constructed from a reference group of prior workers whose performance can no longer be changed, ensuring that peer information is exogenous.}

\subsection{Standard Model}

We begin with a benchmark model in which the worker has standard preferences. Absent peer information, the worker chooses effort to maximize earnings net of effort costs:
\begin{align}
    e_{no-info} = \argmax_{e \in \R_{+}} \left\{ 
     w e - \frac{c}{2} e^2 \right\} = \frac{w}{c},
\end{align}
with the indirect utility given by $V_{no-info}(c) \equiv U(e_{no-info}) = \frac{w^2}{2c}$.  

In this model, peer information neither changes the worker's preferences nor beliefs. Whether information is provided ex ante or ex post, the worker's utility, and hence their optimal effort, remains unchanged: $e_{info, s} = \argmax_{e \in \R_{+}} \left\{w e - \frac{c}{2} e^2 \right\} = \frac{w}{c}$, for $s \in \{ex ante, ex post\}$. The corresponding indirect utility in each case is $V_{info, s}(c) \equiv U(e_{info, s}) = \frac{w^2}{2c}$. 

We allow the worker to be uncertain about their performance relative to others by assuming they hold a belief distribution $p(\bar{e})$ over possible average effort levels $\bar{e} \in \R_{+}$. In the standard model, however, this uncertainty plays no role, since utility depends only on own effort and not on the peer average.

Accordingly, we define the worker's WTP for peer information under each timing scenario $s$ as the difference in utility with and without information:
\begin{align}
    WTP_s(c) \equiv V_{info, s}(c) - V_{no-info}(c), \quad s \in \{ex ante, ex post\}
\end{align}

In the experiment, we elicit $WTP(c)$ conditional on the worker's baseline effort without information, $e_{no-info} = \frac{w}{c}$, which itself is a function of the cost parameter $c$. We summarize the testable predictions below: 

\bigskip
\noindent \textbf{HYPOTHESIS 1.1} (Information Preferences): \textit{Under the standard model, the worker's WTP for peer information is zero in both the ex ante and ex post scenarios and does not vary with baseline effort $e_{no-info}$. 
}

\bigskip
\noindent \textbf{HYPOTHESIS 1.2} (Effort Response):
\textit{Under the standard model, peer information has no effect on effort in either scenario: $e_{info, exante} = e_{info, expost} = e_{no-info}$.}
\bigskip

While the \textit{standard model} offers a useful benchmark, it assumes that workers care only about their own earnings and are unaffected by peers. We next relax this assumption and consider a class of models in which peer information enters utility through relative performance concerns.

\subsection{Social Preferences Model}

The \textit{social preferences model} posits that the provision of peer information primes workers to compare their own effort or performance to others (e.g., the average), thereby activating social preferences such as competitiveness \citep{azmat2010importance, charness2014dark} or inequality aversion \citep{fehr1999theory,bolton2000erc}. We model these preferences flexibly by incorporating an additional term into the worker's utility that depends on relative earnings, following prior work on workplace social comparisons \citep{dellavigna2022estimating, breza2018morale, cullen2022much}. 

{\centering \subsubsection*{\textit{A. Ex ante scenario}}}
If the worker receives peer information before the task, we assume their utility upon learning the average effort level $\bar{e}$ is given by:  
\begin{align}
\widetilde{U}(e; \bar{e}) = m(w e) - c(e) + f(we - w\bar{e}), \label{eq:eq1}
\end{align}
where $f(\cdot)$ captures social preferences and depends on the difference between the worker's own earnings and the average earnings.

For tractability, we specify $f(\cdot)$ as a piecewise linear function:  
$f(x) = \mathds{1}_{\{x \leq 0 \}} \cdot \lambda_1 x + \mathds{1}_{\{x>0\}} \cdot \lambda_2 x,
$
where the parameters $\lambda_1$ and $\lambda_2$ capture the (possibly asymmetric) intensity of social preferences when the worker earns less or more than the average, respectively. This specification nests two key cases: (i) competitiveness or status concerns, with $\lambda_1 \geq \lambda_2 > 0$; and (ii) inequality aversion, with $\lambda_1>0$, $-1<\lambda_2<0$, with $\lambda_1 \geq |\lambda_2|$. 

Maintaining the same functional form assumptions, the worker's optimal effort is:
\begin{align}
e_{info,exante} = \argmax_{e \in \R_{+}} \left\{ w e - \frac{c}{2}e^{2} + \mathds{1}_{\{e \leq \bar{e} \}} \cdot \lambda_1 (we - w\bar{e}) + \mathds{1}_{\{e > \bar{e} \}} \cdot \lambda_2 (we - w\bar{e}) \right\},
\end{align}
with the indirect utility given by $V_{info, exante}(c;\bar{e}) \equiv \widetilde{U}(e_{info,exante}; \bar{e})$.

We assume that without peer information, social preferences are not activated and the worker's utility reduces to the standard model. While this may be a strong assumption, relaxing it to allow weaker (but non-zero) social preferences without peer information yields similar qualitative predictions (see Appendix C.8 for details).

As before, the worker holds beliefs $p(\bar{e})$ over possible average effort levels. Their WTP for peer information delivered ex ante is then defined as:
\begin{align}
WTP_{exante}(c) \equiv \E_p[V_{info, exante}(c;\bar{e})] - V_{no-info}(c),
\label{eq:wtp_exante}
\end{align}
where $\E_p[\cdot]$ denotes the expectation with respect to the worker's beliefs $p(\bar{e})$, and $V_{no-info}$ is the indirect utility previously derived in the \textit{standard model}.

Based on this setup, we derive predictions for the worker's WTP and effort response under two types of social preferences: (i) competitive preferences ($\lambda_1 \geq \lambda_2 > 0$), and (ii) inequality-averse preferences ($\lambda_1>0$, $-1<\lambda_2<0$, with $\lambda_1 \geq |\lambda_2|$). 

The first part of the hypothesis reflects that a competitive worker values peer information more when they expect to outperform others, whereas the second reflects that an inequality-averse worker avoids such information as it induces disutility whether they are above or below average.

\bigskip
\noindent \textbf{HYPOTHESIS 2.1} (Information Preferences):
\textit{Under the social preferences model, the worker's WTP for peer information in the ex ante scenario, $WTP_{exante}$, depends on the type of social preferences. For any belief distribution $p(\bar{e})$:}
\begin{itemize}
\item[(i)] \textbf{Competitive preferences:} $WTP_{exante}$ is increasing in the worker's baseline effort under no information, $e_{no-info}$.
\item[(ii)] \textbf{Inequality-averse preferences:} $WTP_{exante}$ is negative and single-peaked in the worker's baseline effort under no information, $e_{no-info}$.
\end{itemize}

The next hypothesis considers how receiving peer information ex ante affects effort. For competitive preferences, the result follows directly from $\lambda_1, \lambda_2 > 0$: receiving peer information raises the marginal benefit of effort, inducing higher effort. For inequality-averse preferences, the asymmetry between $\lambda_1>0$ and $\lambda_2<0$ implies that the marginal benefit of effort is greater when the worker is performing below average, and lower when the worker is above average. As a result, effort adjusts towards the average, leading to ``bunching'' at $\bar{e}$. 

\bigskip
\noindent \textbf{HYPOTHESIS 2.2} (Effort Response):
\textit{Under the social preferences model, the worker's effort response depends on the type of social preferences. Compared to no peer information:}
\begin{itemize}
\item[(i)] \textbf{Competitive preferences:} The worker will exert greater effort when they receive peer information ex ante: $e_{info, exante} > e_{no-info}$.
\item[(ii)] \textbf{Inequality-averse preferences:} The worker will choose an effort level that is weakly closer to the average $\bar{e}$ with peer information ex ante: $|e_{info,exante} - \bar{e}| \leq |e_{no-info} - \bar{e}|$.
\end{itemize}

{\centering \subsubsection*{\textit{B. Ex post scenario}}}

If the worker is informed that they will receive peer information only after completing the task, we assume they hold the same type of social preferences as in the ex ante scenario. The key distinction is that the worker must now choose their effort without knowing the average $\bar{e}$, and thus maximizes expected utility over their beliefs $p(\bar{e})$. 

Accordingly, the worker's optimal effort is given by:
\begin{align}
    e_{info,expost} &= \argmax_{e \in \R_{+}} \left\{ 
    w e - \frac{c}{2}e^2 + \delta \int_{\R_{+}} \left[ \lambda_1 w (e - \bar{e}) \cdot  \mathds{1}{\{e \leq \bar{e}\}} +  \lambda_2 w (e-\bar{e})\cdot \mathds{1}{\{e>\bar{e}\}} \right] \cdot p(\bar{e}) d\bar{e}
    \right\},
\end{align} 
where the additional parameter $\delta \geq 0$ allows the intensity of social preferences to vary with the timing of information.\footnote{For instance, $\delta=1$ corresponds to the same intensity as in the ex ante scenario, while $\delta<1$ or $\delta>1$ allow for weaker or stronger responses, respectively.} The indirect utility is $V_{info, expost}(c; \delta) \equiv \E_p[\widetilde{U}(e_{info,expost}; \bar{e}, \delta)]$, and the worker's WTP for peer information delivered ex post is defined analogously as:
\begin{align}
WTP_{expost}(c; \delta) = V_{info, expost}(c; \delta) - V_{no-info}(c)
\label{eq:wtp_expost}
\end{align}

We first derive predictions for WTP in the ex post scenario, relative to the ex ante case. Under the \textit{social preferences model}, peer information is more valuable ex ante, as it can be incorporated into the worker's decision to better optimize effort. 

\bigskip
\noindent \textbf{HYPOTHESIS 2.3} (Information Preferences – Ex ante vs. Ex post): \textit{Assume $\delta=1$, i.e., the strength of social preferences is the same in both the ex ante and ex post scenarios. Then, for any type of social preferences $f(\cdot)$, the worker's WTP for peer information is weakly lower when it is provided ex post rather than ex ante: $WTP_{exante} \geq WTP_{expost}$.}
\bigskip

We next consider how receiving peer information ex post affects effort for each type of social preferences, maintaining the assumption of $\delta=1$. For competitive preferences, we do not obtain a general result comparing effort between the ex ante and ex post scenarios, as it depends on the worker's beliefs $p(\bar{e})$.\footnote{Intuitively, the more probability mass the worker places on higher values of $\bar{e}$, the greater the expected marginal benefit of effort, and hence the higher the optimal effort in the ex post scenario.} However, we can compare effort under the ex post scenario to the no-information case: since $\lambda_1, \lambda_2 > 0$, the anticipation of receiving peer information increases the marginal value of effort, inducing higher effort. For inequality-averse preferences, we can compare effort under the ex ante and ex post scenarios. The second part of the hypothesis reflects that effort in the ex post case exhibits less ``bunching'' since the worker does not know $\bar{e}$ at the time of decision and thus cannot condition on it. 

\bigskip
\noindent \textbf{HYPOTHESIS 2.4} (Effort Response):
\textit{Under the social preferences model, the worker's effort response depends on the type of social preferences.}
\bigskip
\begin{itemize}
\item[(i)] \textbf{Competitive preferences:} The worker will exert greater effort when they receive peer information ex post compared to without information: $e_{info, expost} > e_{no-info}$.
\item[(ii)] \textbf{Inequality-averse preferences:} The worker will choose an effort level that is farther from the average $\bar{e}$ compared to the ex ante case: $|e_{info,expost} - \bar{e}| \geq |e_{info,exante} - \bar{e}|$.
\end{itemize}

In sum, the \textit{social preferences model} captures nonstandard preferences arising from comparisons between a worker's own effort (or earnings) and those of their peers. However, peer information may also affect workers through a direct affective channel, inducing stress or anxiety independent of relative performance, which we formalize in the next model.

\subsection{Stress Avoidance Model}

The \textit{stress avoidance model} posits that peer information imposes psychological costs, such as stress arising from implicit pressure to meet a perceived standard or distraction from the task.\footnote{This is supported by recent evidence that social comparisons can negatively affect psychological well-being \citep{senn2023leveraging, buunk2017social, barcena2017social}.} We model stress as a direct utility cost incurred whenever peer information is provided, independent of the worker's own effort. This formulation is consistent with evidence from psychology and neuroscience showing that stress elicits negatively valenced affective states associated with distress and reduced well-being \citep{lazarus1984stress, hammen2005stress, duman2012synaptic}.

{\centering \subsubsection*{\textit{A. Ex ante scenario}}}
If the worker receives peer information before the task, their utility is given by:  
\begin{align}
\widehat{U}(e; \bar{e}) = m(w e) - c(e) - \Theta(\bar{e}), 
\end{align}
where $\Theta(\cdot)$ represents the stress component, assumed to be weakly increasing in $\bar{e}$ with $\Theta(0) = 0$. This reflects that higher peer performance induces greater psychological pressure.

Under the same quadratic effort cost function, the worker's optimal effort is:
\begin{align}
e_{info,exante} = \argmax_{e \in \R_{+}} \left\{ w e - \frac{c}{2}e^{2} - \Theta(\bar{e}) \right\}
\end{align}
The indirect utility is given by $V_{info, exante}(c;\bar{e}) \equiv \widehat{U}(e_{info,exante}; \bar{e})$, and WTP ex ante is defined analogously to \eqref{eq:wtp_exante}. We immediately arrive at the two testable predictions below:

\bigskip
\noindent \textbf{HYPOTHESIS 3.1} (Information Preferences): \textit{Under the stress avoidance model, for any beliefs $p(\bar{e})$ and any stress function $\Theta(\cdot)$, the worker's WTP for peer information delivered ex ante, $WTP_{exante}$, is strictly negative and does not vary with baseline effort $e_{no-info}$.}

\bigskip
\noindent \textbf{HYPOTHESIS 3.2} (Effort Response):
\textit{Under the stress avoidance model, the worker will exert the same effort when peer information is provided ex ante as in the case without information: $e_{info, exante} = e_{no-info}$.}

{\centering \subsubsection*{\textit{B. Ex post scenario}}}
In the ex post scenario, the worker knows they will receive peer information only after completing the task. While the information is not yet available when choosing effort, its anticipated arrival may still impose stress, though likely weaker than in the ex ante case. As in the \textit{social preferences model}, the worker maximizes expected utility over their beliefs $p(\bar{e})$. 

Accordingly, the worker's optimal effort is given by:
\begin{align}
    e_{info,expost} &= \argmax_{e \in \R_{+}} \left\{ 
    w e - \frac{c}{2}e^2 - \delta \int_{\R_{+}} \Theta(\bar{e}) \cdot p(\bar{e}) d\bar{e}
    \right\}, 
\end{align} 
where $\delta \in [0,1]$ captures the intensity of stress from anticipating peer information ex post relative to ex ante.\footnote{When $\delta=1$, the utility cost from stress is the same as in the ex ante case. When $\delta=0$, the worker is completely unaffected by the prospect of receiving peer information after the task, and the model reverts to the \textit{standard model}.} The indirect utility is $V_{info, expost}(c; \delta) \equiv \E_p [\widehat{U}(e_{info,expost}; \bar{e})]$, and WTP ex post is defined analogously to \eqref{eq:wtp_expost}.

We similarly derive two hypotheses comparing information preferences and effort to the ex ante case. The first follows directly from $\delta \leq 1$, while the second mirrors Hypothesis 3.2.

\bigskip
\noindent \textbf{HYPOTHESIS 3.3} (Information Preferences – Ex ante vs. Ex post): \textit{For any stress function $\Theta(\cdot) \geq 0$, the worker's WTP for peer information is weakly higher when it is provided ex post rather than ex ante: $WTP_{exante} \leq WTP_{expost}$.}

\bigskip
\noindent \textbf{HYPOTHESIS 3.4} (Effort Response):
\textit{Under the stress avoidance model, the worker will exert the same effort when they receive peer information ex post as when it is provided ex ante: $e_{info, expost} = e_{info, exante}$.}
\bigskip

The above two models assume peer information affects utility through nonstandard preferences. In contrast, the next model retains standard preferences but introduces uncertainty over strategy productivity. In this case, peer information can have instrumental value by helping workers update their beliefs and decide whether to search for a better task strategy.

\bigskip 

\subsection{Learning Model}

The \textit{learning model} posits that peer information affects behavior by helping workers learn and adopt more effective strategies for the task. We consider a worker who is uncertain about whether they are currently using the most effective strategy. Let the strategy space be $\mathcal{S}$, where each strategy $s \in \mathcal{S}$ is associated with a productivity parameter $\alpha_s \in [\underline{\alpha}, \bar{\alpha}]$, which determines how effort translates into output via a linear production function: $f(e;s) = \alpha_s e$.\footnote{This formulation generalizes the previous models, which implicitly assumed a single strategy with productivity $\alpha = 1$, so that effort and output coincide (i.e., $f(e) = e$).}

The worker begins with a baseline strategy $s$, associated with productivity $\alpha_s > 0$. The worker can learn alternative strategies $s' \in \mathcal{S}$, which are characterized by productivity levels $\alpha_{s'} \in [\underline{\alpha}, \bar{\alpha}]$. However, identifying a new strategy requires incurring a fixed cost $K>0$, which represents search or experimentation costs. If the worker pays this cost and acquires a new strategy $s'$, they adopt it if only if it improves productivity (i.e., $\alpha_{s'} > \alpha_s$); otherwise, they revert to their current strategy $s$.

Absent peer information, we assume the worker retains their current strategy $s$ and does not search.\footnote{Formally, this assumption can be justified if the worker holds a prior belief distribution over $\alpha$ that is sufficiently concentrated near $\alpha_s$, or if the fixed cost of searching $K$ is sufficiently large, such that the expected utility gain from searching is negative.} Under the same functional form assumptions, the worker's optimal effort is:
\begin{align}
    e_{no-info} = \argmax_{e \in \R_{+}} \left\{ w \alpha_s e - \frac{c}{2} e^2 \right\} = \frac{w \alpha_s}{c},
\end{align}
with corresponding indirect utility given by $V_{no-info}(c) \equiv U(e_{no-info}; s) = \frac{w^2 \alpha_s^2}{2c}$.

{\centering \subsubsection*{\textit{A. Ex ante scenario}}}

In the ex ante scenario, the worker chooses whether to receive peer information about the average output $\bar{y}$ of other workers. Observing $\bar{y}$ leads the worker to update their beliefs about the productivity levels $\alpha$ associated with alternative strategies. 

Let $F(\alpha | \bar{y})$ denote the worker's (posterior) belief distribution over $\alpha$ conditional on observing $\bar{y}$, which is continuously differentiable in $\alpha$ and admits a density function $f(\alpha | \bar{y})$. We assume that higher average peer output $\bar{y}$ shifts beliefs toward higher productivity levels. Formally, $F(\alpha | \bar{y})$ first-order stochastically decreases in $\bar{y}$, i.e. $\bar{y}' > \bar{y} \implies F(\alpha| \bar{y}') \leq F(\alpha| \bar{y}), \forall \alpha$.

The worker’s expected utility from searching for a new strategy after observing $\bar{y}$ is:
\begin{align*}
    V_{search}(c;\bar{y}) = \int_{\alpha_s}^{\bar{\alpha}} \frac{w^2 \alpha^2}{2c} dF(\alpha | \bar{y}) + F(\alpha_s | \bar{y}) \cdot \frac{w^2 \alpha_s^2}{2c} - K
\end{align*}
That is, with probability $1 - F(\alpha_s | \bar{y})$, the new strategy is better and yields higher utility; with probability $F(\alpha_s | \bar{y})$, the worker retains the current strategy. 

The worker compares this expected value to the baseline utility without information and chooses to search if $V_{search}(c;\bar{y}) \geq V_{no-info}(c)$. The maximized utility from receiving peer information is therefore:
\begin{align*}
    V_{info}(c;\bar{y}) = \max\left\{ V_{search}(c;\bar{y}), V_{no-info}(c) \right\}
\end{align*}

Finally, let $p(\bar{y})$ denote the worker's belief distribution over the average output level $\bar{y}$.\footnote{For tractability, we assume the worker's beliefs about $\bar{y}$ is independent of their current strategy $s$.} The worker's WTP for peer information delivered ex ante is similarly defined as in \eqref{eq:wtp_exante}:
\begin{align*}
    WTP_{ex ante}(c) \equiv \E_p [V_{info}(c;\bar{y})] - V_{no-info}(c)
\end{align*}

From this setup, we derive the following two testable predictions:

\bigskip
\noindent \textbf{HYPOTHESIS 4.1} (Information Preferences):
\textit{Under the learning model, the worker's WTP for peer information in the ex ante scenario, $WTP_{exante}$, is positive and decreasing in their baseline effort under no information, $e_{no-info}$.}

\bigskip
\noindent \textbf{HYPOTHESIS 4.2} (Effort Response):
\textit{Under the learning model, the worker will exert greater effort when they receive peer information ex ante compared to the case without information. That is, $e_{info, exante} \geq e_{no-info}$.}
\bigskip

{\centering \subsubsection*{\textit{B. Ex post scenario}}}

If the worker receives peer information ex post, then it cannot be used to update their beliefs about the productivity $\alpha$ of alternative strategies prior to choosing effort. The worker thus retains their baseline strategy and chooses the same effort as in the no-information case:  
\begin{align}
    e_{info, expost} = \argmax_{e \in \R_{+}} \left\{ w \alpha_s e - \frac{c}{2} e^2 \right\} = \frac{w \alpha_s}{c}.
\end{align}

That is, peer information has no instrumental value when it arrives after the effort decision has already been made. We thus arrive at the following two predictions: 

\bigskip
\noindent \textbf{HYPOTHESIS 4.3} (Information Preferences – Ex ante vs. Ex post): \textit{Under the learning model, the worker's WTP for peer information is lower when it is provided ex post rather than ex ante, for any prior beliefs over the productivity parameter $\alpha$. Specifically, $WTP_{expost} = 0$ and $WTP_{exante} \geq WTP_{expost}$.}

\bigskip
\noindent \textbf{HYPOTHESIS 4.4} (Effort Response):
\textit{Under the learning model, the worker will exert the same effort when peer information is provided ex post as in the no-information case: $e_{info, expost} = e_{no-info}$.}
\bigskip

\subsection{Summary of Predictions}

In summary, the \textit{standard model} provides a benchmark in which peer information is not valued by workers and has no effect on their effort. The other models depart from it by changing either the worker's preferences or beliefs: the \textit{social preferences} and \textit{stress avoidance} models incorporate nonstandard preferences, while the \textit{learning model} retains standard preferences but introduces belief updating about strategy effectiveness through search and learning. Table \ref{tab:model-predictions} summarizes the key predictions of each model, which we test experimentally.

\medskip 
\textit{Conceptual Note.}---Throughout, our framework assumes that beliefs about peers' performance, $p(\bar{e})$, are acted upon only when peer information is provided (whether ex ante or ex post). We view this as a reasonable assumption as absent peer information, workers perform the task individually for a piece rate and are unlikely to think about relative performance unless such information is explicitly presented. Accordingly, in the no-information case, we model utility such that beliefs about peer average and the nonstandard utility component (i.e., social preferences or stress) do not enter the worker’s decision-making process. Notwithstanding, in Appendix C.8, we relax this assumption by allowing beliefs to play a role even when peer information is not provided. This extended formulation preserves all predictions presented above and further predicts how belief precision affects WTP independently of worker type.

{
\singlespacing
\begin{table}[ht]
\centering 
\caption{Summary of Theoretical Predictions}
\label{tab:model-predictions}
\begin{threeparttable}
\resizebox{\textwidth}{!}{%
\begin{tabular}{llcccc}
\toprule
\toprule
& & \multicolumn{2}{c}{\textbf{WTP}} & \multicolumn{2}{c}{\textbf{\shortstack{Effort Response \\ (relative to no-info case)}}} \\
\cmidrule(lr){3-4} \cmidrule(lr){5-6} 
& \textbf{Model} & \textbf{Ex ante} & \textbf{Ex post} & \textbf{Ex ante} & \textbf{Ex post} \\
\midrule \addlinespace

1. & \textbf{Standard} & Zero & Zero & No change & No change \\



\midrule

2a. & \textbf{\shortstack[l]{Social Preferences: \\ Competitive}} & Increasing in $e_{no-info}$ & $\leq WTP_{exante}$ & Increases & Increases \\

\midrule

2b. & \textbf{\shortstack[l]{Social Preferences: \\ Inequality-Averse}} & \shortstack{Negative; \\ Single-peaked} & $\leq WTP_{exante}$ & Bunches at $\bar{e}$ & Depends on beliefs \\

\midrule

3. &  \textbf{Stress Avoidance} & \shortstack{Negative; \\ Independent of $e_{no-info}$} & $\geq WTP_{exante}$ & No change & No change \\

\midrule

4. &  \textbf{Learning} & \shortstack{Positive; \\ Decreasing in $e_{no-info}$} & Zero & Increases & No change \\

\bottomrule
\end{tabular}
}

\begin{tablenotes}
\scriptsize\vspace{0.1cm} 
\parbox{.95\textwidth}{
\item \emph{Notes.} This table summarizes each model's predictions for the worker's WTP and its impact on effort, depending on whether peer information is provided ex ante or ex post. $e_{no-info}$ denotes the worker's effort choice without peer information. For the social preferences model, we maintain the assumption that $\delta=1$. 
}
\end{tablenotes}
\end{threeparttable}
\end{table}
}

\section{Experimental Design} \label{sec:experiment}

\subsection{Design Overview}

In the experiment, workers perform an effort-intensive task that requires them to deduce the missing number in a row of numbers based on the underlying numerical pattern. Workers receive a fixed participation wage of \$4, plus a piece rate of 1 cent for each correct answer and no payment for incorrect answers. After providing informed consent, workers review the task instructions and complete a practice round lasting up to 30 seconds. They have to perform the task for two consecutive work periods (Period 1 and Period 2), each lasting up to 5 minutes, with the option to end early.\footnote{This follows previous work documenting that labor supply in real-effort settings is more elastic when participants can choose the `extensive margin' of their effort \citep{dellavigna2018motivates, dellavigna2022estimating, butera2022measuring}.} Importantly, workers are only informed about the part of the experiment they are currently completing and are not told in advance that there will be a second work period. The experiment proceeds as follows:

\vspace{3mm}
\noindent
\textbf{Part 1 (Period 1):} Workers complete the real-effort task without receiving any feedback on their performance. Afterward, we elicit measures of their experience with the task and well-being (i.e., stress, motivation, and perceived meaning of work) using a 0--10 scale. Performance in this period provides a clean baseline measure of each worker’s productivity before any treatment assignment.

\vspace{3mm}
\noindent
\textbf{Part 2 (Belief Elicitation):} 
Before learning about the second work period, workers report their prior beliefs about the performance of other Prolific workers. Specifically, they are asked to provide their best guess of how likely (percent chance) it is that the average performance falls within each of the following eight possible ranges: 0-10 rows, 11-20 rows, ..., 61-70 rows, and 70+ rows. This part is unincentivized. 

\vspace{3mm}
\noindent
\textbf{Part 3 (Peer Information \& WTP Elicitation):} After reporting their beliefs, workers are told they will perform the task again and will receive real-time feedback on their performance this time. Before proceeding, they are informed that a previous version of the study was conducted and are given the opportunity to receive peer information about how these 100 prior workers performed over the same 5-minute period. Specifically, they can learn the average performance (i.e., average number of rows solved) of this ``reference population.''

Workers are then presented with two scenarios in which they decide whether to receive peer information. In the ex ante scenario, information about average performance is provided before they begin the second work period. In the ex post scenario, the same information is provided only after they have completed the second work period. The only difference between the two is the timing of information receipt, if workers choose to receive it. 

We use an incentive-compatible BDM procedure to elicit workers' WTP for peer information, conditional on every possible realization of their performance in the first work period \citep{becker1964measuring, butera2022measuring}. For each scenario, ex ante and ex post, WTP is elicited for nine possible performance levels, defined relative to the average: 20+ rows below average, 11-20 rows below average, 6-10 rows below average, 2-5 rows below average, within 1 row of average, 2-5 rows above average, 6-10 rows above average, 11-20 rows above average, and 20+ rows above average.\footnote{A potential concern with using the strategy method to elicit WTP is that workers may have been less attentive when answering questions associated with performance levels that were outside the range of what they thought were likely \citep{butera2022measuring}. To mitigate this issue, we frame performance in relative terms rather than absolute values. This approach makes it more likely that workers assign a nonzero probability to their performance falling within each specified range, ensuring that the elicitation is incentive-compatible for all questions.}

Workers are first asked: ``If your earlier performance is [e.g., 2–5 rows below average], do you want information about the average performance?'' After choosing Yes or No, they then indicate how much of their 50-cent bonus they would be willing to use to ensure their preferred choice is implemented. Specifically, they respond to: ``If your earlier performance is [e.g., 2–5 rows below average], how much of your 50-cent bonus would you be willing to use to receive (or not receive) information about the average performance?'' This elicitation is implemented for both scenarios, with clear instructions on whether the information would be provided before or after the task in Period 2.

\vspace{3mm}
\noindent
\textbf{Part 4 (Treatment Assignment):} To preserve random assignment, workers are informed that their WTP would determine their assignment with a 10\% chance. For the remaining chance, assignment is random. Workers are assigned to one of four experimental groups: 
\begin{enumerate}
    \item \textbf{Control group} (with 30\% chance): No peer information is provided at any point.
    \item \textbf{Ex ante Info group} (with 30\% chance): Workers receive information about average performance before starting the task in Period 2. 
    \item \textbf{Ex post Info group} (with 30\% chance): Workers are told they will receive information about average performance, but only after completing the task in Period 2.
    \item \textbf{Choose-Your-Info group} (with 10\% chance): Assignment is determined using the BDM procedure, based on workers' WTP for the question matched to their actual Period 1 performance.
\end{enumerate}

For workers in the Choose-Your-Info group, one of the two scenarios is randomly selected as the ``scenario-that-counts,'' and their actual Period 1 performance is matched to the corresponding WTP question to determine whether they receive peer information in that scenario (ex ante or ex post). Specifically, with 50\% chance, their Yes/No choice in the matched question is implemented directly, and they keep the full bonus. Otherwise, the BDM procedure is applied: if the randomly drawn amount is less than or equal to their stated WTP, their choice is implemented and they pay that amount out of their bonus; otherwise, their choice is not implemented and they keep the full bonus. This ensures that it is in workers' best interests to truthfully report their preferences.

\vspace{3mm}
\noindent
\textbf{Part 5 (Period 2):} Workers perform the task a second time for up to 5 minutes, following the same structure as Period 1, this time with real-time feedback on their own performance. Those in the Ex ante Info group also see the average performance while completing the task. Upon completion, we re-elicit the same measures as in Period 1.

\vspace{3mm}
\noindent
\textbf{Part 6 (Endline Survey):} Before learning their total earnings, workers complete a survey collecting information on demographics (e.g., gender, age, and education) and personality traits (e.g., competitiveness, neuroticism, and risk-taking). They also provide open-ended explanations for their choices in the ex ante and ex post scenarios, explaining why they choose to receive or not receive peer information at each contingent performance level.

\vspace{3mm}
With our design, one concern is that sophisticated workers might strategically game the WTP elicitation to infer their own Period 1 performance rather than reveal their true valuation of peer information. For example, a worker who always bids their full bonus to avoid information if they are below average and to receive it if above average could then infer their relative performance based on whether they receive peer information. We view this as unlikely for three reasons. First, there is only a 10\% chance their WTP will be implemented. Assignment to peer information is otherwise random, preventing inference from information receipt. Second, even if workers attempted such inference, it would require highly strategic WTP responses across nine contingencies per scenario,\footnote{Indeed, we find no evidence of such strategic WTP bids in our data.} and still be unlikely to pin down any single performance range. Third, workers receive real-time feedback on their performance in Period 2 and can easily deduce their Period 1 performance from their bonus post-study, further reducing any incentive to game the elicitation.

\subsection{Implementation and Experimental Sample}
 
We programmed the experiment using oTree \citep{chen2016otree} and pre-registered the design on AsPredicted.org (\#210555).\footnote{Pre-registration link: https://aspredicted.org/5hwf-4967.pdf} The main study was conducted on Prolific in February 2025, where we recruited 693 workers.\footnote{We initially targeted recruitment of 700 workers, as pre-registered, but fell slightly short because a few participants submitted incorrect identifiers or completion codes, preventing their responses from being matched to our database.} Prior to this, we collected data from an initial 100 workers who formed the ``reference population'' used to construct the peer information.\footnote{These 100 workers completed a similar version of the experiment, performing the same real-effort task for a piece rate over a 5-minute period. Peer information was based on their performance in the first work period.}

We recruited workers who (i) reside in the US, (ii) have completed at least 100 prior studies on Prolific, and (iii) have an approval rating of at least 95\%. To check understanding, workers are required to answer comprehension questions correctly before they can advance to the main study sections. Instead of screening workers out, they are given as many attempts as needed to answer them correctly. Total earnings ranged from \$4 to \$5.77 for an average session duration of about 20 minutes.\footnote{Earnings consisted of a \$4 fixed wage, plus earnings from both work periods, and any bonuses earned based on their WTP and the BDM procedure.} The implied hourly earnings therefore range from approximately \$12 to \$17 per hour, which is well above Prolific's minimum payment rate. The piece-rate incentives are comparable to those used in other real-effort experiments studying social preferences (e.g., \citealp{dellavigna2018motivates, senn2023leveraging}). Full experimental instructions are provided in the Appendix. 

Table \ref{tab:sum_stats} presents descriptive statistics for the main sample in column (1). The average age was 40, and 40\% of workers identified as female. Average effort, measured by the number of rows solved, was 25.9 in Period 1 and 29.8 in Period 2. Columns (2) to (5) report sample statistics by experimental group, and column (6) reports balance tests, showing no significant differences in baseline characteristics. 

\section{Results} \label{sec:results}

We organize our results around a central hypothesis: workers differ in whether and how they value peer information, and that these differences help predict their effort responses. We begin by documenting aggregate patterns in information demand and effort outcomes when peer information is provided ex ante versus ex post. As we show, however, these averages mask substantial heterogeneity in both worker preferences and behavior. 

To study this heterogeneity, we adopt two complementary approaches. First, we group workers according to their WTP profiles, guided by the theoretical predictions. Second, we analyze their open-ended explanations for choosing peer information using both manual coding and a data-driven text analysis.

\subsection{Aggregate Information Demand and Effort Responses}

\subsubsection{Average Preferences for Peer Information}

We first summarize workers' average preferences for peer information across the performance distribution. Overall, 70.6\% and 69.6\% of workers display a nonzero WTP at one or more performance levels in the ex ante and ex post scenarios, respectively. That is, the majority of workers have a significant demand to either seek or avoid peer information.

\begin{figure}[ht]
	\centering
\caption{Average Preferences for Peer Information}
 \vspace{1.0em}
	\begin{subfigure}{0.495\textwidth}
 \caption{1 if prefer information}
 \includegraphics[width=1.01\textwidth]{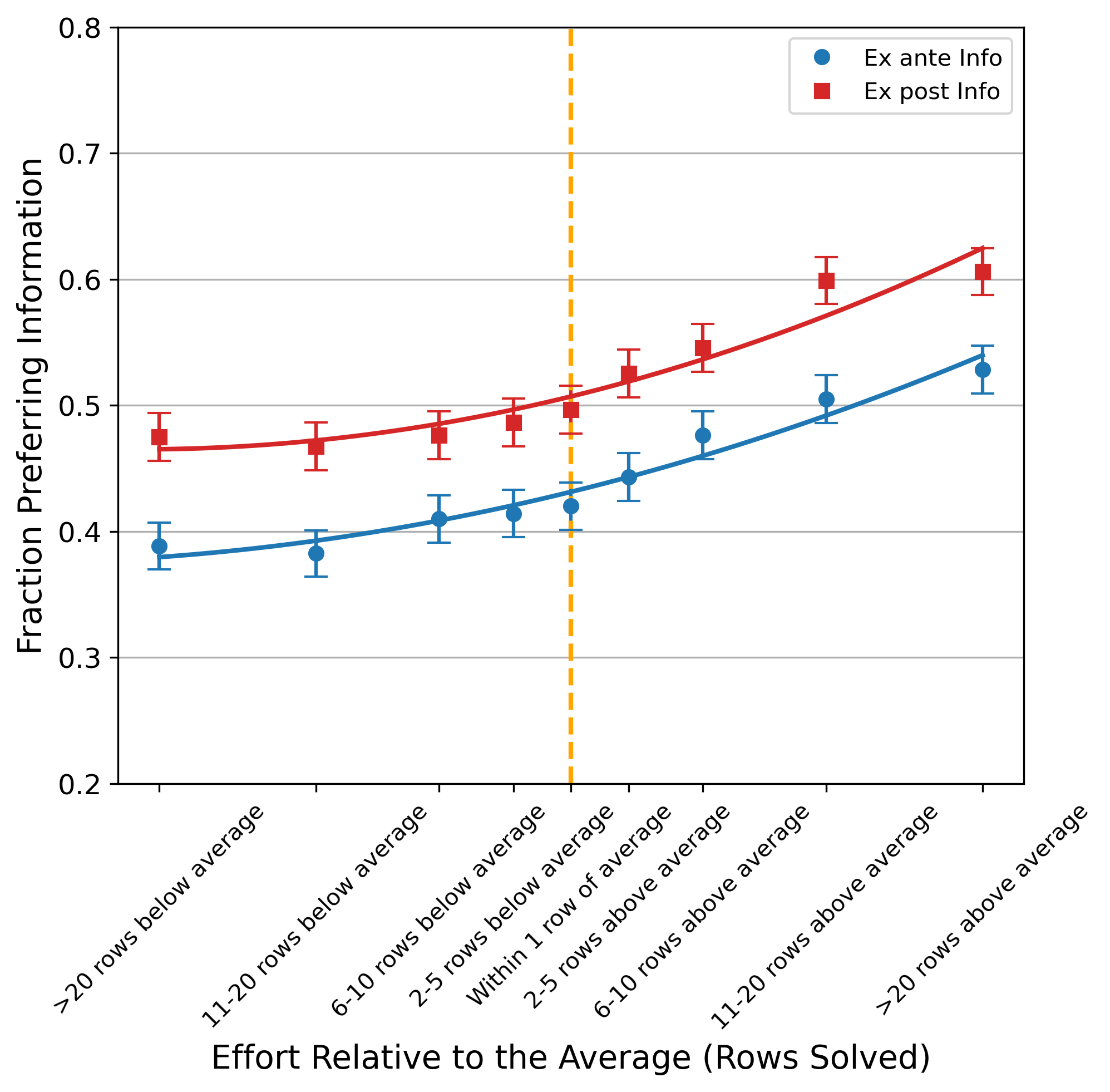} 
  \label{fig:choiceInfo_byScenario} 
  
  \vspace{0.5em}
	\end{subfigure}
	\begin{subfigure}{0.495\textwidth}

 \caption{WTP (¢)}
  \includegraphics[width=1.01\textwidth]{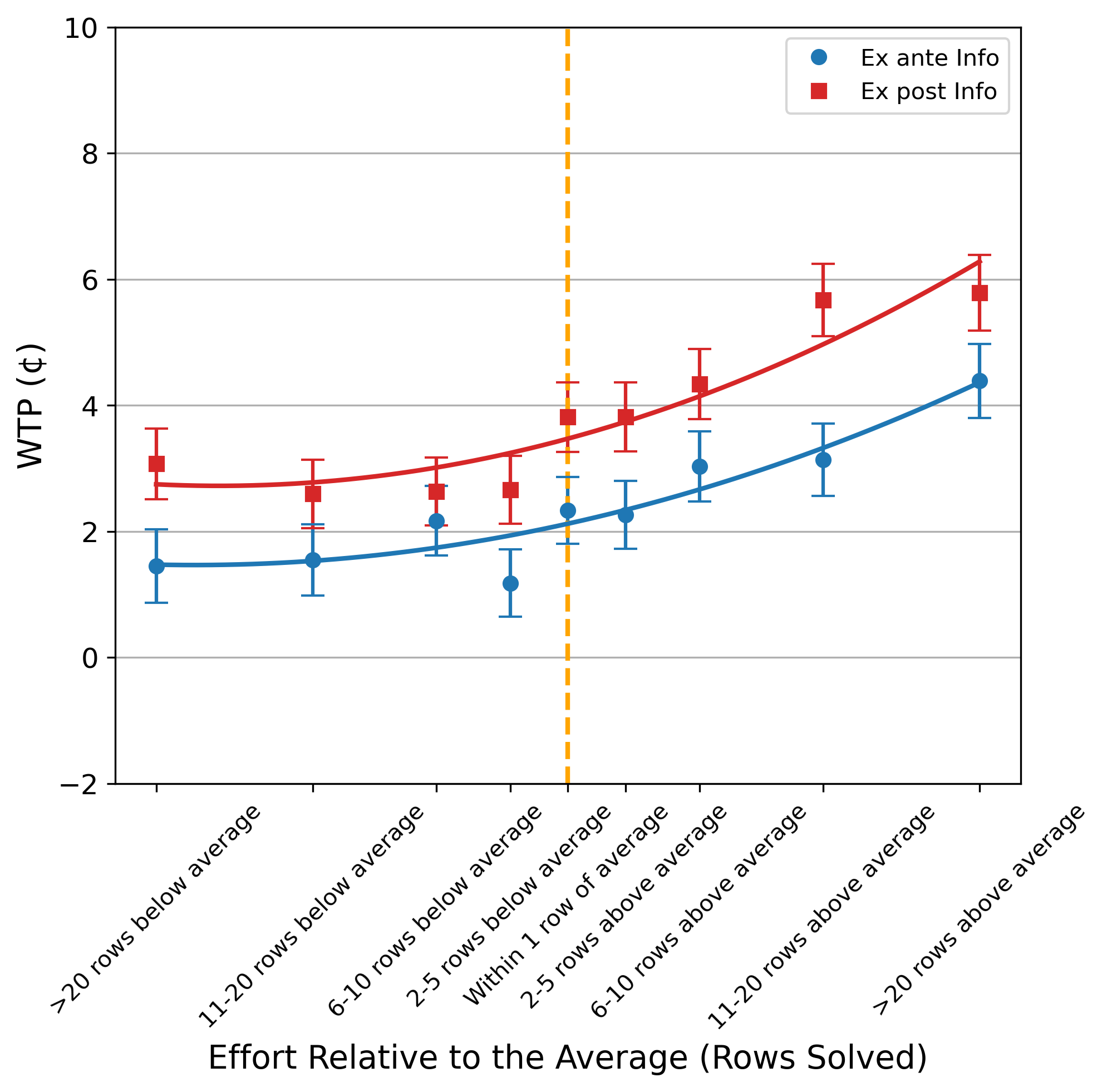} 
  \label{fig:avgWTP_byScenario} 
  
	\end{subfigure}
	\begin{minipage}{15cm}
		\scriptsize \singlespacing{\emph{Notes.} These figures depict the average preferences for peer information that is provided ex ante or ex post by each of the 9 possible realizations of rows solved. Panel (a) plots the fraction of workers who stated they want to receive peer information. Panel (b) plots the average WTP for peer information. The vertical dashed line corresponds to the average rows solved, i.e., if their performance was within 1 row of the true average. The error bars display $+/-$ one standard error of the mean.} 
	\end{minipage}
\label{fig:preferences_byScenario} 
\vspace{1.0em}
\end{figure}

Figure \ref{fig:preferences_byScenario} depicts workers' information preferences across the nine contingent performance levels. Panel (a) shows the fraction of workers who prefer to receive peer information; Panel (b) plots their average WTP under each scenario. On average, workers are more likely to seek information and have higher WTP when it is provided ex post rather than ex ante. In addition, the fitted curves reveal an increasing trend: workers are more likely to want information, and are willing to pay more for it, as their performance improves relative to the average. This pattern holds under both the ex ante and ex post scenarios.

Table \ref{tab:table-wtp1} in the Appendix reports OLS estimates that confirm these patterns. Two key findings emerge across all specifications. First, at every contingent performance level, a significantly larger share of workers opt to receive peer information ex post rather than ex ante. Second, average WTP is also consistently higher when information is provided ex post.

\vspace{-1.0em}
\subsubsection{Average Impact on Worker Effort}

\begin{figure}[ht]
	\centering
\caption{Average Effect of Peer Information on Worker Effort}
 \vspace{1.0em}
	\begin{subfigure}{0.48\textwidth}
 \caption{Change in Effort (Period 2 – Period 1)}
 \includegraphics[width=1.01\textwidth]{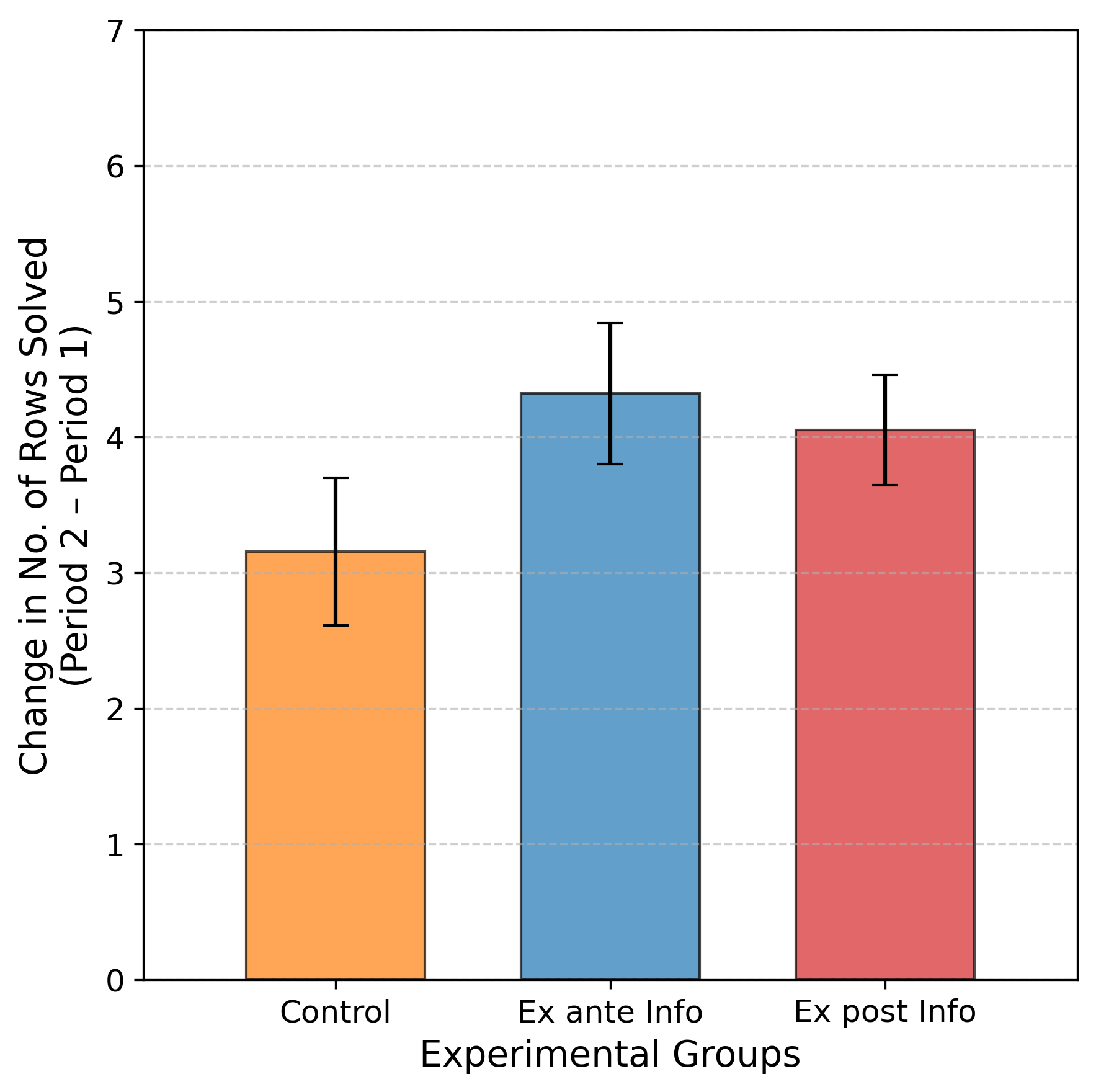} 
  \label{fig:effortDiff_byGroup} 
  
  \vspace{0.5em}
	\end{subfigure}
	\begin{subfigure}{0.48\textwidth}

 \caption{CDF of Change in Effort}
  \includegraphics[width=1.04\textwidth]{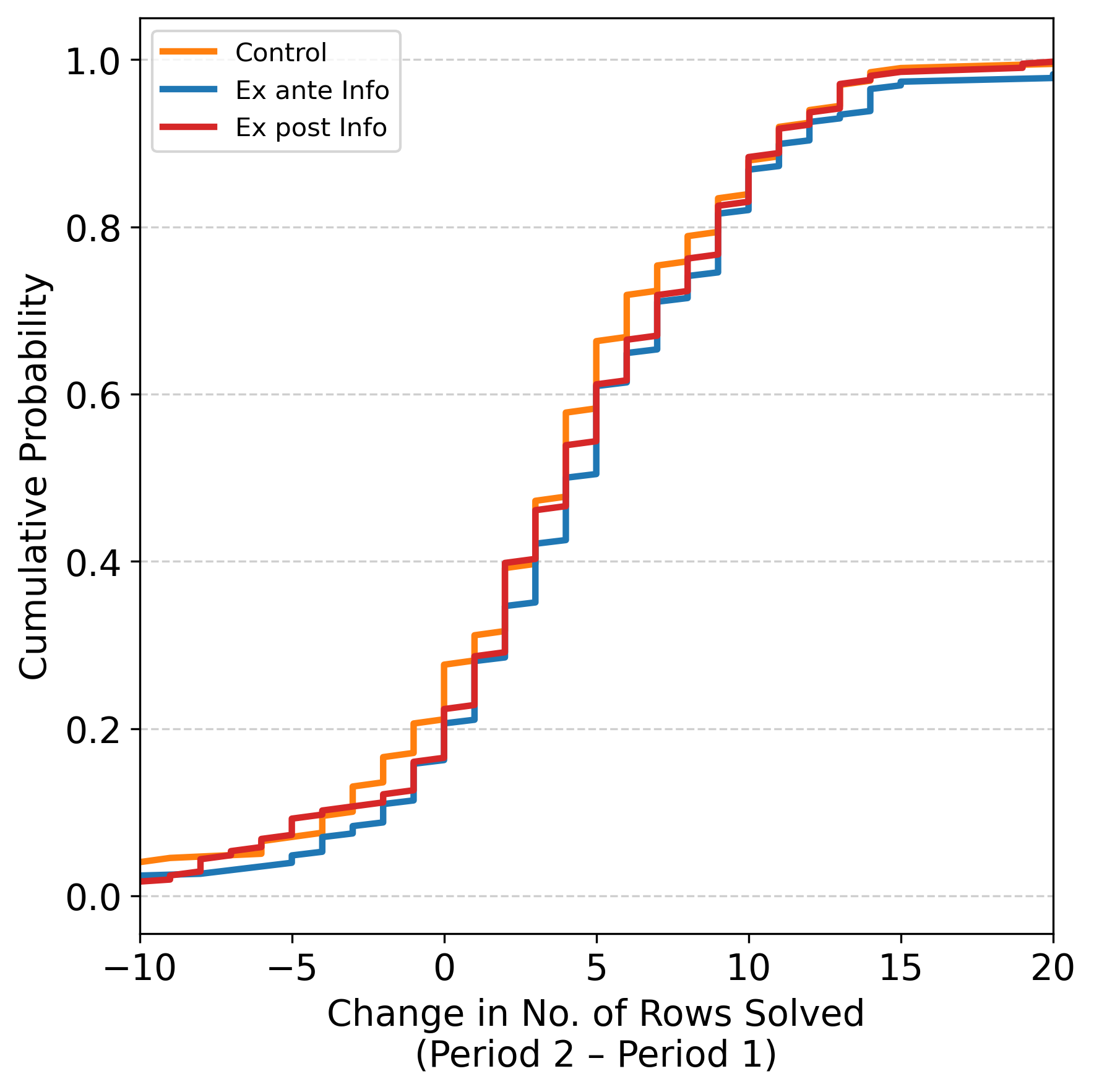} 
  \label{fig:effortCDF_byGroup} 
  
	\end{subfigure}
	\begin{minipage}{15cm}
		\scriptsize \singlespacing{\emph{Notes.} Panel (a) plots the average change in worker effort from Period 1 to Period 2, measured by the number of rows solved, by experimental group. The error bars display $+/-$ one standard error of the mean. Panel (b) plots the CDFs of the change in effort (Period 2 – Period 1) by the different experimental groups.} 
	\end{minipage}
\label{fig:effort_byGroup} 
\vspace{1.0em}
\end{figure}

We next examine the average effect of receiving peer information on effort.\footnote{As preregistered, we exclude the 60 workers in the Choose-Your-Info group, for whom the receipt of peer information was endogenous. This yields a final sample of $N=633$ for the analysis.} In Figure \ref{fig:effort_byGroup}, Panel (a) shows the average change in effort from Period 1 to Period 2 by experimental group. The Control group exhibits a modest increase, likely reflecting learning or the real-time feedback provided during the task. In contrast, the increase in effort is larger in both the Ex ante Info and Ex post Info groups, suggesting a positive treatment effect. Panel (b) plots the cumulative distribution functions (CDFs) of effort change. Notably, the CDF for the Ex ante Info group first-order stochastically dominates that of the Control group, indicating a positive effect across the entire distribution.

Table \ref{tab:table-effort1} in the Appendix reports the corresponding OLS estimates. As preregistered, we focus interpretation on our primary effort measure, the number of rows solved. For robustness, we also report two additional, non-preregistered outcomes: the number of rows attempted and worker's self-assessed effort. Column (1) shows that receiving peer information ex ante increases effort by 1.26 rows (4.9\% relative to baseline), while receiving it ex post increases effort by 0.98 rows (3.8\% relative to baseline). These effects are consistent across all effort measures, though the ex post estimates are imprecisely estimated and not statistically significant in most specifications.\footnote{We also examine the impact of receiving peer information on workers' self-reported well-being (i.e., stress, motivation, and perceived task meaning). Full results are reported in Table \ref{tab:table-wellbeing1} in the Appendix.} 

Importantly, if workers are heterogeneous in how they value and respond to peer information, consistent with different model predictions, then averaging across individuals with different-signed responses can attenuate estimated effects and lead to power issues. This may help explain the mixed findings in the performance feedback literature.

\subsection{Heterogeneity in Information Demand and Effort Responses}

We now turn to the central contribution of the paper: examining whether heterogeneity in workers’ demand for peer information predicts heterogeneity in effort responses. Guided by the theoretical framework, we first classify workers into different ``types'' based on their WTP profiles and then examine whether these types differ in their subsequent effort responses.

\subsubsection{Theory-driven Classification: WTP Profiles}

We begin by classifying workers based on their WTP in the ex ante scenario. Specifically, we use WTP responses at three contingent performance levels representing the bottom, middle, and top of the distribution: (i) if their performance is 20 or more rows below average, (ii) within 1 row of the average, and (iii) 20 or more rows above average. This three-point summary of the WTP profile captures meaningful variation across the performance distribution while limiting potential misclassification due to noisy responses. In Appendix B, we provide a benchmark analysis of how noise can affect classification by varying the number and choice of WTP points along the performance distribution. We show that the estimated shares of the resulting types are highly robust to a range of alternative specifications.

Importantly, the classification is based only on WTP in the ex ante scenario. We then examine whether WTP in the ex post scenario and subsequent effort responses align with the predictions for these types (see Table \ref{tab:model-predictions}). Based on ex ante WTP, workers are grouped into four mutually exclusive types:

\begin{enumerate}[label=\textbf{Type \arabic*}:, align=left, left=10pt, labelsep=0.5em, itemsep=0.8em]
    \item \textbf{(Indifferent)} WTP is zero at all three performance levels, consistent with the \textit{standard model}, in which workers are indifferent to peer information.
    
    \item \textbf{(Stress-avoidant)} WTP is negative at all three performance levels, with at least one value strictly negative, consistent with the \textit{stress avoidance model}.
    
    \item \textbf{(Competitive)} WTP increases with performance, with at least one value strictly positive, consistent with the \textit{social preferences (competitive) model}.\footnote{The requirement of a strictly positive value ensures that workers are not simultaneously classified into Type 2 (Stress-avoidant), since types are defined to be mutually exclusive.} 
    
    \item \textbf{(Learning-oriented/Residual)} WTP profiles that do not fit any of the above types. This group includes workers with decreasing WTP (consistent with the \textit{learning model}) as well as others with non-monotonic patterns.
\end{enumerate}
\smallskip

While our theoretical framework considers five mechanisms---standard preferences, social preferences (competitive and inequality-averse), stress avoidance, and learning---the data provide clear support for only three. The \textit{standard}, \textit{stress avoidance}, and \textit{social preferences (competitive)} models map directly onto Types 1--3. We find little support for inequality-averse preferences: only 2\% of workers ($N=14$) exhibit the predicted single-peaked, negative WTP profile, where demand is highest (least negative) near the middle of the performance distribution and lowest at the extremes. Given the small number, we pool these workers with Type 2 (stress-avoidant).\footnote{When we analyze the workers' stated motivations later in the paper, none express concerns consistent with inequality aversion.}   

For the \textit{learning model}, we identify a subset of workers (8\% of the sample) whose ex ante WTP profile decreases with performance. The remaining unclassified workers exhibit non-monotonic patterns. On average, their average WTP follows a U-shaped profile that initially decreases (suggestive of a learning motive), before rising with performance. To avoid overfitting, we group all workers whose WTP profiles do not match Types 1--3 into the broader Type 4 (learning-oriented/residual) category.

\begin{figure}[htpb]
	\centering
\caption{WTP for Peer Information by Type}
 \vspace{1.0em}

 \begin{subfigure}{0.45\textwidth}
	\caption{Type 1: Indifferent}
\includegraphics[width=1.01\textwidth]{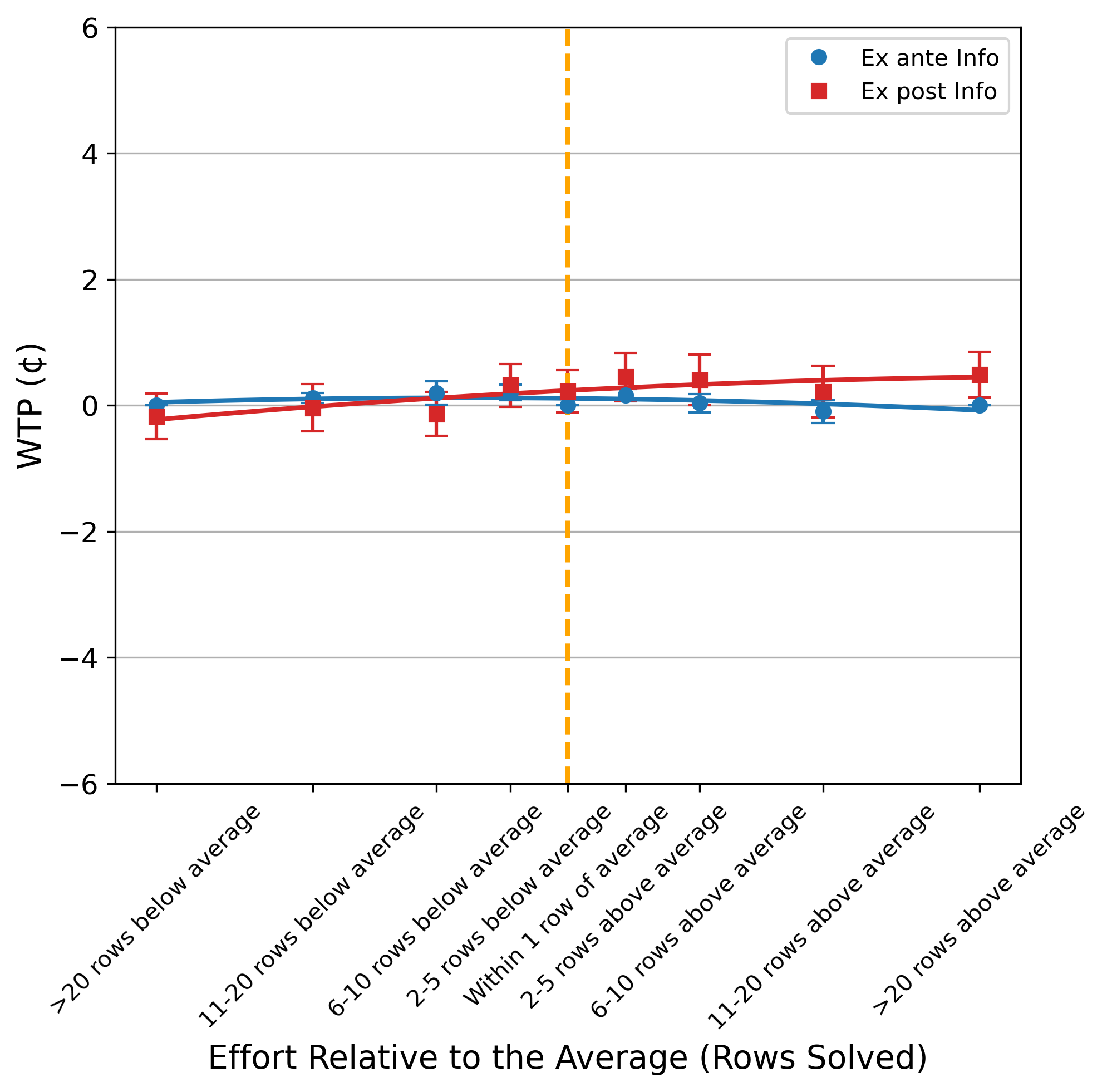} 
\label{fig:type1} 
\vspace{0.5em}
	\end{subfigure}
	\begin{subfigure}{0.45\textwidth}
	\caption{Type 2: Stress-avoidant}
\includegraphics[width=1.01\textwidth]{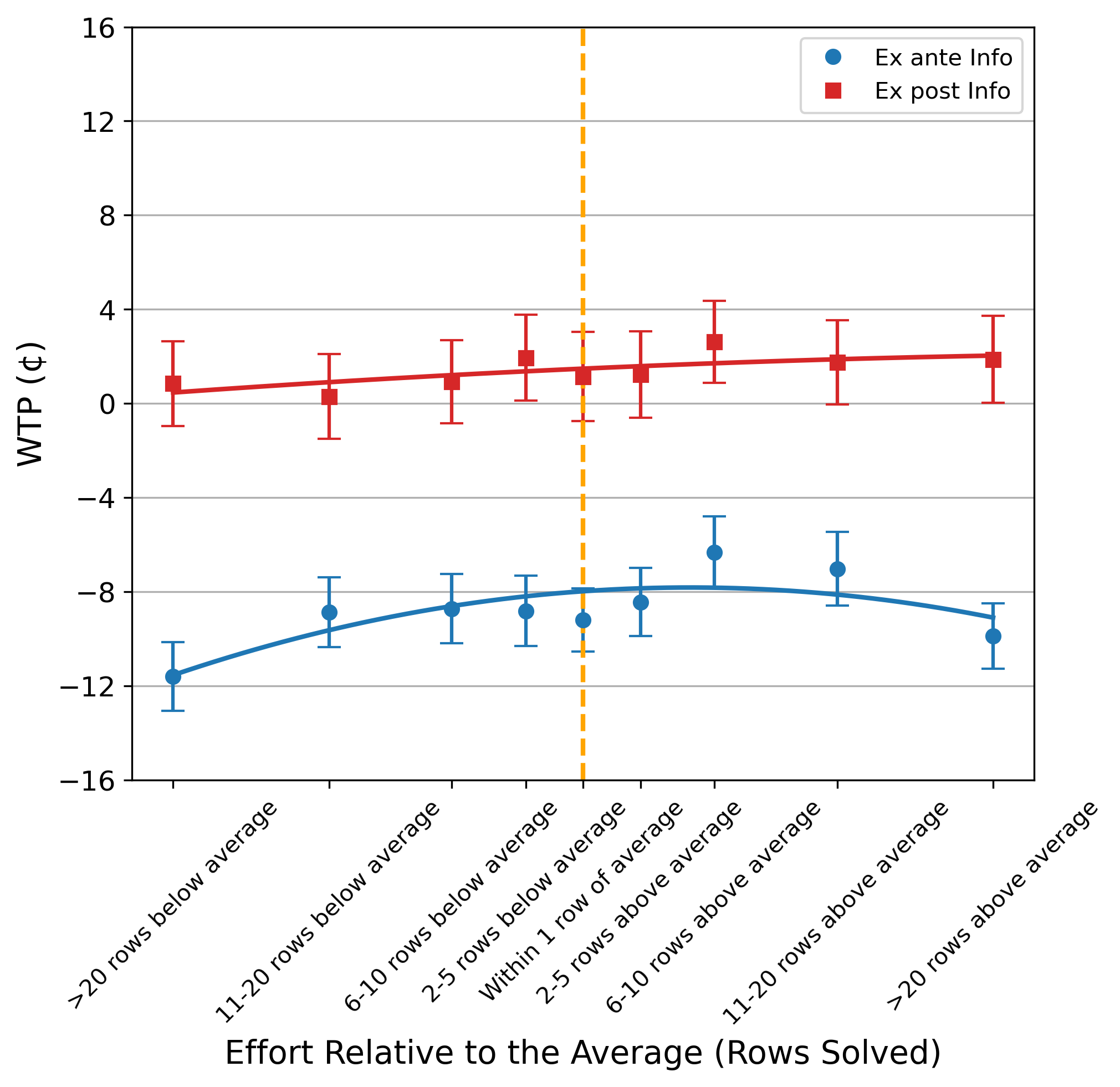} 
\label{fig:type2} 
\vspace{0.5em}
	\end{subfigure}
    
\vspace{0.5em}
    \begin{subfigure}{0.45\textwidth}
		\caption{Type 3: Competitive}
\includegraphics[width=1.01\textwidth]{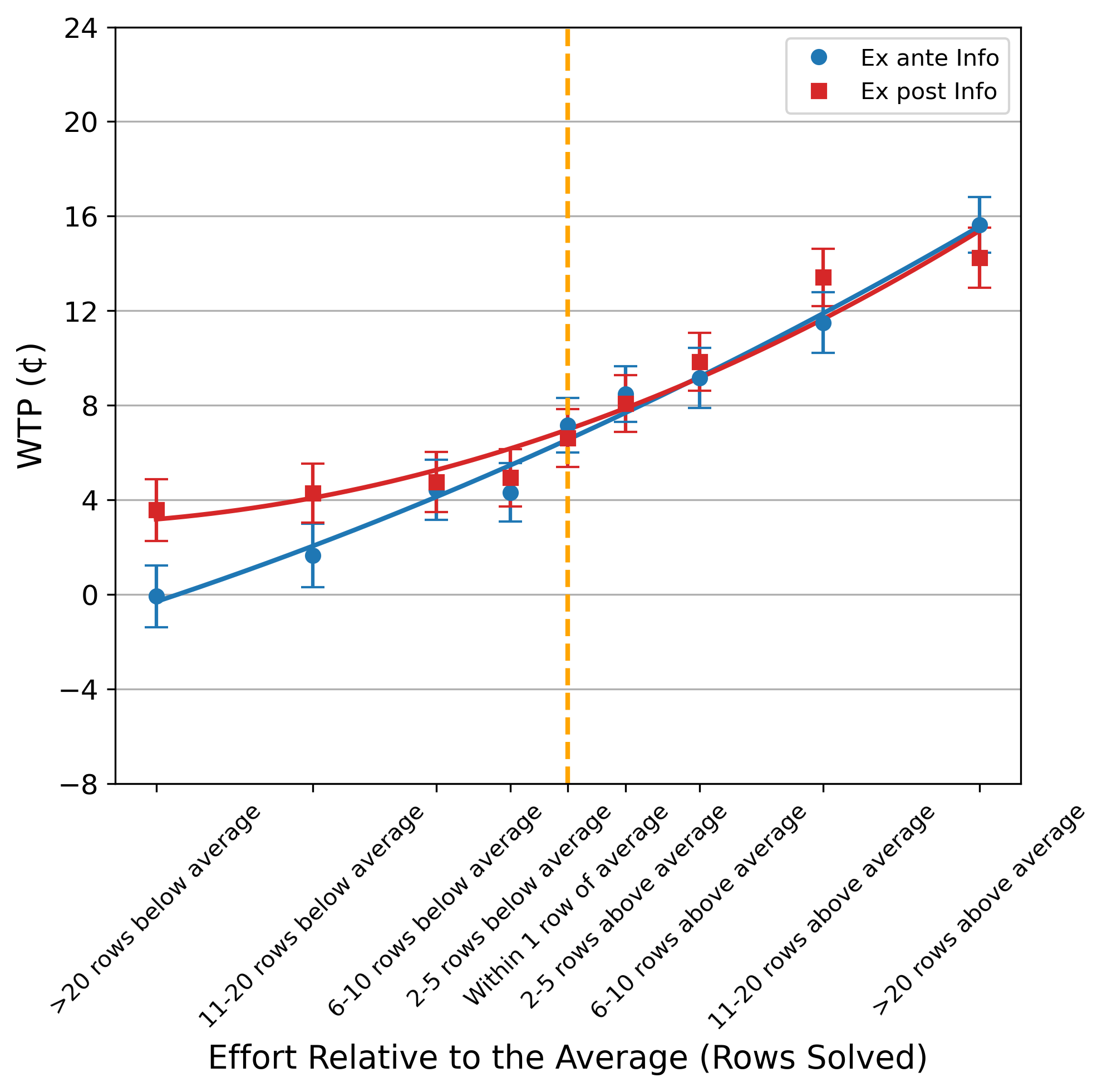} 
\label{fig:type3} 
\vspace{0.5em}
	\end{subfigure}
	\begin{subfigure}{0.45\textwidth}
		\caption{Type 4: Learning / Residual}
\includegraphics[width=1.01\textwidth]{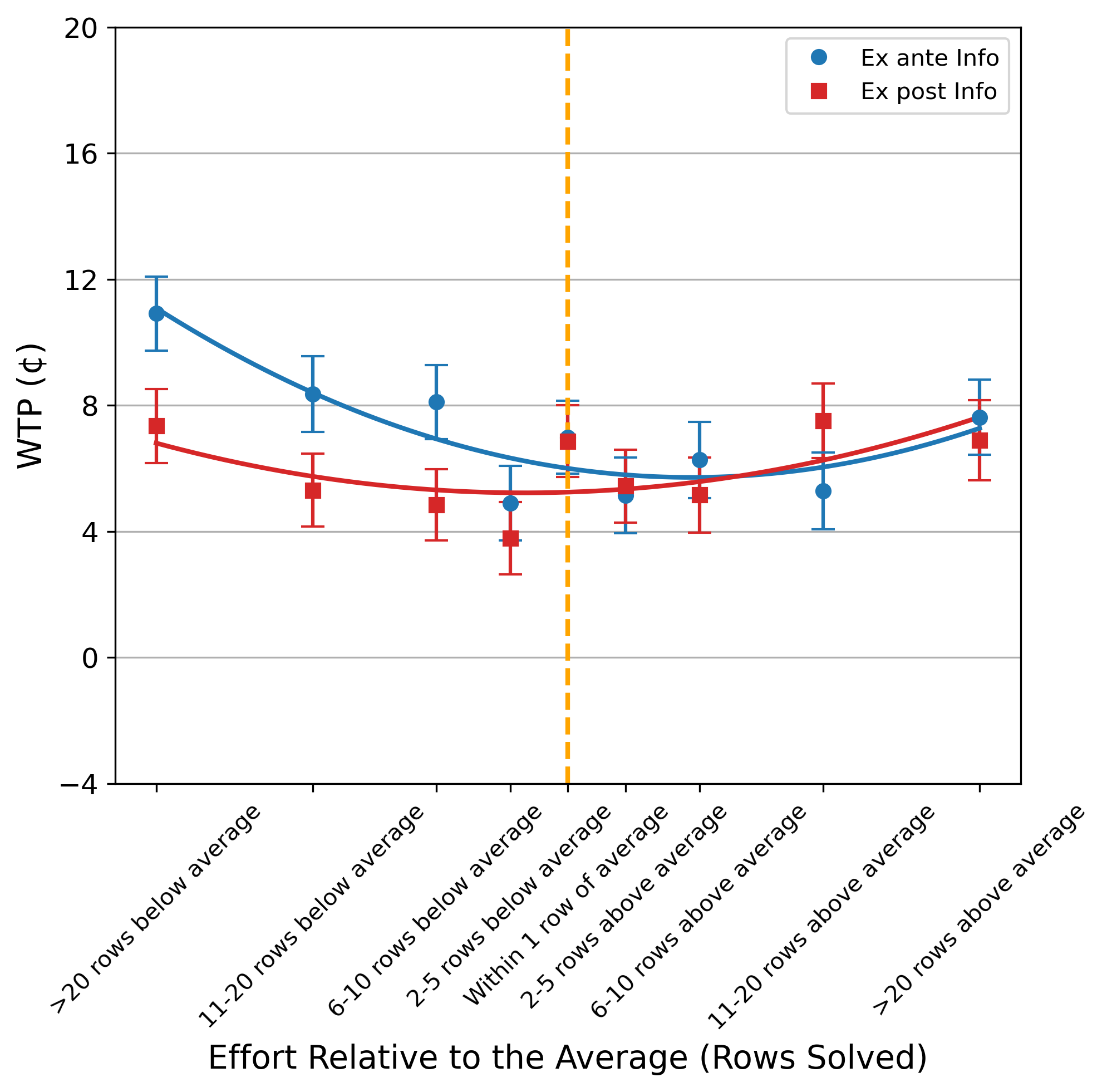} 
\label{fig:type4} 
\vspace{0.5em}
	\end{subfigure}

	\begin{minipage}{15cm}
		\scriptsize \singlespacing{\emph{Notes.} These figures depict the average WTP for peer information that is provided ex ante or ex post for each of the 9 possible realizations of rows solved, separately for each worker type (Types 1--4). The vertical dashed line corresponds to the average rows solved, i.e., if their performance was within 1 row of the true average. The error bars display $+/-$ one standard error of the mean.}
	\end{minipage}
\label{fig:WTP_by_preference_types} 
\vspace{1.0em}
\end{figure}

Figure \ref{fig:WTP_by_preference_types} displays the average WTP for peer information in the ex ante and ex post scenarios, separately by type. Based on this classification, 32\% of workers fall into Type 1 (indifferent), 15\% into Type 2 (stress-avoidant), 23\% into Type 3 (competitive), and the remaining 30\% into Type 4 (learning-oriented/residual).

As an out-of-sample test, we first examine workers' WTP in the ex post scenario across types. Panel (a) shows that Type 1 (indifferent) workers exhibit zero WTP for peer information ex ante (by construction), and their ex post WTP remains similarly near-zero across all performance levels. This is consistent with the \textit{standard model}, which predicts no demand for peer information regardless of timing (Hypothesis 1.1).  

Panel (b) shows that Type 2 (stress-avoidant) workers exhibit strictly negative WTP in the ex ante scenario, with a relatively flat profile across performance levels. In the ex post scenario, their WTP is close to zero throughout. This pattern aligns with the \textit{stress avoidance model}, which predicts aversion to information before the task but to a lesser degree once the task is completed (Hypothesis 3.3).

Panel (c) shows that Type 3 (competitive) workers exhibit an ex ante WTP profile that increases with performance (by construction), and their ex post WTP follows a similarly increasing trend. This is consistent with the \textit{social preferences (competitive) model}, which posits that utility from peer information increases with relative performance. On average, ex post WTP is comparable in magnitude to ex ante WTP, though slightly lower at the bottom of the distribution. This pattern contrasts with Hypothesis 2.3, which predicts that WTP should always be weakly higher ex ante than ex post, assuming the strength of social preferences is invariant across timing ($\delta=1$). The data suggest that competitive preferences may be attenuated ex post (i.e., $\delta < 1$), or that other mechanisms (e.g., stress or distraction) may have lowered workers' ex ante WTP.

Panel (d) shows that Type 4 (learning-oriented/residual) workers exhibit an ex ante WTP profile that decreases with performance, a pattern suggestive of a learning motive in which workers value information more when underperforming. However, their ex post WTP remains consistently positive across all performance levels, contrary to the \textit{learning model}'s sharp prediction of zero WTP in that scenario. This suggests that other motives, such as curiosity, may also contribute to their demand for peer information.

We next test whether effort responses to peer information vary systematically across these WTP-based types, as predicted by the theoretical framework. Table \ref{tab:table-effort2} presents OLS estimates of the impact of receiving peer information ex ante and ex post, separately by type. 

We begin by pooling Types 1 and 2 workers, the two groups predicted to have no effort response to peer information. Consistent with this prediction, column (2) shows that receiving peer information ex ante or ex post has no significant effect on effort for these workers. By contrast, column (3) pools Types 3 and 4 workers, the groups predicted to respond positively to peer information, particularly when it is provided ex ante. For these workers, receiving peer information ex ante leads to a significant increase in effort by 2.50 rows ($p<0.01$), or approximately 10.3\% relative to baseline. The corresponding effect of ex post information is also positive, though not statistically significant. This contrast reveals a key insight: the average treatment effect in column (1) masks substantial heterogeneity, and effort responses only become evident once we condition on workers' information preferences.

Columns (4) through (7) further unpack these responses by type. For Type 1 (indifferent) workers, peer information has no significant impact on effort, whether ex ante or ex post. This is consistent with Hypothesis 1.2 of the \textit{standard model}. Likewise, Type 2 (stress-avoidant) workers show no effort response in either timing scenario, consistent with Hypothesis 3.2 of the \textit{stress avoidance model}. For Type 3 (competitive) workers, effort increases by 2.45 rows ($p<0.05$) when information is received ex ante, and by 1.33 rows ex post, though the latter is not significant. This aligns with Hypotheses 2.2 and 2.4 of the \textit{social preferences model} under competitive concerns. For Type 4 (learning-oriented/residual) workers, we find a marginally significant increase in effort by 2.49 rows ($p<0.1$) when information is received ex ante, but no significant effect ex post, a pattern broadly consistent with Hypotheses 4.2 and 4.4 of the \textit{learning model}.

{
\singlespacing
\begin{table}[ht]
\centering 
	\def\sym#1{\ifmmode^{#1}\else\(^{#1}\)\fi}
	\caption{Heterogeneous Impact on Worker Effort by WTP-based Type} \label{tab:table-effort2}
	\begin{threeparttable}
    \resizebox{\textwidth}{!}{
    \begin{tabular}{l*{7}{c}}
    \toprule
    \toprule
    \textit{Dependent variable:} & \multicolumn{7}{c}{Effort (\# Rows Solved)} \\ 
    \cmidrule(lr){2-8} 
& \multicolumn{1}{c}{\textbf{Full Sample}}
& \multicolumn{1}{c}{\makecell{\textbf{Types 1 \& 2}}}
& \multicolumn{1}{c}{\makecell{\textbf{Types 3 \& 4}}}
& \multicolumn{1}{c}{\makecell{\textbf{Type 1:} \\ Indifferent}}
& \multicolumn{1}{c}{\makecell{\textbf{Type 2:} \\ Stress-avoidant}}
& \multicolumn{1}{c}{\makecell{\textbf{Type 3:} \\ Competitive}}
& \multicolumn{1}{c}{\makecell{\textbf{Type 4:} \\ Learning/Residual}} \\
    & (1) & (2) & (3) & (4) & (5) & (6) & (7) \\ 
    \midrule \addlinespace 
    
    $ \text{1 if receive info ex ante} $ 
    &        1.26*  &       --0.02   &        2.50***&       --0.45   &        0.45   &        2.45** &        2.49*  \\
            &      (0.75)   &      (1.22)   &      (0.91)   &      (1.64)   &      (1.73)   &      (1.14)   &      (1.39)   \\ \addlinespace
            
$ \text{1 if receive info ex post} $
&        0.98   &        0.71   &        1.24   &        1.03   &        0.01   &        1.33   &        1.16   \\
            &      (0.68)   &      (1.10)   &      (0.84)   &      (1.39)   &      (1.74)   &      (1.22)   &      (1.15)   \\ \addlinespace

            Baseline mean &  25.73 &  27.35 &  24.30 &  29.11 &  23.69 &  28.27 &  21.26 \\
            & (13.83) & (13.61) & (13.89) & (14.38) & (11.07) & (13.88) & (13.14)  \\ \addlinespace
			\midrule \addlinespace
            	Controls & \multicolumn{1}{c}{\ding{51}} & \multicolumn{1}{c}{\textit{\ding{51}}} & \multicolumn{1}{c}{\textit{\ding{51}}} & \multicolumn{1}{c}{\textit{\ding{51}}} & \multicolumn{1}{c}{\textit{\ding{51}}} & \multicolumn{1}{c}{\textit{\ding{51}}} & \multicolumn{1}{c}{\textit{\ding{51}}} \\ 
            $p$-value: $\beta_{\text{ex ante}} = \beta_{\text{ex post}}$ & 0.677 & 0.487 & 0.134 & 0.305 & 0.756 & 0.285 & 0.306 \\ \addlinespace
			\midrule \addlinespace
			\(R^{2}\)   &       0.095   &       0.113   &       0.093   &       0.105   &       0.131   &       0.100   &       0.130   \\

            No. of obs. &        1238   &         582   &         656   &         394   &         188   &         284   &         372   \\
    No. of workers & 619 & 291 & 328  & 197 & 94   & 142 & 186  \\ 
    
			\bottomrule
		\end{tabular}
        }
		\begin{tablenotes}
			\scriptsize\vspace{0.1cm} 
            \parbox{.95\textwidth}{
			\item \emph{Notes.} This table reports OLS estimates of the average effects of receiving peer information ex ante and ex post on effort, separately for each WTP-based type.
            The controls include gender (1 if female), age, education attainment (1 if college degree), and the log of time taken to complete the study. Standard errors clustered at the worker level in parentheses.
			\item * $p<0.10$, ** $p<0.05$, *** $p<0.01$
            }
		\end{tablenotes}
	\end{threeparttable}	
\end{table}

}

Together, these results show that the WTP-based types predict workers' effort responses to peer information. To our knowledge, this is the first experimental study to move beyond average treatment effects toward a mechanism-informed understanding of heterogeneity in how workers value and respond to peer information.

\subsubsection{Data-driven Classification: Open-ended Responses}

To provide complementary evidence on underlying mechanisms, we asked workers to explain their contingent choices in the ex ante and ex post scenarios. Specifically, workers were prompted with the following: ``Below is a summary of the choices you made for each scenario. Please briefly explain why you made those choices for Scenario 1 (Scenario 2), where information is provided before (after) the task.'' We analyze these open-ended responses using two approaches: (i) manual classification based on a coding scheme, and (ii) automated classification using BERT embeddings combined with k-means clustering.

{\centering \subsubsection*{\textit{I. Manual coding of responses}}}

We begin by analyzing the responses using a hand-coded scheme that classifies each explanation into one or more of seven non-mutually exclusive categories. \textbf{``Do Not Care''} responses indicate that the worker has no interest in the information or is unwilling to pay for it. \textbf{``Stress or Distraction''} responses reflect concerns that receiving information, especially before the task, would induce stress or be a source of distraction. \textbf{``Social Comparison''} responses indicate an interest in comparing one's performance with others, including affective reactions to being ahead or behind peers. \textbf{``Goal or Motivation''} responses describe viewing the average as a target to beat or a source of motivation when performing the task. \textbf{``Curiosity''} responses reflect a desire to learn the results out of curiosity. \textbf{``Value at Extremes''} responses indicate that the worker values peer information more when their performance falls at the extremes of the distribution, especially when underperforming, as a way to adjust or improve their strategy. Finally, \textbf{``No Value After Task''} responses indicate that the information is perceived as not useful if it is only received after task completion. Table \ref{tab:table-codingscheme} presents the coding scheme with example responses for each category.

\begin{figure}[ht]
	\centering
\caption{Distribution of Worker Motivations for Choosing Information}
 \vspace{0.5em}
 \includegraphics[width=.65\textwidth]{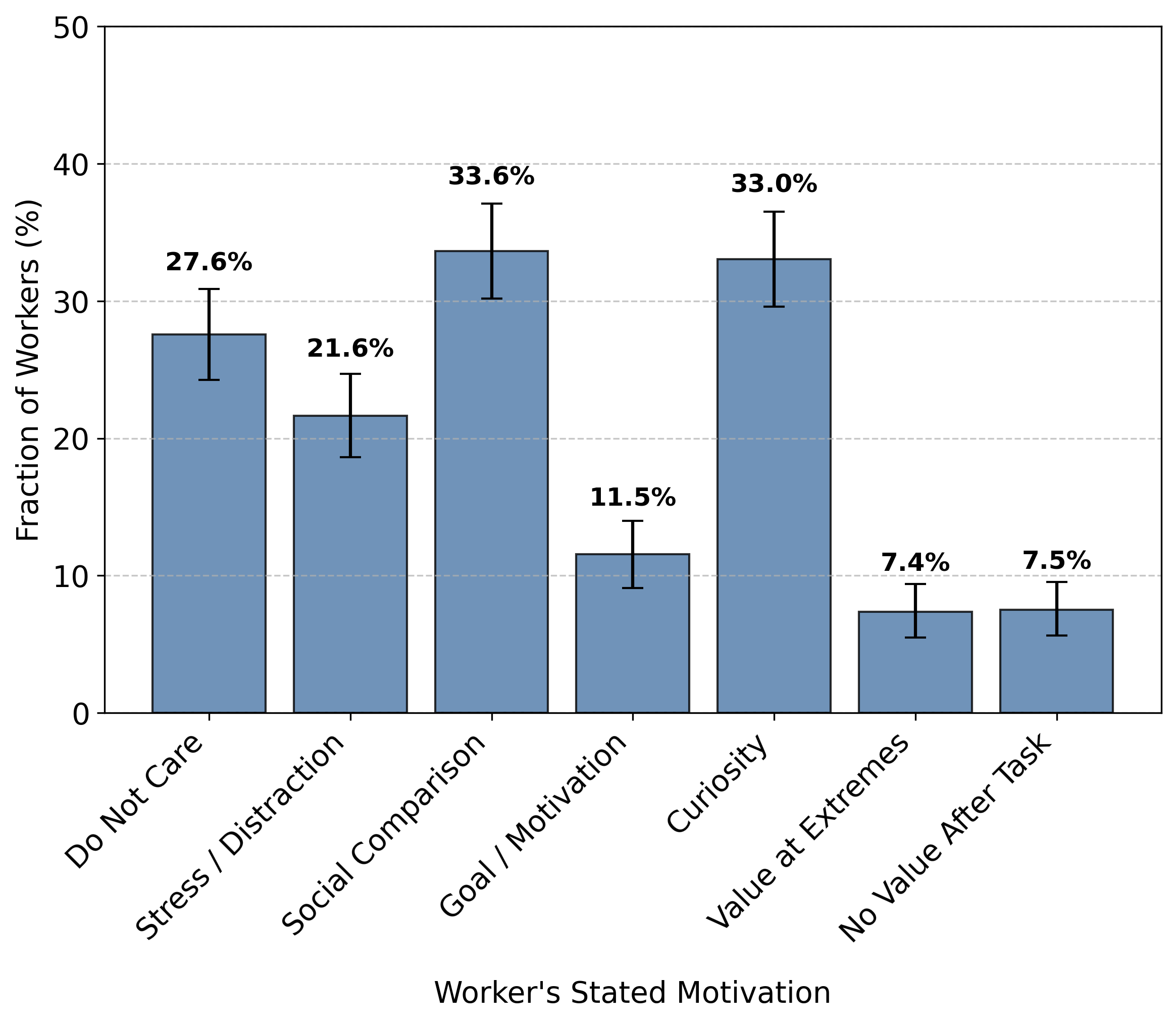} 
  \label{fig:worker_rationales} 
  \vspace{0.2em}
	\begin{minipage}{15cm}
		\scriptsize \singlespacing{\emph{Notes.} This figure shows the fraction of workers citing different motivations for seeking (or avoiding) peer information. Each worker is asked to provide open-ended explanations for their choices in both the ex ante and ex post scenarios. Their responses are hand-coded into one or more of seven non-mutually exclusive categories (see Table \ref{tab:table-codingscheme} for details). The error bars represent bootstrapped 95\% confidence intervals.} 
	\end{minipage}
\vspace{1.0em}
\end{figure}

Figure \ref{fig:worker_rationales} shows the distribution of workers' hand-coded explanations for why they chose to seek or avoid peer information. We observe that workers' motivations vary considerably: 27.6\% of responses indicate they do not care about peer information, 21.6\% cite stress or distraction as a concern, and 33.6\% (the largest category) reference some form of social comparison. Additionally, 11.5\% of responses describe using peer information as a goal or source of motivation to work harder. Another frequently cited rationale is curiosity, appearing in 33.0\% of responses; however, the majority of these ``curiosity'' responses also indicate other motivations. Finally, 7.4\% of workers report valuing information more if they fall at the extreme ends of the performance range, and 7.5\% explicitly mention that information is not useful if it arrives after the task.

To examine how workers' self-reported motivations relate to their elicited preferences, Figure \ref{fig:types_byWTP} plots workers’ WTP in both the ex ante and ex post scenarios, separately by each hand-coded category. The results again highlight substantial heterogeneity in how workers value peer information, with stated motives closely aligning with revealed preferences. Workers who report not caring about the information display near-zero WTP across all performance levels (panel (a)); those citing stress or distraction exhibit highly negative WTP when information is provided ex ante (panel (b)); and workers motivated by social comparisons show upward-sloping WTP profiles regardless of information timing (panel (c)). Those motivated by goal-setting or curiosity exhibit positive WTP across the entire performance range (panels (d) and (e)). Workers who mention valuing information at the extremes of the performance distribution exhibit a ``U-shaped'' WTP pattern (panel (f)), while those indicating that information is not useful ex post only seek it ex ante. 

In addition, we link these self-reported motivations to our earlier WTP-based types. Figure \ref{fig:worker_rationales_byTypes} shows that workers' stated motives align closely with the mechanism underlying each type, providing additional validation of our theory-driven classification.

Given the substantial heterogeneity in workers' stated motivations, we next examine whether receiving information ex ante or ex post differentially affects effort across the seven hand-coded categories. Table \ref{tab:table-effort3} presents OLS estimates by category. The key takeaway is that workers' effort responses vary systematically with stated motivations. Below, we highlight the categories that map most directly onto the mechanisms in our theoretical framework.

In columns (1) and (2), we find no significant impact of receiving peer information (both ex ante and ex post) on effort among workers who either do not care about the information or choose to avoid it due to stress or distraction, consistent with the \textit{standard model} and \textit{stress avoidance model}, respectively. By contrast, column (3) shows that workers motivated by social comparisons increase effort by 2.6 rows ($p<0.05$) when information is provided ex ante, and by 2.1 rows ($p<0.1$) when provided ex post. Similarly, in column (4), those who use peer information as a goal or motivation for themselves increase effort by 3.3 rows ($p<0.05$) when information is provided ex ante. Both groups are consistent with the \textit{social preferences model} where workers display competitive preferences and use the peer average as a benchmark to evaluate their own performance.

Taken together, these results show that heterogeneity in workers' motivations for seeking or avoiding peer information helps explain heterogeneity in effort responses. In particular, distinguishing between workers who are uninterested in peer information and those who value it for different reasons is highly predictive of its impact on effort.

{
\singlespacing
\begin{table}[ht]
\centering 
	\def\sym#1{\ifmmode^{#1}\else\(^{#1}\)\fi}
	\caption{Heterogeneous Impact on Worker Effort by Hand-coded Category} \label{tab:table-effort3}
	\begin{threeparttable}
    \resizebox{\textwidth}{!}{
    \begin{tabular}{l*{7}{c}}
    \toprule
    \toprule
    \textit{Dependent variable:} & \multicolumn{6}{c}{Effort (\# Rows Solved)} \\ 
    \cmidrule(lr){2-8} 
    & \multicolumn{1}{c}{\textbf{\shortstack{Do Not\\Care}}}
& \multicolumn{1}{c}{\textbf{\shortstack{Stress or \\Distraction}}}
& \multicolumn{1}{c}{\textbf{\shortstack{Social \\ Comparison}}}
& \multicolumn{1}{c}{\textbf{\shortstack{Goal or \\Motivation}}}
& \multicolumn{1}{c}{\textbf{\shortstack{ \\Curiosity}}}
& \multicolumn{1}{c}{\textbf{\shortstack{Value at \\Extremes}}}
& \multicolumn{1}{c}{\textbf{\shortstack{No Value\\After Task}}} \\

    & (1) & (2) & (3) & (4) & (5) & (6) & (7) \\ 
    \midrule \addlinespace 

        $ \text{1 if receive info ex ante} $   &       --0.36   &        1.12   &        2.60** &        3.29** &        1.12   &        5.17** &        3.54   \\
            &      (1.77)   &      (1.29)   &      (1.13)   &      (1.59)   &      (1.02)   &      (2.19)   &      (3.22)   \\ \addlinespace
$ \text{1 if receive info ex post} $ &        1.09   &        1.81   &        2.08*  &       --1.10   &        0.17   &        5.37** &        5.25   \\
            &      (1.37)   &      (1.33)   &      (1.11)   &      (1.71)   &      (0.98)   &      (2.60)   &      (3.50)   \\ \addlinespace

        Baseline mean &  27.94 &  27.02 &  29.72 &  26.57 &  29.66 &  27.02 &  23.96 \\
		& (14.72) & (11.89) & (13.63) & (12.62) & (14.71) & (12.93) & (12.91)  \\ \addlinespace
			\midrule \addlinespace
            	Controls & \multicolumn{1}{c}{\ding{51}} & \multicolumn{1}{c}{\textit{\ding{51}}} & \multicolumn{1}{c}{\textit{\ding{51}}} & \multicolumn{1}{c}{\textit{\ding{51}}} & \multicolumn{1}{c}{\textit{\ding{51}}} & \multicolumn{1}{c}{\textit{\ding{51}}} & \multicolumn{1}{c}{\textit{\ding{51}}} \\ 
            $p$-value: $\beta_{\text{ex ante}} = \beta_{\text{ex post}}$ & 0.353 & 0.526 & 0.595 & 0.081 & 0.318 & 0.930 & 0.490 \\ \addlinespace
			\midrule \addlinespace
			\(R^{2}\)   &       0.157   &       0.155   &       0.156   &       0.230   &       0.149   &       0.182   &       0.134   \\

            No. of obs.  &         340   &         272   &         412   &         134   &         414   &          84   &          92   \\
    No. of workers & 170 & 136   & 206   & 67 & 207 & 42 & 46 \\ 
    
			\bottomrule
		\end{tabular}
        }
		\begin{tablenotes}
			\scriptsize\vspace{0.1cm} 
            \parbox{.95\textwidth}{
			\item \emph{Notes.} This table reports OLS estimates of the average effects of receiving peer information ex ante and ex post on effort (measured by the number of rows solved), disaggregated by the seven hand-coded categories. Each column corresponds to a category. The controls include gender (1 if female), age, education attainment (1 if college degree), and the log of time taken to complete the study. Standard errors clustered at the worker level in parentheses.
			\item * $p<0.10$, ** $p<0.05$, *** $p<0.01$
            }
		\end{tablenotes}
	\end{threeparttable}
\end{table}

}

\vspace{-1.0em}
{\centering \subsubsection*{\textit{II. BERT-based coding of responses}}}

To complement the manual coding exercise, we implement an automated classification approach that combines natural language processing with unsupervised machine learning to uncover latent clusters of workers based on the similarity of their self-reported motivations. We first use the Bidirectional Encoder Representations from Transformers (BERT), a pre-trained deep learning model, to convert each worker's open-ended explanations into numerical embeddings that capture the semantic content \citep{devlin2019bert}.\footnote{Specifically, we use the \texttt{all-MiniLM-L6-v2} variant of the BERT model.} For each worker, we generate separate embeddings for their responses to the ex ante and ex post scenarios and concatenate them into a single high-dimensional feature vector. We then apply k-means clustering to these embeddings to group workers into distinct clusters \citep{jain1988algorithms}. The optimal number of clusters is selected using the silhouette score, which measures both tightness and separation of clusters across candidate values \citep{rousseeuw1987silhouettes}. This procedure yields two optimal clusters in our data.

\begin{figure}[ht]
	\centering
\caption{WTP for Peer Information by BERT-based Cluster}
 \vspace{1.0em}
	\begin{subfigure}{0.495\textwidth}
 \caption{Cluster 1}
 \includegraphics[width=1.01\textwidth]{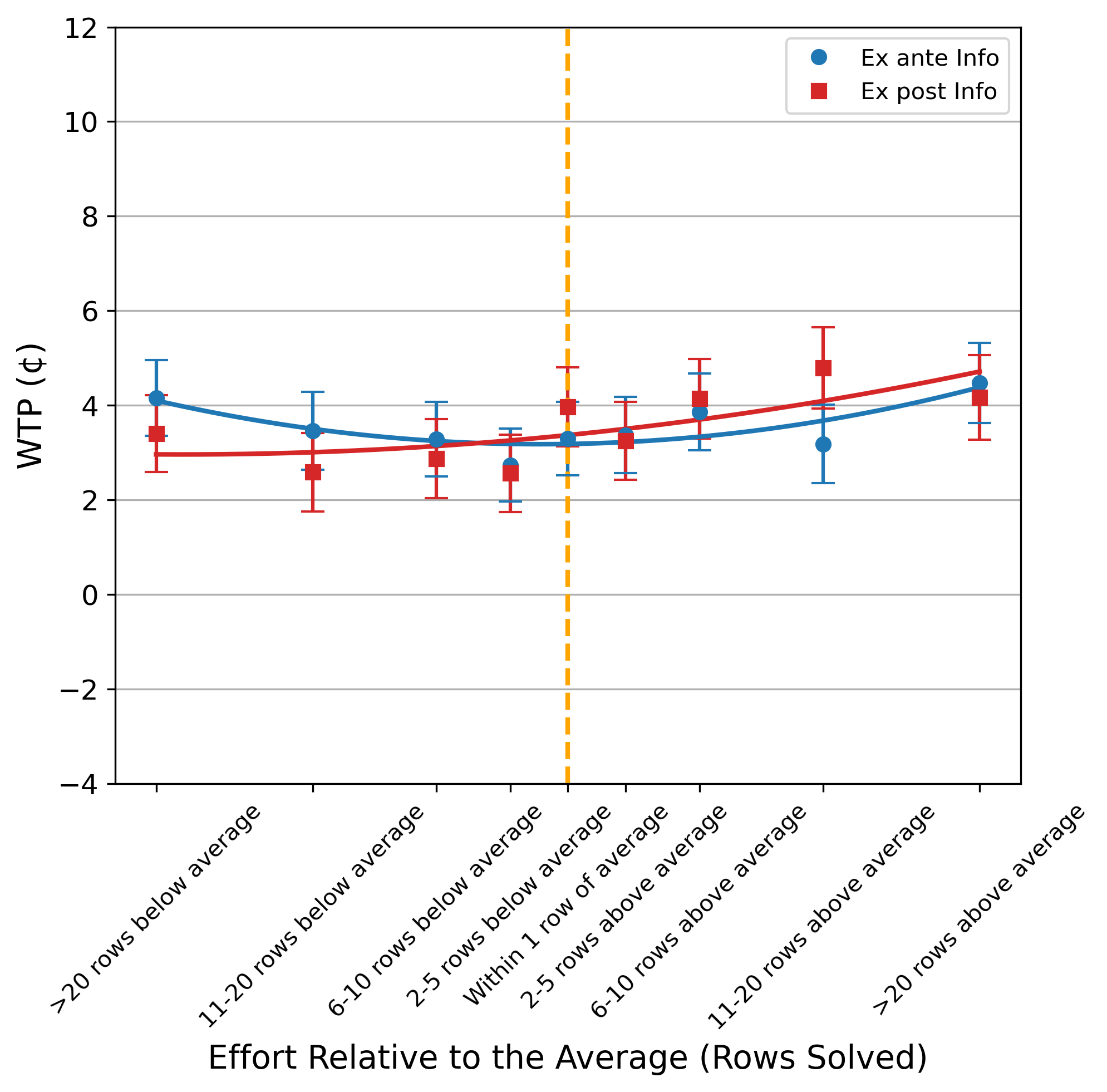} 
  \label{fig:wtp_cluster1} 
  
  \vspace{0.5em}
	\end{subfigure}
	\begin{subfigure}{0.495\textwidth}

 \caption{Cluster 2}
  \includegraphics[width=1.01\textwidth]{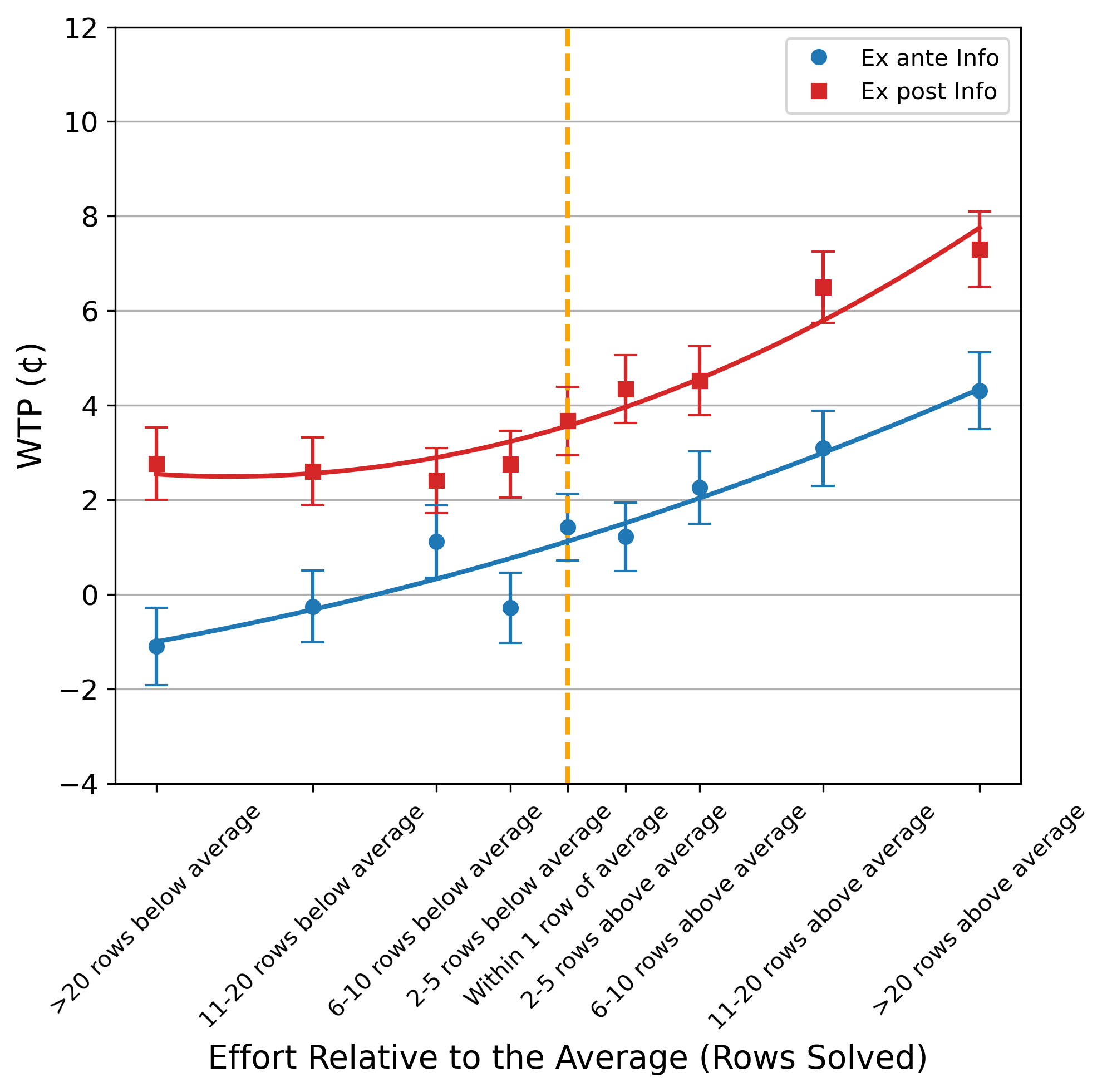} 
  \label{fig:wtp_cluster2} 
  
	\end{subfigure}
	\begin{minipage}{15cm}
		\scriptsize \singlespacing{\emph{Notes.} This figure plots the average WTP for peer information that is provided ex ante or ex post by each cluster. The vertical dashed line corresponds to the average rows solved, i.e., if their performance was within 1 row of the true average. The error bars display $+/-$ one standard error of the mean.} 
	\end{minipage}
\label{fig:wtp_byClusters} 
\vspace{1.0em}
\end{figure}

Figure \ref{fig:wtp_byClusters} presents workers' WTP for peer information in both the ex ante and ex post scenarios by cluster. Panel (a) shows Cluster 1, which comprises 48\% of the sample ($N=336$). These workers exhibit a relatively flat WTP profile, with no significant difference between the ex ante and ex post scenarios. Panel (b) shows Cluster 2, which comprises the remaining 52\% of the sample ($N=357$). In contrast, these workers display an upward-sloping WTP profile, valuing information more as their performance increases, and display significantly higher WTP for receiving information ex post than ex ante across the performance range.

While the clustering exercise reveals meaningful heterogeneity in preferences for information, the two clusters are inherently a ``black box'' from unsupervised machine learning. To interpret them, we examine the distribution of workers’ self-reported motivations using the hand-coded categories introduced earlier. Figure \ref{fig:worker_rationales_byCluster} provides this breakdown by cluster. We find that Cluster 1 consists mainly of workers who report either not caring about peer information or being simply curious about the results. In contrast, Cluster 2 comprises a large share of workers who use peer information for social comparisons or as a goal, along with a significant share citing stress or distraction from receiving information (ex ante). This provides a useful lens for interpreting the different types of workers each cluster captures.  

With this interpretation in mind, we next examine whether workers in each cluster respond differently in effort provision when receiving peer information. Table \ref{tab:table-effort4} presents OLS estimates separately for each BERT-based cluster. Column (1) shows that workers in Cluster 1 do not significantly change their effort whether information is provided ex ante or ex post, and if anything, the point estimates are negative. This is consistent with the earlier observation that Cluster 1 contains many workers who reported having no interest in receiving peer information. In contrast, column (2) reveals that workers in Cluster 2 respond strongly to peer information: receiving information ex ante increases effort by 3.5 rows ($p<0.01$), and receiving it ex post increases effort by 2.6 rows ($p<0.01$). These findings align with the earlier observation that Cluster 2 comprises workers who either use peer information for social comparisons or as a goal to motivate themselves. 

In sum, our results reveal marked differences in how workers value and respond to peer information. This heterogeneity helps explain why the average treatment effects are only marginally significant when pooling all workers (see column (1) of Table \ref{tab:table-effort1}). A central contribution of the paper is to show that workers hold systematically distinct information preferences that strongly predict their effort responses.\footnote{As exploratory analysis, Table \ref{tab:table-wellbeing2} in the Appendix reports heterogeneous effects of peer information on three self-reported measures of worker well-being (i.e., stress level, motivation, and perceived task meaning), which again reveals substantial heterogeneity across clusters.}

{
\singlespacing
\begin{table}[ht]
\centering 
\footnotesize
	\def\sym#1{\ifmmode^{#1}\else\(^{#1}\)\fi}
	\caption{Heterogeneous Impact on Worker Effort by BERT-based Cluster} \label{tab:table-effort4}
	\begin{threeparttable}
    \setlength{\tabcolsep}{18pt} 
    \begin{tabular}{l*{2}{c}}
    \toprule
    \toprule
    \textit{Dependent variable:} & \multicolumn{2}{c}{Effort (\# Rows Solved)} \\ 
    \cmidrule(lr){2-3} 
    & \multicolumn{1}{c}{\textbf{Cluster 1}} & \multicolumn{1}{c}{\textbf{Cluster 2}} \\
    & (1) & (2)  \\ 
    \midrule \addlinespace 

            $ \text{1 if receive info ex ante} $ &       --0.94   &        3.46***\\
            &      (1.18)   &      (0.93)   \\ \addlinespace
$ \text{1 if receive info ex post} $ &       --0.67   &        2.57***\\
            &      (0.99)   &      (0.92)   \\ \addlinespace

        Baseline mean &  23.51 &  27.84 \\
		& (14.15) & (13.21)  \\ \addlinespace
			\midrule \addlinespace
            	Controls & \multicolumn{1}{c}{\ding{51}} & \multicolumn{1}{c}{\textit{\ding{51}}} \\ 
            $p$-value: $\beta_{\text{ex ante}} = \beta_{\text{ex post}}$ & 0.804 & 0.256  \\ \addlinespace
			\midrule \addlinespace
			\(R^{2}\)   &       0.103   &       0.145   \\
            No. of obs. &         610   &         628   \\
    No. of workers & 305 & 314 \\ 
    
			\bottomrule
		\end{tabular}
		\begin{tablenotes}
			\scriptsize\vspace{0.1cm} 
			\item \emph{Notes.} This table reports OLS estimates of the average effects of receiving peer information ex ante and ex post on effort (measured by the number of rows solved), disaggregated by clusters. Each column corresponds to a BERT-based cluster. The controls include gender (1 if female), age, education attainment (1 if college degree), and the log of time taken to complete the study. Standard errors clustered at the worker level in parentheses.
			\item * $p<0.10$, ** $p<0.05$, *** $p<0.01$
		\end{tablenotes}
	\end{threeparttable}
\end{table}

}

\vspace{-1.0em}
\subsection{Welfare Effects of Peer Information}

\begin{figure}[ht]
	\centering
 \vspace{1.0em}
\caption{Realized Payoffs Induced by Peer Information}
 \vspace{0.5em}
 \includegraphics[width=.65\textwidth]{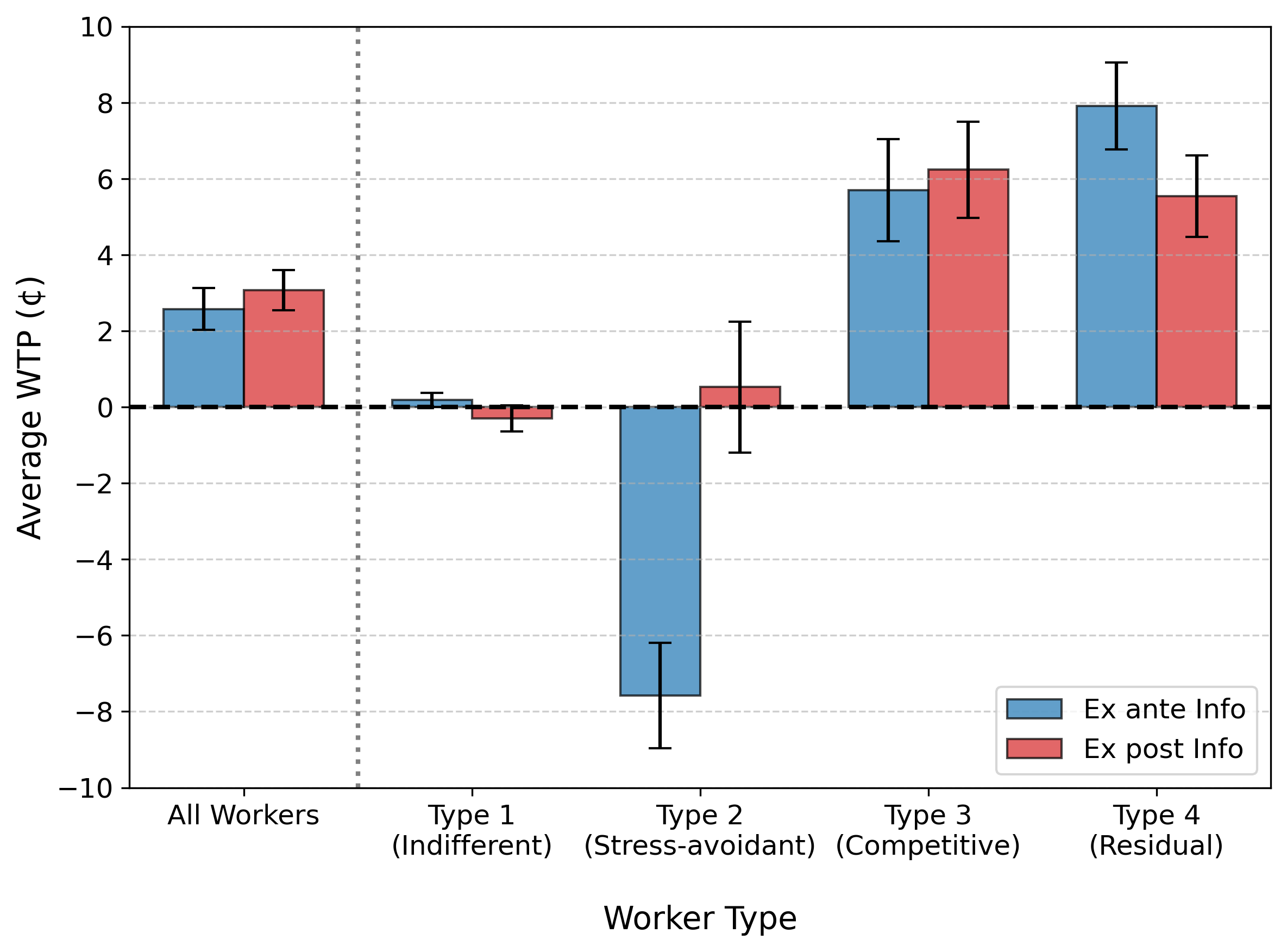} 
  \label{fig:welfare} 
  \vspace{0.2em}
	\begin{minipage}{15cm}
		\scriptsize \singlespacing{\emph{Notes.} This figure plots the average realized payoff of workers assigned to peer information either ex ante or ex post. A worker's payoff is their elicited WTP matched to their actual period 1 performance. The figure shows the payoff for the full sample, as well as a breakdown by the four WTP-based types. The error bars display $+/-$ one standard error of the mean.} 
	\end{minipage}
\vspace{1.0em}
\end{figure}

We conclude our analysis by estimating the realized welfare effects of a uniform policy that provides peer information to all workers either ex ante or ex post. To compute each worker's realized payoff, we use their elicited WTP matched to their actual Period 1 performance. Figure \ref{fig:welfare} presents the average payoff under each policy for the full sample and separately for each WTP-based type.

On average, peer information increases workers' welfare, with slightly higher realized payoffs under ex post provision than ex ante, though the difference is not statistically significant. However, this average payoff masks substantial heterogeneity across types. In particular, Type 2 workers experience strictly negative payoffs from receiving peer information ex ante, reflecting psychological costs such as stress and distraction. By contrast, Type 3 and Type 4 workers derive consistently positive payoffs under both timing regimes. These results underscore that uniform feedback policies may be suboptimal when workers differ in their information preferences.

In our models, the worker's WTP internalizes both the earnings gain (or loss) from effort adjustments induced by peer information and any associated non-monetary costs or benefits. Assuming workers are paid a piece rate equal to their marginal product, these realized payoffs map directly to social welfare effects. In settings where workers’ marginal product exceeds their wage rate, additional surplus would arise from productivity gains accruing to the firm. Thus, our estimates represent a lower bound on the social welfare effects of peer information. 

Finally, to quantify the potential gains from tailoring information, we simulate a counterfactual policy that varies the timing of feedback by type: peer information is provided ex ante to Types 1, 3, and 4, while Type 2 (stress-avoidant) workers only receive it ex post. Assuming the same distribution of types as in our sample, this simple targeting rule increases welfare by 47.6\% relative to a uniform ex ante policy, and by 23.9\% relative to a uniform ex post policy. These estimates suggest that accounting for heterogeneity in information preferences can materially improve the design of feedback policies.

\section{Conclusion} 
\label{sec:conclusion}

This paper studies heterogeneity in workers' demand for peer information and how such heterogeneity shapes worker effort. We develop a conceptual framework that generates predictions for workers' WTP and subsequent effort responses, which we use to organize heterogeneity in the data. Using ex ante WTP measures, we identify four distinct types of information demand: indifferent workers, stress-avoidant workers, competitive workers, and a broader learning-oriented/residual group. The central result is that these differences strongly predict effort responses to receiving peer information. Open-ended survey responses provide complementary evidence on the motives underlying these WTP-based types.

Accounting for heterogeneity in information preferences is critical for two reasons. First, it helps reconcile mixed findings in the performance feedback literature by showing that average treatment effects can mask substantial heterogeneity across workers. Failing to account for this heterogeneity risks understating both the efficacy and unintended costs of peer information. Second, our methodology offers a tractable way to identify distinct types using WTP measures and tailor the timing of feedback. In our experimental setting, such targeting can improve welfare by up to 48\% compared to a uniform policy.

More broadly, our findings have important managerial implications for organizations that deploy peer performance feedback. While such policies are often implemented uniformly across workers, our results show that a one-size-fits-all approach may be suboptimal. The same information can motivate some workers, leave others unaffected, and impose psychological costs on a meaningful subset. Designing feedback policies that account for these heterogeneous preferences and responses can improve worker welfare and potentially organizational performance.

Our experimental setting is deliberately stylized to isolate the underlying mechanisms, abstracting from features such as unequal pay, team-based production, and repeated exposure to feedback. At the same time, our sample of Prolific workers provides a useful benchmark, as these are experienced platform workers who engage in paid, performance-based tasks similar in spirit to gig-economy settings. Notably, 15\% of workers in our study chose to avoid peer information out of stress or distraction even in a relatively low-stakes environment. This suggests that similar or larger shares of stress-avoidant workers (who are most consequential for the welfare assessment of peer information) may arise in settings where performance feedback carries higher stakes.

In a companion study, we adapt our WTP elicitation approach into a survey module deployable in field settings and find similar patterns of heterogeneous preferences and responses to peer information among rideshare drivers. Beyond the workplace, peer information is prevalent in domains such as prosocial behavior, resource conservation, and job choice.\footnote{See, for example, \citet{frey2004social} on charitable giving, \citet{allcott2011social} and \citet{allcott2014short} on energy use, and \citet{coffman2017can} on career choice.} Extending our framework to these settings could reveal new dimensions of heterogeneity and inform the design of more personalized information interventions.

\newpage
\singlespacing
\bibliographystyle{apalike}
\bibliography{ref}

\newpage
\setcounter{table}{0} 
\setcounter{figure}{0}
\renewcommand{\thetable}{A\arabic{table}}
\renewcommand{\thefigure}{A\arabic{figure}}

\section*{\centering \LARGE{Online Appendix}} 
\label{sec:appendix}

\medskip
\section*{A. Supplementary Figures and Tables}

\begin{figure}[H]
	\centering
\caption{WTP for Peer Information by Hand-coded Categories}
\label{fig:types_byWTP} 
 \vspace{1.0em}

 \begin{subfigure}{0.45\textwidth}
	\caption{``Do Not Care''}
\includegraphics[width=1.01\textwidth]{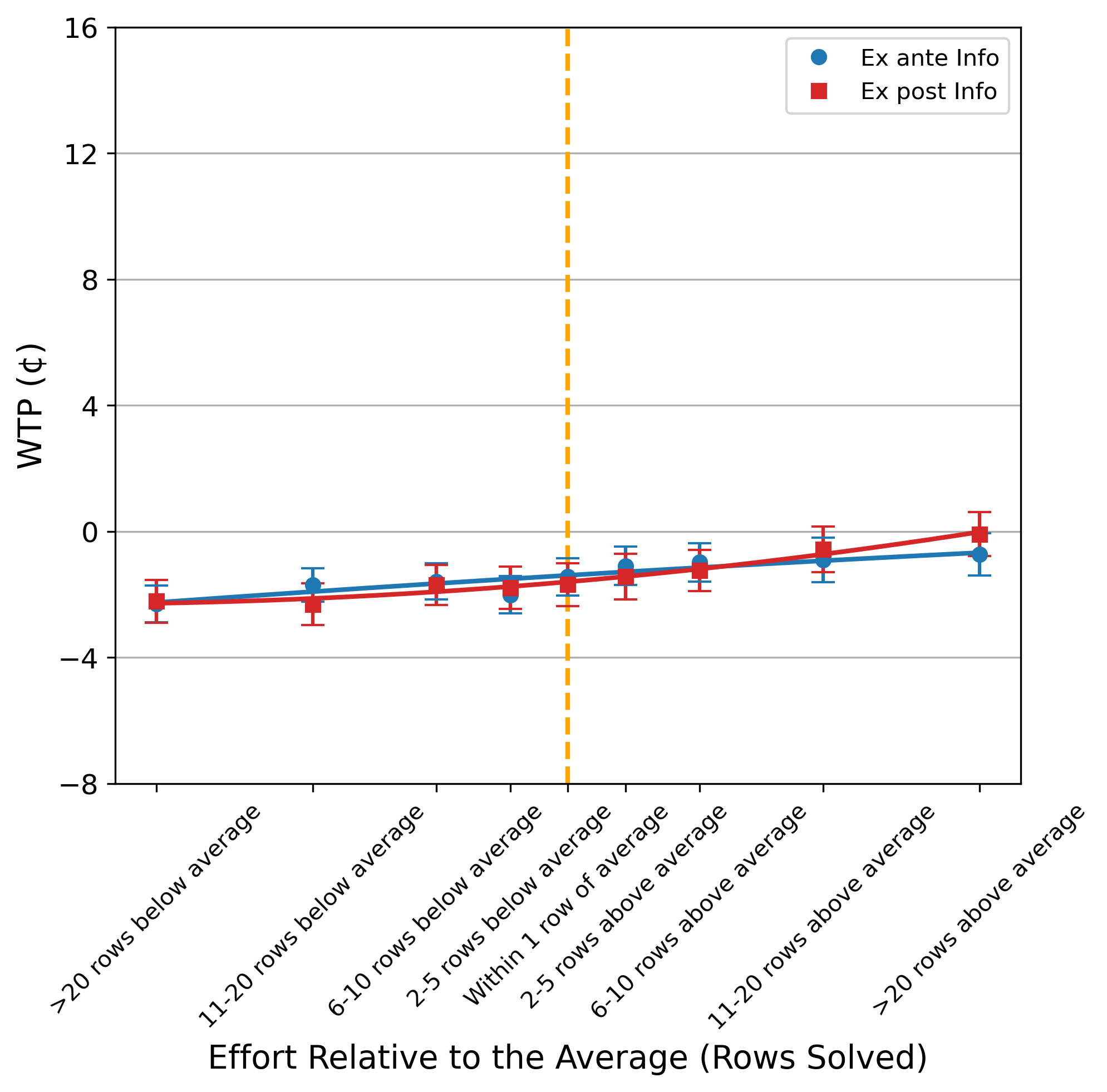} 
\vspace{0.5em}
	\end{subfigure}
	\begin{subfigure}{0.45\textwidth}
	\caption{``Stress or Distraction''}
\includegraphics[width=1.01\textwidth]{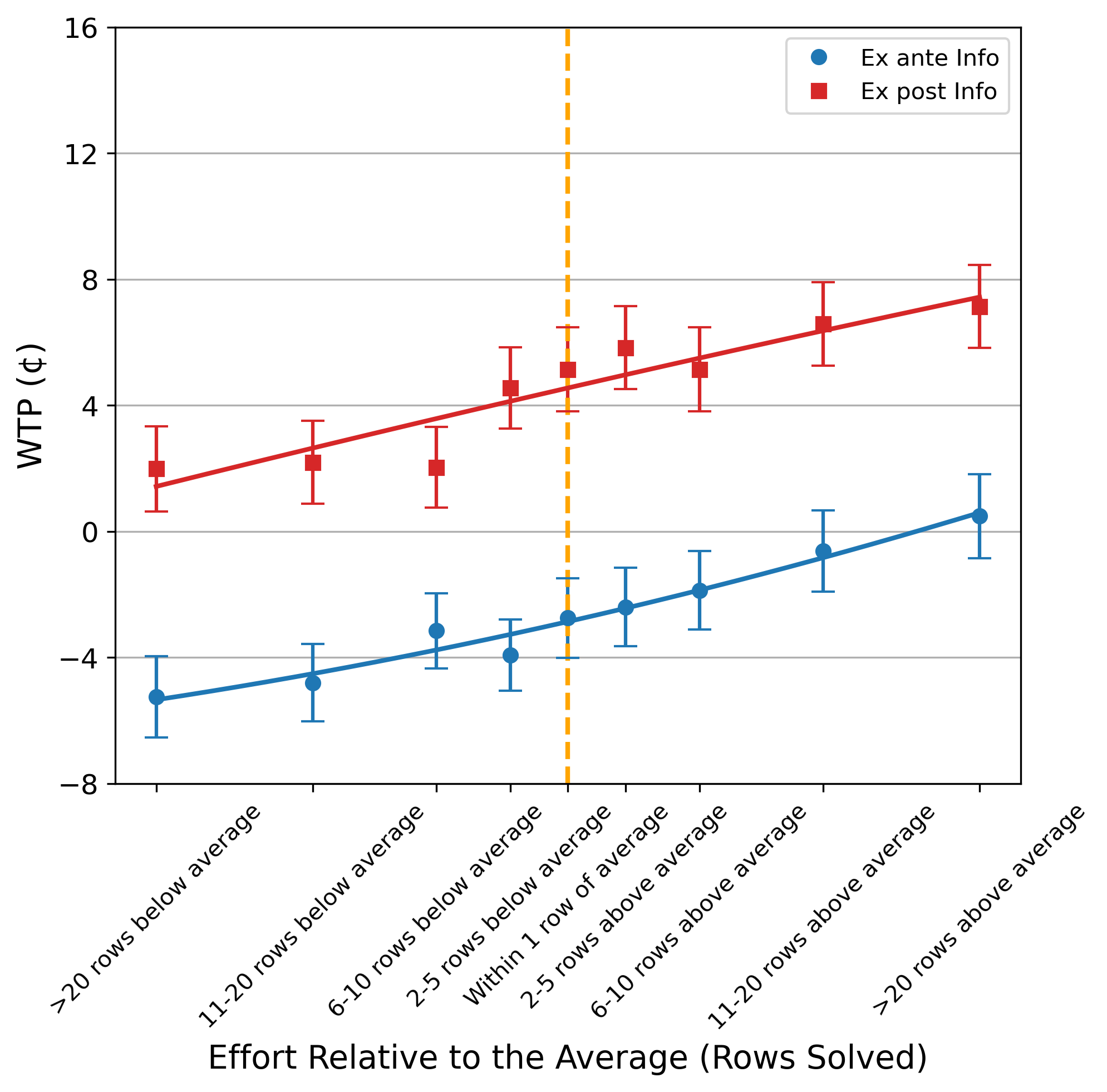} 
\vspace{0.5em}
	\end{subfigure}
    
\vspace{0.5em}
    \begin{subfigure}{0.45\textwidth}
\caption{``Social Comparison''}
\includegraphics[width=1.01\textwidth]{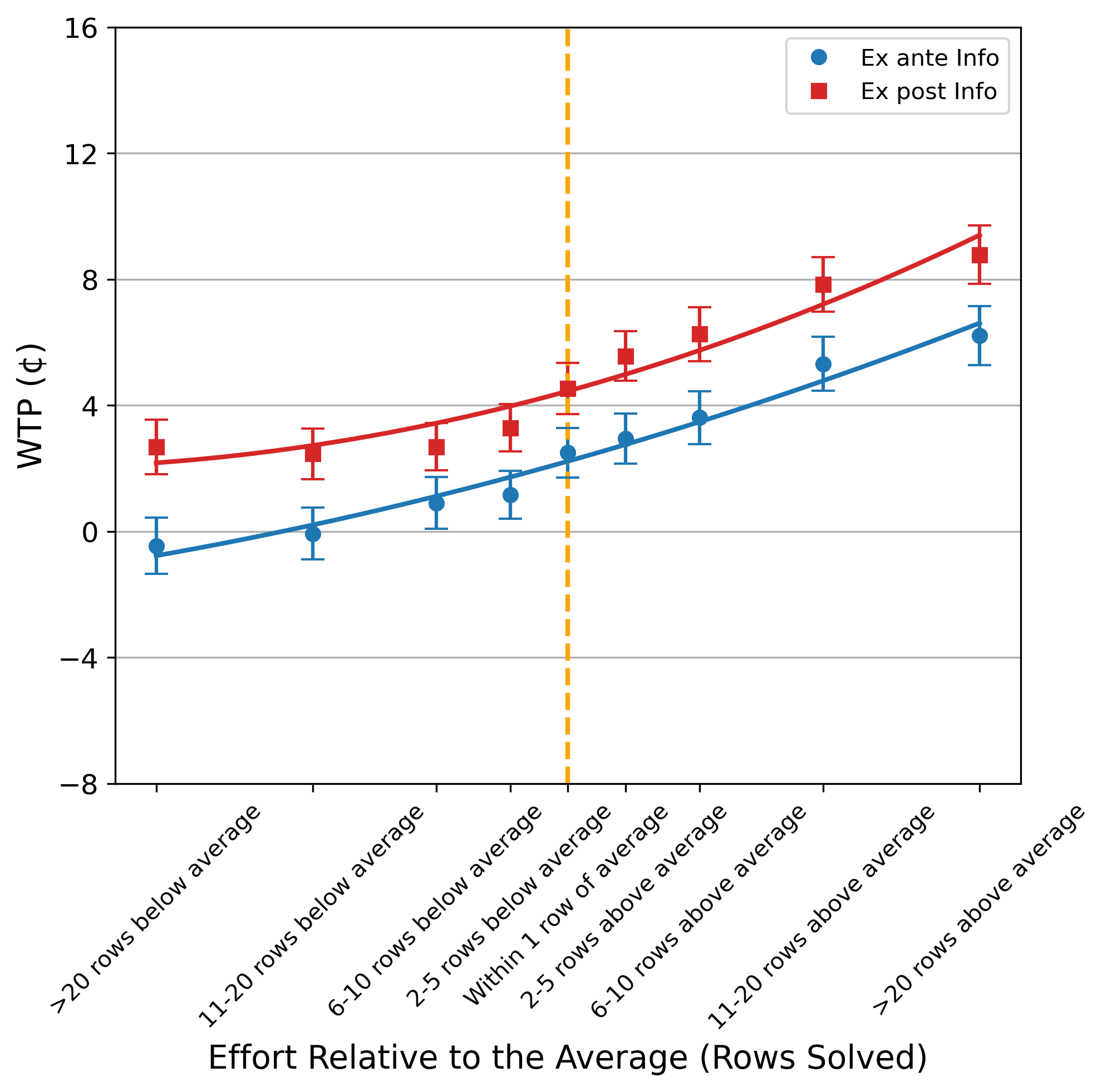} 
\vspace{0.5em}
	\end{subfigure}
	\begin{subfigure}{0.45\textwidth}
	\caption{``Goal or Motivation''}
\includegraphics[width=1.01\textwidth]{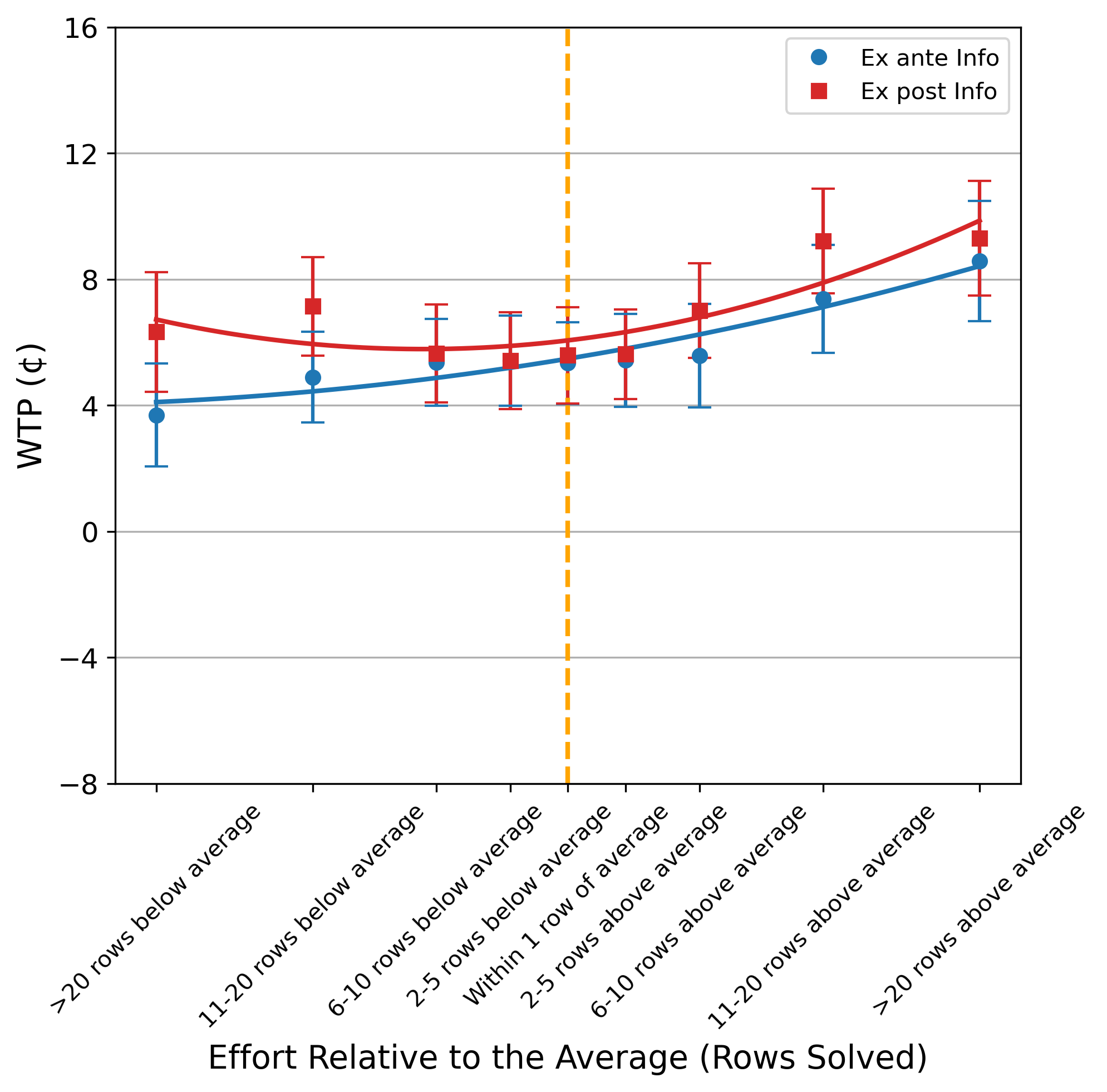} 
\vspace{0.5em}
	\end{subfigure}

	\begin{minipage}{15cm}
		\scriptsize \singlespacing{\emph{Notes.} This figure plots the average WTP for peer information that is provided ex ante or ex post by each hand-coded category. The vertical dashed line corresponds to the average rows solved, i.e., if their performance was within 1 row of the true average. The error bars display $+/-$ one standard error of the mean.} 
	\end{minipage}

\vspace{1.0em}
\end{figure}

\begin{figure}[H]
	\centering
 \vspace{1.5em}
 \ContinuedFloat
\caption{WTP for Peer Information by Hand-coded Categories (\textit{continued})}
 \vspace{1.0em}

 \begin{subfigure}{0.45\textwidth}
	\caption*{(e) ``Curiosity''}
\includegraphics[width=1.01\textwidth]{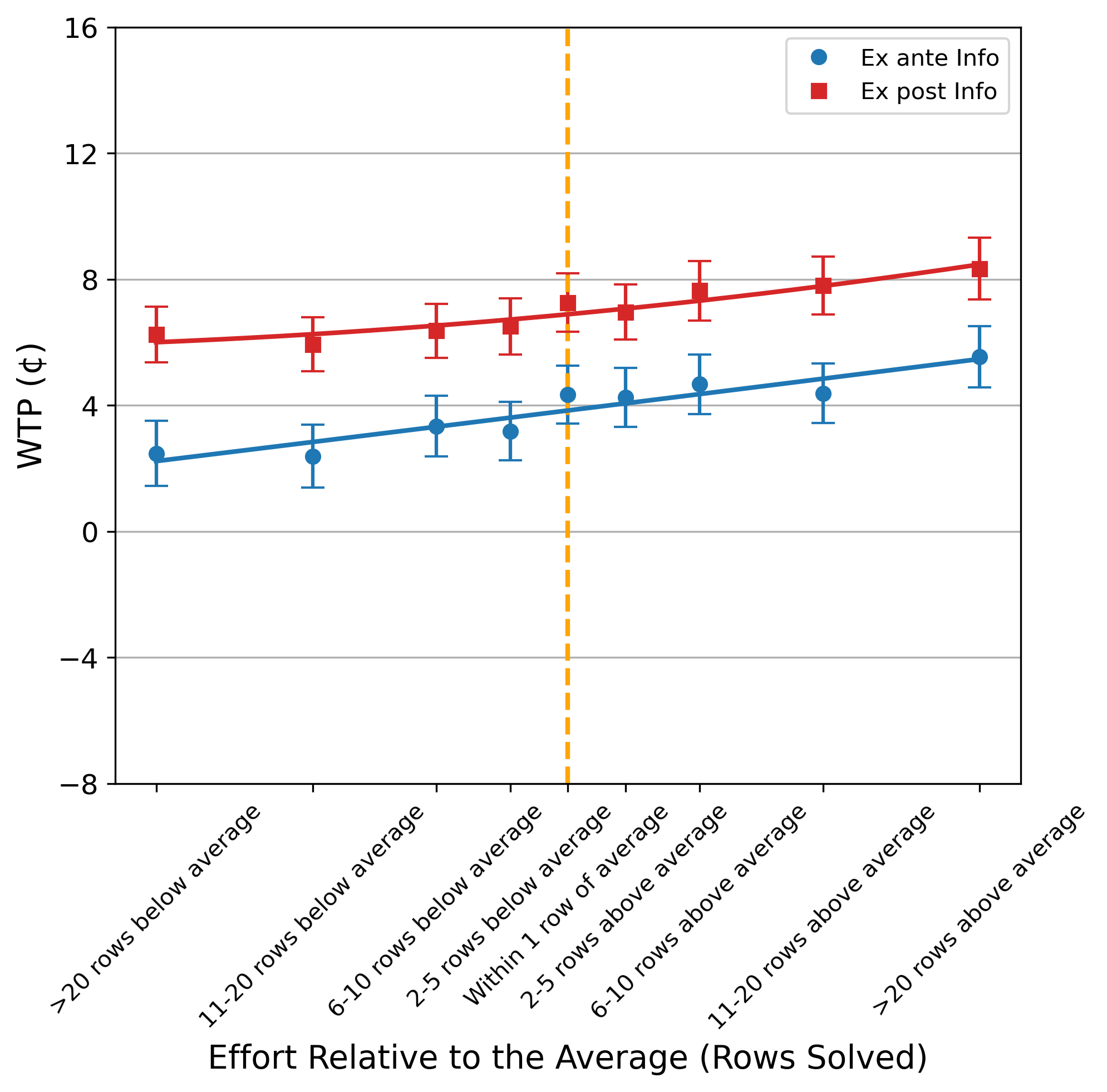} 
\vspace{0.5em}
	\end{subfigure}
	\begin{subfigure}{0.45\textwidth}
	\caption*{(f) ``Value at Extremes''}
\includegraphics[width=1.01\textwidth]{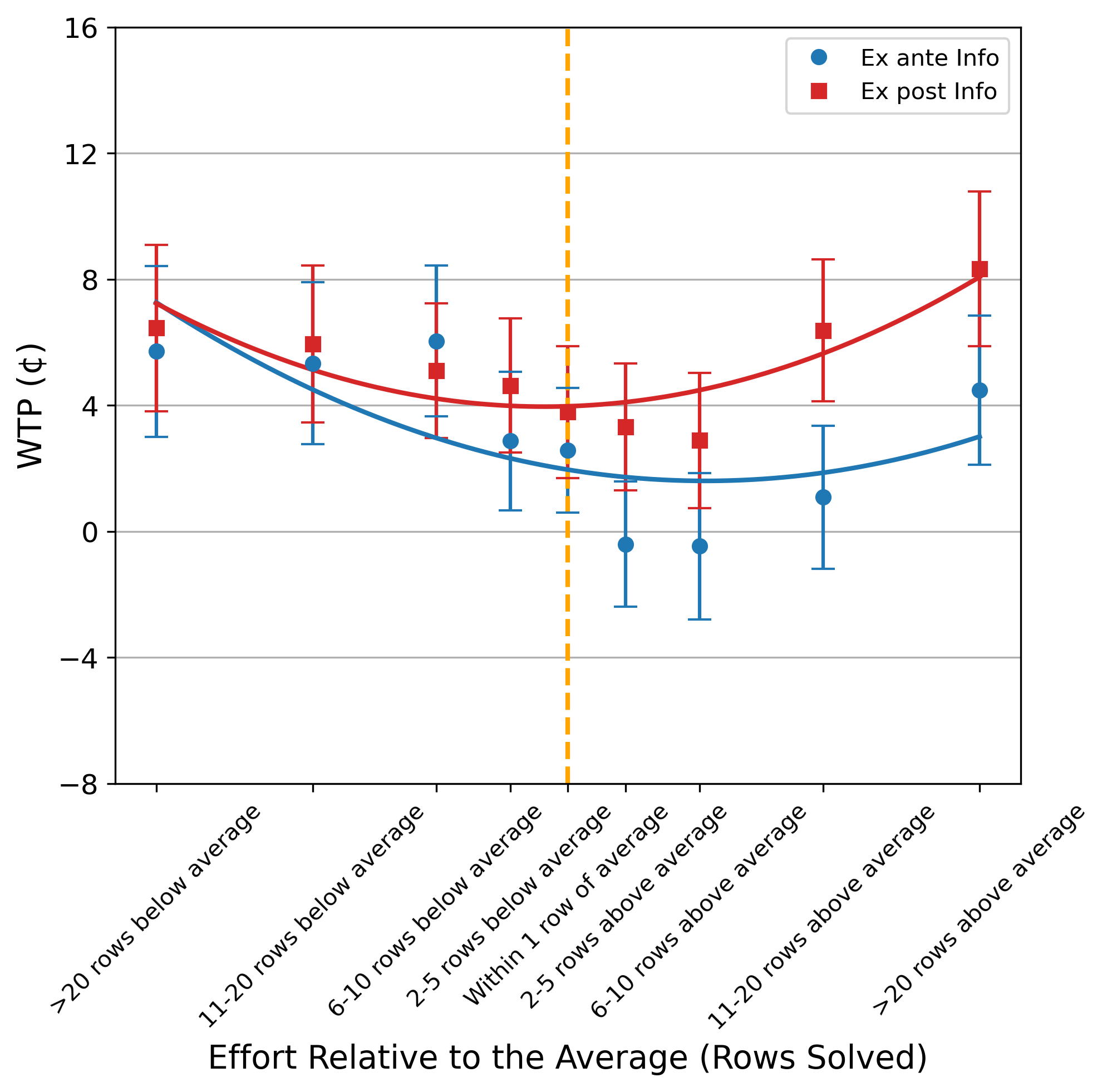} 
\vspace{0.5em}
	\end{subfigure}
    
\vspace{0.5em}
    \begin{subfigure}{0.45\textwidth}
\caption*{(g) ``No Value After Task''}
\includegraphics[width=1.01\textwidth]{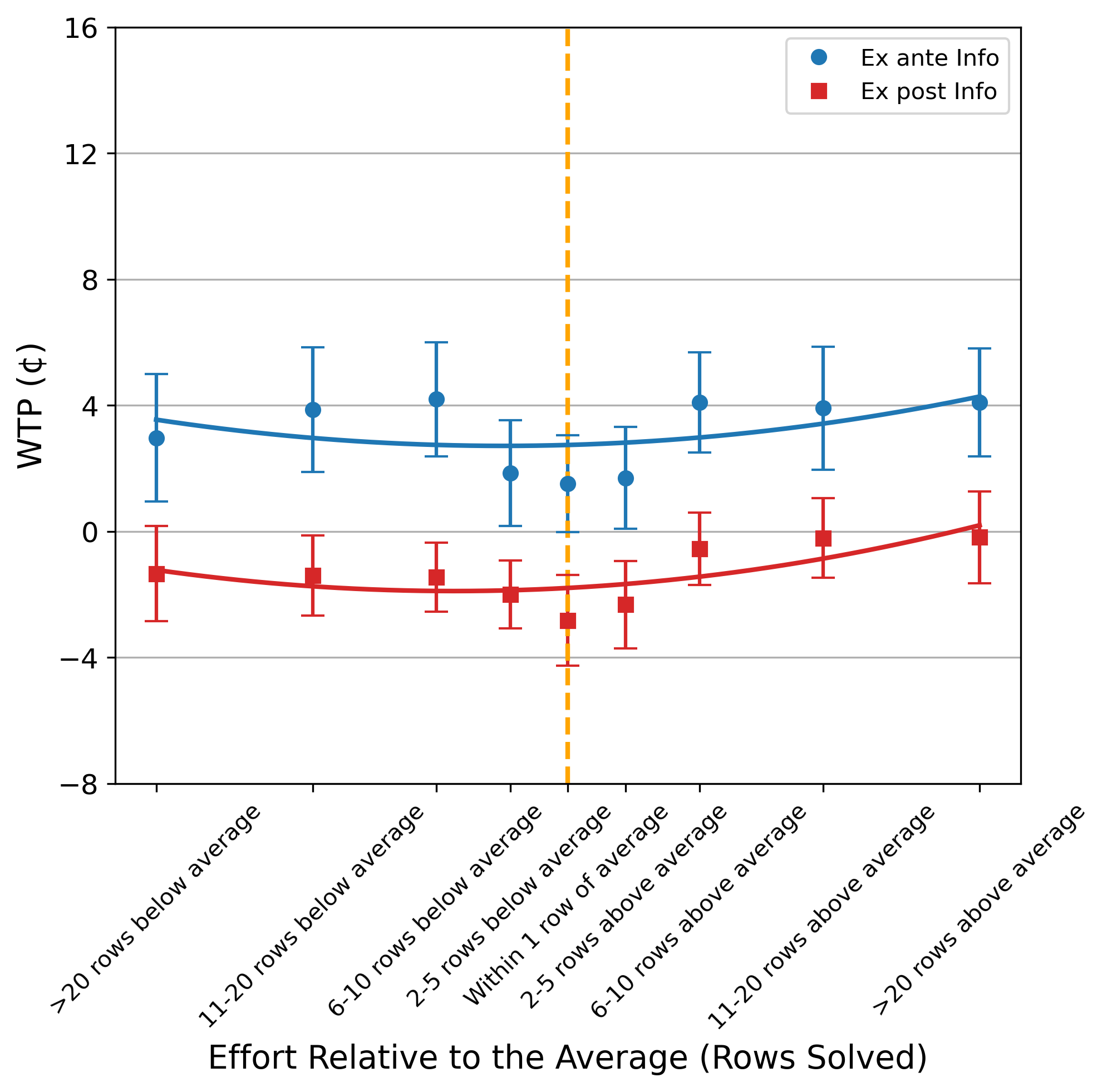} 
\vspace{0.5em}
	\end{subfigure}

	\begin{minipage}{15cm}
		\scriptsize \singlespacing{\emph{Notes.} This figure plots the average WTP for peer information that is provided ex ante or ex post by each hand-coded category. The vertical dashed line corresponds to the average rows solved, i.e., if their performance was within 1 row of the true average. The error bars display $+/-$ one standard error of the mean.}
	\end{minipage}
\vspace{1.0em}
\end{figure}

\begin{figure}[H]
	\centering
 \vspace{1.5em}
\caption{Distribution of Worker Motivations by WTP Preference Type}
 \vspace{0.5em}
 \includegraphics[width=.75\textwidth]{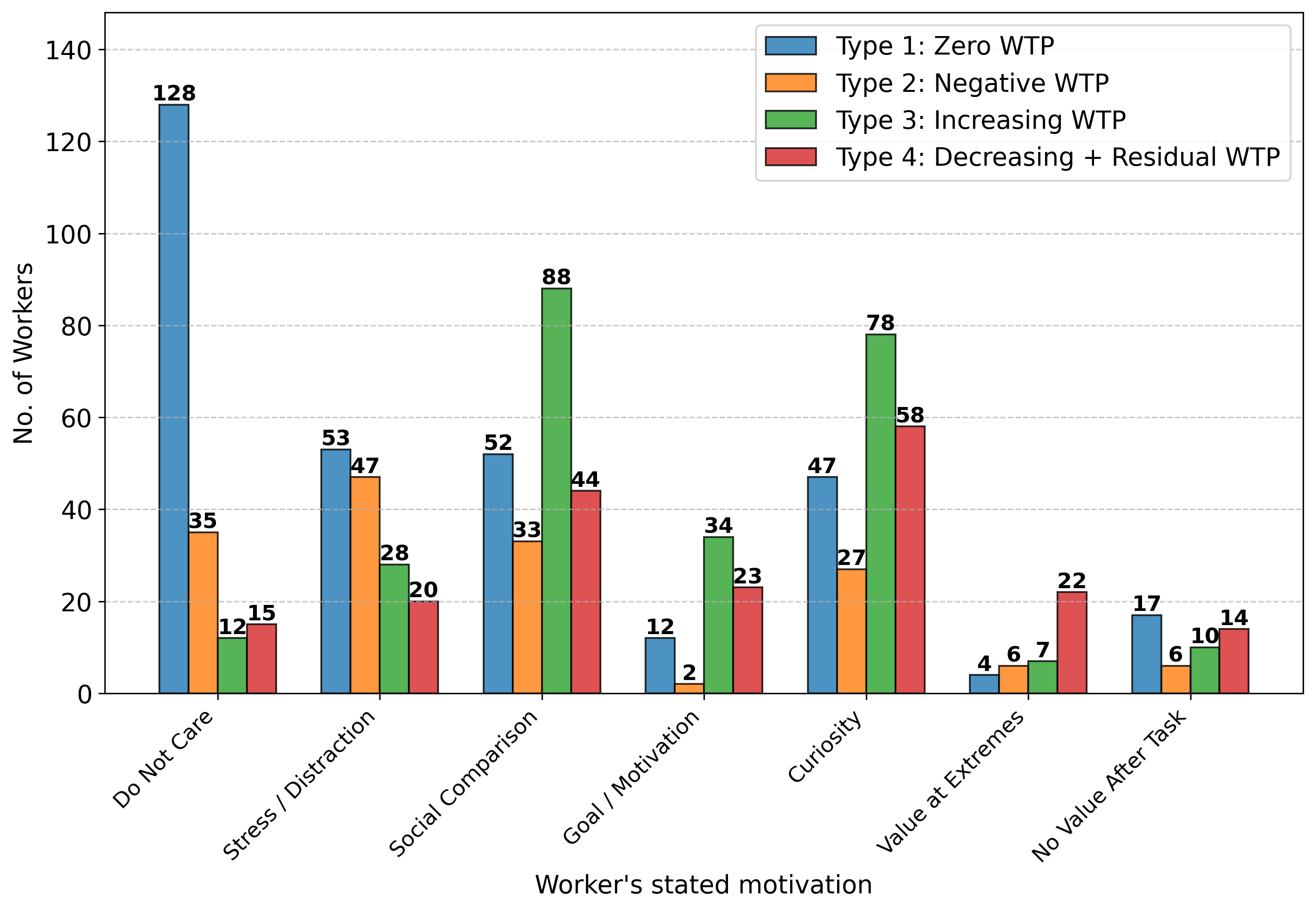} 
  \label{fig:worker_rationales_byTypes} 
  \vspace{0.2em}
	\begin{minipage}{15cm}
		\scriptsize \singlespacing{\emph{Notes.} This figure presents the distribution of workers mentioning different motivations for seeking or avoiding peer information, disaggregated by preference types. Each worker is asked to provide open-ended responses explaining their choice of information in both the ex ante and ex post scenarios. Their responses are categorized into one or more of seven (non-mutually exclusive) categories using a hand-coded scheme.} 
	\end{minipage}
\end{figure}

\newpage
\begin{figure}[H]
	\centering
 \vspace{1.5em}
\caption{Distribution of Worker Motivations by BERT-based Cluster}
 \vspace{0.5em}
 \includegraphics[width=.75\textwidth]{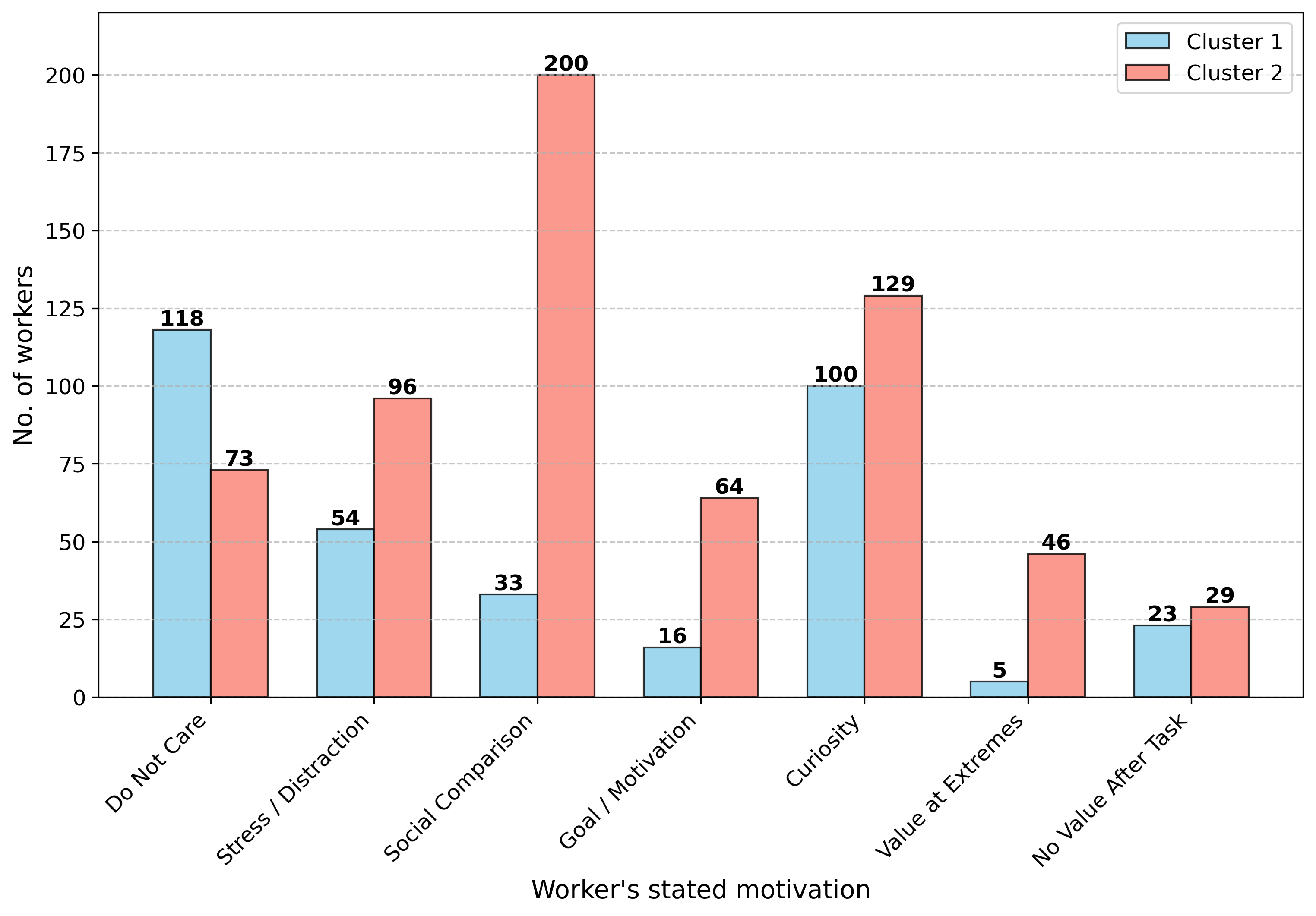} 
  \label{fig:worker_rationales_byCluster} 
  \vspace{0.2em}
	\begin{minipage}{15cm}
		\scriptsize \singlespacing{\emph{Notes.} This figure presents the distribution of workers mentioning different motivations for seeking or avoiding peer information, disaggregated by clusters. Each worker is asked to provide open-ended responses explaining their choice of information in both the ex ante and ex post scenarios. Their responses are categorized into one or more of seven (non-mutually exclusive) categories using a hand-coded scheme.} 
	\end{minipage}
\end{figure}



{
\singlespacing
\begin{table}[H]
    \centering 
        \def\sym#1{\ifmmode^{#1}\else\(^{#1}\)\fi}
	\caption{Sample and Balance} \label{tab:sum_stats}
	\begin{threeparttable}
        \resizebox{\textwidth}{!}{
        \begin{tabular}{l*{6}{>{\centering\arraybackslash}p{0.12\textwidth}}}
			\toprule
            & \textbf{Full Sample} & \textbf{Control} & \textbf{Ex ante Info} & \textbf{Ex post Info} & \textbf{Choose-Your-Info} & \textbf{p-value ($F$-test)} \\ 
            & \textbf{(1)} & \textbf{(2)} & \textbf{(3)} & \textbf{(4)} & \textbf{(5)} & \textbf{(6)}\\ 
			\midrule \addlinespace 
                \textbf{Demographics}   \\ \addlinespace
			1 if Female  & 0.40  &   0.37 &  0.43 & 0.39  & 0.47 & 0.35  \\ \addlinespace
			Age &  39.67 & 38.29 & 39.57  &  41.13   & 39.68 &  0.16  \\ \addlinespace
			1 if College  & 0.55 &  0.59 & 0.49 & 0.59   &  0.54 & 0.11  \\ \midrule \addlinespace
            \textbf{Personality Traits}   \\ \addlinespace
            Risk Taking & 5.10 & 5.12  & 5.13 & 5.06  & 5.08 &  0.99  \\ \addlinespace
		Competitiveness & 6.08 & 5.95 & 6.19 & 6.19 & 5.75 & 0.54 \\  \addlinespace
            Conscientiousness  & 8.20 & 8.26 & 8.25 & 8.10 & 8.2 & 0.76 \\ \addlinespace
            Extrinsic & 7.46 & 7.54 & 7.41 & 7.35 & 7.8 & 0.49 \\ \addlinespace
            Neuroticism & 4.02 & 4.14 & 4.04 & 3.91 & 3.98 & 0.85 \\ \midrule \addlinespace
            \textbf{Study Metrics}   \\  \addlinespace
            Time Taken (in logs) & 7.39 & 7.39 & 7.39 & 7.41  & 7.32 &  0.27  \\ \addlinespace
		\# Rows Attempted & 28.69 & 27.93 & 28.55 & 29.19 & 30.0 & 0.68 \\  \addlinespace
            \# Rows Solved  & 25.87 & 25.60 & 25.65 & 25.93 & 27.38 & 0.84 \\ \addlinespace
			\bottomrule
		\end{tabular}
        }
		\begin{tablenotes}
			\scriptsize\vspace{0.1cm} 
			\item \emph{Notes.} Each p-value is from an F-test of joint significance in an OLS regression of the variable on treatment group indicators.
		\end{tablenotes}
	\end{threeparttable}	
\end{table}
}
\bigskip
\begin{table}[H] 
\centering
	\def\sym#1{\ifmmode^{#1}\else\(^{#1}\)\fi}
	\caption{Preferences for Peer Information by Timing of Receipt} 
 \label{tab:table-wtp1}
	\begin{threeparttable}
		\begin{tabular}{l*{4}{c}}
			\toprule
			\toprule
            \textit{Dependent variable:} & \multicolumn{2}{c}{1 if prefer information} & \multicolumn{2}{c}{WTP (¢)}   \\ 
            \cmidrule(lr){2-3} \cmidrule(lr){4-5} 
			& \multicolumn{1}{c}{(1)} & \multicolumn{1}{c}{(2)} & \multicolumn{1}{c}{(3)} & \multicolumn{1}{c}{(4)}  \\ 
			\midrule \addlinespace 

            	$ \text{Effort range (coded: 1--9)} $ &        0.01***&        0.02***&        0.24***&        0.32***\\
            &      (0.00)   &      (0.00)   &      (0.08)   &      (0.08)   \\ \addlinespace

            \text{Effort range} $\times$ \text{$\mathbf{1}$[ex post info]} &        0.01***&               &        0.24***&               \\
              &      (0.00)   &               &      (0.07)   &               \\ \addlinespace

              \text{$\mathbf{1}$[20+ below avg.]} $\times$ \text{$\mathbf{1}$[ex post info]} &               &        0.11***&               &        1.97***\\
            &               &      (0.02)   &               &      (0.62)   \\ \addlinespace
  \text{$\mathbf{1}$[11--20 below avg.]} $ \times $ \text{$\mathbf{1}$[ex post info]} &               &        0.08***&               &        1.17** \\
            &               &      (0.02)   &               &      (0.59)   \\ \addlinespace
  \text{$\mathbf{1}$[6--10 below avg.]} $ \times$ \text{$\mathbf{1}$[ex post info]} &               &        0.07***&               &        0.89   \\
            &               &      (0.02)   &               &      (0.56)   \\ \addlinespace
  \text{$\mathbf{1}$[2--5 below avg.]} $ \times$ \text{$\mathbf{1}$[ex post info]} &               &        0.06***&               &        0.60   \\
            &               &      (0.02)   &               &      (0.58)   \\ \addlinespace
  \text{$\mathbf{1}$[within 1 of avg.]} $ \times$ \text{$\mathbf{1}$[ex post info]} &               &        0.06***&               &        1.42** \\
            &               &      (0.02)   &               &      (0.56)   \\ \addlinespace
  \text{$\mathbf{1}$[2--5 above avg.]} $ \times$ \text{$\mathbf{1}$[ex post info]} &               &        0.07***&               &        1.10*  \\
            &               &      (0.02)   &               &      (0.56)   \\ \addlinespace
  \text{$\mathbf{1}$[6--10 above avg.]} $ \times$ \text{$\mathbf{1}$[ex post info]} &               &        0.07***&               &        1.30** \\
            &               &      (0.02)   &               &      (0.54)   \\ \addlinespace
  \text{$\mathbf{1}$[11--20 above avg.]} $ \times$ \text{$\mathbf{1}$[ex post info]} &               &        0.10***&               &        2.32***\\
            &               &      (0.02)   &               &      (0.55)   \\ \addlinespace
  \text{$\mathbf{1}$[20+ above avg.]} $ \times$ \text{$\mathbf{1}$[ex post info]} &               &        0.09***&               &        2.11***\\
            &               &      (0.02)   &               &      (0.58)   \\ \addlinespace

			Constant   &        0.39***&        0.35***&        1.27***&        0.77   \\
			  &      (0.02)   &      (0.02)   &      (0.48)   &      (0.58)   \\
            \addlinespace
			\midrule \addlinespace
			\(R^{2}\)   &       0.014   &       0.016   &       0.006   &       0.007   \\
			No. of obs.  &       12474   &       12474   &       12474   &       12474   \\
            No. of workers &        693   &       693   &     693      &      693   \\
		\bottomrule
		\end{tabular}
		\begin{tablenotes}
			\scriptsize\vspace{0.1cm} 
			\item \emph{Notes.} This table reports regression estimates of workers' preferences for peer information and their WTP. 
            Standard errors clustered at worker level in parentheses.
			\item * $p<0.10$, ** $p<0.05$, *** $p<0.01$
		\end{tablenotes}
	\end{threeparttable}	
\end{table}
{
\singlespacing

\begin{table}[H]
\centering 
	\def\sym#1{\ifmmode^{#1}\else\(^{#1}\)\fi}
	\caption{Average Impact of Peer Information on Worker Effort} \label{tab:table-effort1}
	\begin{threeparttable}
		\begin{tabular}{l*{3}{c}}
			\toprule
			\toprule
            \textit{Dependent variable:} & \multicolumn{3}{c}{Effort Level}  \\ 
            \cmidrule(lr){2-4}
            	& \multicolumn{1}{c}{\# Rows Solved} & \multicolumn{1}{c}{\# Rows Attempted} & \multicolumn{1}{c}{Self-Assessed} \\ 
                	& \multicolumn{1}{c}{(1)} & \multicolumn{1}{c}{(2)} & \multicolumn{1}{c}{(3)} \\
			\midrule \addlinespace 

            1 if receive info ex ante &        1.26*  &        1.61** &        0.33** \\
            &      (0.75)   &      (0.76)   &      (0.16)   \\ \addlinespace
1 if receive info ex post  &        0.98   &        1.30*  &        0.10   \\
            &      (0.68)   &      (0.68)   &      (0.16)   \\ \addlinespace
    
			Baseline mean &  25.73 &  28.57 &   9.00 \\
			   & (13.83) & (13.47) & (1.67) \\ \addlinespace
			\midrule \addlinespace
            	Controls & \multicolumn{1}{c}{\ding{51}} & \multicolumn{1}{c}{\textit{\ding{51}}} & \multicolumn{1}{c}{\textit{\ding{51}}} \\ 
            $p$-value: $\beta_{\text{ex ante}} = \beta_{\text{ex post}}$ & 0.677 & 0.650 & 0.134 \\ \addlinespace
			\midrule \addlinespace
			\(R^{2}\)   &       0.095   &       0.103   &       0.019   \\
			No. of obs. &        1238   &        1238   &        1238   \\
            No. of workers &        619   &       619   &     619     \\ 
			\bottomrule
		\end{tabular}
		\begin{tablenotes}
			\scriptsize\vspace{0.1cm} 
			\item \emph{Notes.} This table reports regression estimates of the average effects of receiving peer information ex ante and ex post on effort. Column 1 uses the number of rows solved as the dependent variable, Column 2 uses the number of rows attempted, and Column 3 uses the worker's self-assessed effort on a 0--10 scale. The controls include gender (1 if female), age, education attainment (1 if college degree), and the log of time taken to complete the study. Standard errors clustered at the worker level in parentheses.
			\item * $p<0.10$, ** $p<0.05$, *** $p<0.01$
		\end{tablenotes}
	\end{threeparttable}	
\end{table}

}
{
\singlespacing

\begin{table}[H]
\centering 
	\def\sym#1{\ifmmode^{#1}\else\(^{#1}\)\fi}
	\caption{Average Impact of Peer Information on Worker Well-Being} \label{tab:table-wellbeing1}
	\begin{threeparttable}
    \setlength{\tabcolsep}{12pt} 
		\begin{tabular}{l*{3}{c}}
			\toprule
			\toprule
            \textit{Dependent variable:} & \multicolumn{3}{c}{Measures of Well-Being (0--10 scale)}  \\ 
            \cmidrule(lr){2-4}
            	& \multicolumn{1}{c}{Stress Level} & \multicolumn{1}{c}{Motivation} & \multicolumn{1}{c}{Work Meaning} \\ 
                	& \multicolumn{1}{c}{(1)} & \multicolumn{1}{c}{(2)} & \multicolumn{1}{c}{(3)} \\
			\midrule \addlinespace 

            $ \text{1 if receive info ex ante} $ & 0.29   & 0.37** & 0.39** \\
        & (0.20)   & (0.18)   & (0.18)   \\ \addlinespace
        $ \text{1 if receive info ex post} $ & $-$0.16   & 0.04   & $-$0.10   \\
        & (0.20)   & (0.19)   & (0.17)   \\ \addlinespace
        Baseline mean   & 3.86 & 8.87 & 6.08 \\
        & (3.01)   & (1.84)   & (3.00)    \\ 
            \addlinespace
            \midrule \addlinespace
            	Controls & \multicolumn{1}{c}{\ding{51}} & \multicolumn{1}{c}{\textit{\ding{51}}} & \multicolumn{1}{c}{\textit{\ding{51}}} \\ 
            $p$-value: $\beta_{\text{ex ante}} = \beta_{\text{ex post}}$ & 0.029 & 0.049 & 0.006 \\ \addlinespace
			\midrule \addlinespace
			\(R^{2}\)   & 0.127   & 0.219   & 0.118   \\
			No. of obs. &        1238   &        1238   &        1238   \\
            No. of workers &        619   &       619   &     619     \\ 
			\bottomrule
		\end{tabular}
		\begin{tablenotes}
			\scriptsize\vspace{0.1cm} 
			\item \emph{Notes.} This table reports regression estimates of the average effects of receiving peer information ex ante and ex post on three different measures of well-being. Column 1 uses the reported stress level as the dependent variable, Column 2 uses the motivation level, and Column 3 uses the worker's perceived meaning of work on a 0--10 scale. The controls include gender (1 if female), age, education attainment (1 if college degree), and the log of time taken to complete the study. Standard errors clustered at the worker level in parentheses.
			\item * $p<0.10$, ** $p<0.05$, *** $p<0.01$
		\end{tablenotes}
	\end{threeparttable}	
    \vspace{1.0em}
\end{table}

}
\bigskip
\begin{table}[H] 
\centering
\caption{Coding Scheme for Open-ended Responses With Examples}
\label{tab:table-codingscheme}

\resizebox{\textwidth}{!}{
\begin{threeparttable}
    \begin{tabular}{@{}p{0.25\textwidth} p{0.35\textwidth} p{0.4\textwidth}@{}}
        \toprule
        \textbf{Category} & \textbf{Explanation} & \textbf{Example} \\
        \midrule

        Do Not Care & The participant specified that they do not care or want to know what the average score is. & ``I was not interested in seeing the average performance metrics''; ``I did not care about seeing the average performance.'' \\

        \midrule
        
        Stress or Distraction & The participant was worried about how knowing the information would impact their stress levels or did not want to be distracted. & ``I would prefer not having the information as it would probably make me more anxious ... ''; ``I want to take the test without any stress ... ''; ``I didn't want the results to affect my mindset going into the 2nd task'' \\

        \midrule
        
        Social Comparison & The participant specified that they only wanted to know the average if they were doing well, or they wanted to avoid knowing the information if they were doing poorly. & ``I would want the information after if I was above average ... This will make me feel good about myself ... ''; ``It would give me an ego boost if I happened to be above average.''; ``I would like to know how I stacked up to everyone else.''\\

        \midrule
        




        Goal or Motivation & The participant thought of the average as a goal to achieve or a source of motivation to do better in the next round. & ``It would have been an extra motivating factor if I knew what number I was trying to beat the whole time''; ``I think that seeing the average performance before the task can help to motivate me.''\\

        \midrule

        Curiosity & The participant specified that they were simply curious or interested to see the results. & ``I was curious to know how I compared after I completed the task.''; ``i was just curious afterwards to know how i did''\\

        \midrule

        Value at Extremes & The participant specified that they value information more when their performance is far from the average, especially when they perform much worse than average so they can improve their strategy. & ``I wanted information if I was below average to adjust my approach.''; ``If I was very below average I would want to know the information before the task so I could do better.''; ``I'd want to know if I performed well above or below the mean score''\\

        \midrule

        No Value After Task & The participant mentioned that they did not think it was useful to see the average after they completed the second run. & ``The information is of no value to me after the task ... ''; ``I felt like it was sort of useless to find out after the task how close I was ... ''\\
        
        \bottomrule
    \end{tabular}
    
    \begin{tablenotes}
        \scriptsize\vspace{0.1cm}
        \item \emph{Notes.} This table provides an overview of the qualitative coding scheme used for categorizing workers' open-ended responses. 
    \end{tablenotes}
    
\end{threeparttable}
}
\end{table}
{
\singlespacing
\begin{table}[H]
\centering 
	\def\sym#1{\ifmmode^{#1}\else\(^{#1}\)\fi}
	\caption{Heterogeneous Impact on Worker Well-Being by BERT-based Cluster} \label{tab:table-wellbeing2}
	\begin{threeparttable}
    \resizebox{\textwidth}{!}{
    \begin{tabular}{l*{6}{c}}
    \toprule
    \toprule
    \textit{Dependent variable:} & \multicolumn{6}{c}{Measures of Well-Being (0--10 scale)}  \\ 
    \cmidrule(lr){2-7}
     & \multicolumn{2}{c}{Stress Level} & \multicolumn{2}{c}{Motivation} & \multicolumn{2}{c}{Work Meaning} \\ 
    \cmidrule(lr){2-3} \cmidrule(lr){4-5} \cmidrule(lr){6-7} 
    & \multicolumn{1}{c}{\textbf{Cluster 1}} & \multicolumn{1}{c}{\textbf{Cluster 2}} & \multicolumn{1}{c}{\textbf{Cluster 1}} & \multicolumn{1}{c}{\textbf{Cluster 2}} & \multicolumn{1}{c}{\textbf{Cluster 1}} & \multicolumn{1}{c}{\textbf{Cluster 2}} \\
    & (1) & (2) & (3) & (4) & (5) & (6) \\ 
    \midrule \addlinespace 

    $ \text{1 if receive info ex ante} $
    &        0.55** &        0.01   &        0.09   &        0.65***&       --0.13   &        0.91***\\
            &      (0.27)   &      (0.29)   &      (0.28)   &      (0.24)   &      (0.27)   &      (0.24)   \\  \addlinespace
$ \text{1 if receive info ex post} $  
&        0.32   &       --0.64** &       --0.33   &        0.39   &       --0.46*  &        0.25   \\
            &      (0.29)   &      (0.29)   &      (0.26)   &      (0.28)   &      (0.25)   &      (0.22)   \\  \addlinespace

			Baseline mean &   4.10 &   3.64 &   8.80 &   8.94 &   6.23 &   5.94 \\
			   & (3.24) & (2.75) & (1.99) & (1.69) & (3.07) & (2.92)  \\ \addlinespace
			\midrule \addlinespace
            	Controls & \multicolumn{1}{c}{\ding{51}} & \multicolumn{1}{c}{\textit{\ding{51}}} & \multicolumn{1}{c}{\textit{\ding{51}}} & \multicolumn{1}{c}{\textit{\ding{51}}} & \multicolumn{1}{c}{\textit{\ding{51}}} & \multicolumn{1}{c}{\textit{\ding{51}}} \\ 
            $p$-value: $\beta_{\text{ex ante}} = \beta_{\text{ex post}}$ & 0.449 & 0.020 & 0.120 & 0.184 & 0.227 & 0.004 \\ \addlinespace
			\midrule \addlinespace
			\(R^{2}\)   &       0.066   &       0.058   &       0.054   &       0.035   &       0.112   &       0.069   \\

            No. of obs. &         610   &         628   &         610   &         628   &         610   &         628   \\
    No. of workers & 305 & 314  & 305 & 314 & 305 & 314  \\ 
    
			\bottomrule
		\end{tabular}
        }
		\begin{tablenotes}
			\scriptsize\vspace{0.1cm} 
            \parbox{.95\textwidth}{
			\item \emph{Notes.} This table reports regression estimates of the average effects of receiving peer information ex ante and ex post on three different measures of well-being, separately for each BERT-based cluster.
            Columns 1 and 2 use the reported stress level as the dependent variable, Columns 3 and 4 use the motivation level, and Column 5 and 6 use the worker's perceived meaning of work on a 0--10 scale.
            The controls include gender (1 if female), age, education attainment (1 if college degree), and the log of time taken to complete the study. Standard errors clustered at the worker level in parentheses.
			\item * $p<0.10$, ** $p<0.05$, *** $p<0.01$
            }
		\end{tablenotes}
	\end{threeparttable}	
    \vspace{1.0em}
\end{table}

}
\bigskip
\begin{table}[H] 
\centering
	\def\sym#1{\ifmmode^{#1}\else\(^{#1}\)\fi}
	\caption{WTP for Peer Information and Beliefs About Peers' Performance} 
 \label{tab:table-wtp2}
	\begin{threeparttable}
		\begin{tabular}{l*{4}{c}}
			\toprule
			\toprule
            \textit{Dependent variable:} & \multicolumn{4}{c}{Absolute value of WTP (¢)}   \\ 
            \cmidrule(lr){2-5}
            & \multicolumn{2}{c}{Ex ante} & \multicolumn{2}{c}{Ex post} \\ 
            \cmidrule(lr){2-3} \cmidrule(lr){4-5} 
			& \multicolumn{1}{c}{(1)} & \multicolumn{1}{c}{(2)} & \multicolumn{1}{c}{(3)} & \multicolumn{1}{c}{(4)}  \\ 
			\midrule \addlinespace 

            	$ \text{Belief variance} $ &        0.01***&               &        0.01***&               \\
                   &      (0.00)   &               &      (0.00)   &               \\ \addlinespace

	$ \text{1 if above-median belief variance} $ &               &        2.61***&               &        2.65***\\
              &               &      (0.95)   &               &      (0.99)   \\ \addlinespace

            	$ \text{Belief mean} $   &        0.03   &        0.06   &        0.03   &        0.06   \\
             &      (0.04)   &      (0.04)   &      (0.04)   &      (0.04)   \\ \addlinespace

            $ \text{Effort range (coded: 1--9)} $ &        0.10** &        0.10** &        0.24***&        0.24***\\
            &      (0.05)   &      (0.05)   &      (0.04)   &      (0.04)   \\ \addlinespace
            
			Constant    &        3.60***&        4.35***&        2.87***&        3.67***\\
			   &      (1.11)   &      (1.08)   &      (1.08)   &      (1.06)   \\
            \addlinespace
			\midrule \addlinespace
			\(R^{2}\)    &       0.031   &       0.019   &       0.035   &       0.021   \\
			No. of obs.  &        6237   &        6237   &        6237   &        6237   \\
            No. of workers &        693   &       693   &     693      &      693   \\
		\bottomrule
		\end{tabular}
		\begin{tablenotes}
			\scriptsize\vspace{0.1cm} 
			\item \emph{Notes.} This table tests the prediction from Appendix C.8 that the magnitude (absolute value) of workers' WTP for peer information increases with belief variance, $\sigma_p^2$. Beliefs $p(\bar{e})$ are elicited as the percent chance that the average performance of other workers falls within eight predefined intervals (0-10 rows, 11-20 rows, ..., 61-70 rows, and 70+ rows). Each column reports OLS estimates of WTP (ex ante or ex post) on belief variance, controlling for the mean of beliefs and the effort range at which WTP is elicited. Odd-numbered columns use belief variance as a continuous measure, while even-numbered columns use an indicator for above-median belief variance.
            Standard errors clustered at worker level in parentheses.
			\item * $p<0.10$, ** $p<0.05$, *** $p<0.01$
		\end{tablenotes}
	\end{threeparttable}	
\end{table}

\newpage
\setcounter{table}{0} 
\setcounter{figure}{0}
\renewcommand{\thetable}{B\arabic{table}}
\renewcommand{\thefigure}{B\arabic{figure}}

\section*{B. Robustness of WTP-based Classification}

Recall that for each worker, we elicit willingness-to-pay at nine contingent performance levels, denoted by $WTP_i$ for $i \in \{1, ... 9\}$. Specifically, $WTP_1$ corresponds to performance $\geq$ 20 rows below average, $WTP_2$ to performance 11-20 rows below average, and so on up to $WTP_9$, which corresponds to performance $\geq$ 20 rows above average. 

In the main text, we type workers based on their ex ante WTP at three representative points along the performance distribution: $WTP_1$ (below average),  $WTP_5$ (average), and $WTP_9$ (above average). Including additional points increases the number of logical conditions that must hold (e.g., $WTP = 0$ for all $i$ to be Type 1), which can lead to an underestimation of each worker type and inflate the residual category. Table \ref{tab:robustness_typing} confirms this pattern: as we increase the number of WTP points from three to five, seven, and nine, the combined shares of Types 1--3 declines steadily, while the residual share (Type 4) increases mechanically from 30 to 42 percent. The relative shares of Types 1 and 2 remain remarkably stable, while the share of Type 3 declines slightly since the monotonicity condition is more likely to be violated when additional WTP points are included. Importantly, the relative ordering of types is preserved, indicating that the typology (the underlying distribution of information preferences) is robust to using stricter classification criteria. We thus use the three-point classification for parsimony, as it provides meaningful variation across the full range of relative performance while limiting potential misclassification arising from noisy WTP responses.

\bigskip
\begin{table}[htbp]
\centering
\caption{Worker Type Shares under Alternative Classification Schemes}
\label{tab:robustness_typing}
\begin{threeparttable}
\begin{tabular}{lcccc}
\toprule
 & \multicolumn{4}{c}{\textbf{Classification Scheme}} \\
 \cmidrule(lr){2-5}
\textbf{Worker Type} & \textbf{3-point} & \textbf{5-point} & \textbf{7-point} & \textbf{9-point} \\
\midrule
Type 1 (Indifferent)                  & 32.0\% & 30.9\% & 29.9\% & 29.4\% \\ 
Type 2 (Stress-avoidant)              & 15.2\% & 13.3\% & 13.9\% & 13.9\% \\ 
Type 3 (Competitive)                  & 23.2\% & 16.9\% & 15.7\% & 14.4\% \\ 
Type 4 (Learning + Residual) & 29.6\% & 39.0\% & 40.6\% & 42.3\% \\
\bottomrule
\end{tabular}
\begin{tablenotes}[flushleft]
\footnotesize
\item \textit{Notes.} This table reports the relative shares of worker types (Types 1--4) under alternative classification schemes. The 3-point classification scheme uses WTP responses at (1,5,9), the 5-point classification scheme at (1,3,5,7,9), the 7-point classification scheme at (1,2,3,5,7,8,9), and the 9-point classification scheme at all points. 
\end{tablenotes}
\end{threeparttable}
\end{table}
\bigskip

Among all possible three-point classifications, our preferred choice is to use WTP responses at $(1,5,9)$ as it most saliently captures the the bottom, middle, and top of the distribution when thinking about relative performance. To verify that our typology is not sensitive to the specific choice of performance intervals, we consider all three-point combinations that include the midpoint of the distribution ($WTP_5$) and pair it with one interval below and one above the midpoint, each at least two bins away. This procedure yields nine possible WTP combinations: (1,5,9), (1,5,8), (1,5,7), (2,5,9), (2,5,8), (2,5,7), (3,5,9), (3,5,8), and (3,5,7).

Table \ref{tab:robustness_triplets} reports the estimated shares of worker types under each combination. Across all combinations, the estimated shares remain highly stable. As a measure of inter-rater reliability, we compute the Krippendorff’s $\alpha$ by treating each three-point combination as an independent ``rater'' assigning workers to one of four nominal categories (Types 1--4). The resulting value of $\alpha = 0.81$ indicates a good level of agreement, confirming that our preferred three-point WTP summary captures stable underlying heterogeneity in worker types rather than artifacts of interval selection.

\bigskip
\begin{table}[H]
\centering
\caption{Worker Type Shares under Alternative 3-Point Combinations}
\label{tab:robustness_triplets}
\begin{threeparttable}
\resizebox{\textwidth}{!}{%
\begin{tabular}{lccccccccc}
\toprule
& \multicolumn{9}{c}{\textbf{3-point Combination}} \\
 \cmidrule(lr){2-10}
\textbf{Worker Type} & (1,5,9) & (1,5,8) & (1,5,7) & (2,5,9) & (2,5,8) & (2,5,7) & (3,5,9) & (3,5,8) & (3,5,7) \\
\midrule
Type 1 (Indifferent)     & 32.0\% & 31.9\% & 33.5\% & 32.6\% & 32.8\% & 34.5\% & 32.6\% & 33.3\% & 35.5\% \\ 
Type 2 (Stress-avoidant)  & 15.2\% & 15.0\% & 15.3\% & 14.7\% & 14.7\% & 15.9\% & 15.0\% & 15.9\% & 16.9\% \\ 
Type 3 (Competitive)      & 23.2\% & 22.9\% & 22.4\% & 24.7\% & 23.5\% & 22.4\% & 26.6\% & 24.8\% & 22.5\% \\ 
Type 4 (Learning + Residual) & 29.6\% & 30.2\% & 28.9\% & 28.0\% & 29.0\% & 27.3\% & 25.8\% & 26.0\% & 25.1\% \\
\bottomrule
\end{tabular}
}
\begin{tablenotes}[flushleft]
\footnotesize
\parbox{.95\textwidth}{
\item \textit{Notes.} This table reports the relative shares of worker types (Types 1--4) under alternative three-point combinations of WTP responses.
}
\end{tablenotes}
\end{threeparttable}
\end{table}

Finally, we note that our results on the average WTP and the heterogeneous effects on worker effort by preference type are qualitatively and quantitatively similar across classification schemes (results available upon request).

\newpage
\section*{C. Theoretical Appendix}
\medskip
\subsection*{C.1 Proof of Hypothesis 2.1}

(i) Assume the worker is competitive (i.e., $\lambda_1 \geq \lambda_2 > 0$). We show that $WTP_{exante}$ is decreasing in the cost parameter $c$, and therefore increasing in baseline effort $e_{no-info} = \frac{w}{c}$.
\\

\noindent Take any $\bar{e} \in \R_+$. The worker's optimal effort upon learning the average effort $\bar{e}$ is:
\begin{align*}
e_{info,exante} &= \argmax_{e \in \R_{+}} \left\{ w e - \frac{c}{2}e^{2} + \mathds{1}_{\{e \leq \bar{e} \}} \cdot \lambda_1 (we - w\bar{e}) + \mathds{1}_{\{e > \bar{e} \}} \cdot \lambda_2 (we - w\bar{e}) \right\} \\[2mm]
&= \begin{cases}
    \frac{w (1+\lambda_2)}{c}  &\text{if } \bar{e} < \frac{w (1+\lambda_2)}{c} \\[1mm]
        \quad \bar{e} &\text{if } \bar{e} \in [\frac{w (1+\lambda_2)}{c},\frac{w (1+\lambda_1)}{c}] \\[1.5mm]
        \frac{w (1+\lambda_1)}{c}  &\text{if } \bar{e} > \frac{w (1+\lambda_1)}{c} \\[1.5mm]
    \end{cases}
\end{align*}
\\
\noindent It follows that the indirect utility is
\begin{align*}
    V_{info, exante}(\bar{e}) \equiv \widetilde{U}(e_{info,exante}; \bar{e}) &= 
    \begin{cases}
    \frac{w^2 (1+\lambda_2)^2}{2c} - \lambda_2 w \bar{e}  &\text{if } \bar{e} < \frac{w (1+\lambda_2)}{c} \\[1mm]
        \quad w\bar{e} - \frac{c}{2}\bar{e}^2 &\text{if } \bar{e} \in [\frac{w (1+\lambda_2)}{c},\frac{w (1+\lambda_1)}{c}] \\[1.5mm]
        \frac{w^2 (1+\lambda_1)^2}{2c} - \lambda_1 w \bar{e} &\text{if } \bar{e} > \frac{w (1+\lambda_1)}{c} \\[1.5mm]
\end{cases}
\end{align*}
\\
\noindent Let $\Delta V(\bar{e}) \equiv V_{info, exante}(\bar{e}) - V_{no-info} $. Since $V_{no-info} = \frac{w^2}{2c}$, we have
\begin{align*}
    \Delta V(\bar{e}) &= 
    \begin{cases}
    \frac{w^2 [(1+\lambda_2)^2 - 1]}{2c} - \lambda_2 w \bar{e} &\text{if } \bar{e} < \frac{w (1+\lambda_2)}{c} \\[1mm]
        \quad w\bar{e} - \frac{c}{2}\bar{e}^2 - \frac{w^2}{2c} &\text{if } \bar{e} \in [\frac{w (1+\lambda_2)}{c},\frac{w (1+\lambda_1)}{c}] \\[1.5mm]
        \frac{w^2 [(1+\lambda_1)^2 - 1]}{2c} - \lambda_1 w \bar{e} &\text{if } \bar{e} > \frac{w (1+\lambda_1)}{c} \\[1.5mm]
\end{cases}
\end{align*}
\\
\noindent Taking the derivative, we have
\begin{align*}
    \frac{\partial \Delta V(\bar{e})}{\partial c} &= 
    \begin{cases}
    - \frac{w^2}{2c^2} [(1+\lambda_2)^2 - 1] &\text{if } \bar{e} < \frac{w (1+\lambda_2)}{c} \\[1mm]
        \quad - \frac{1}{2}\bar{e}^2 + \frac{w^2}{2c^2} &\text{if } \bar{e} \in [\frac{w (1+\lambda_2)}{c},\frac{w (1+\lambda_1)}{c}] \\[1.5mm]
        - \frac{w^2}{2c^2} [(1+\lambda_1)^2 - 1]  &\text{if } \bar{e} > \frac{w (1+\lambda_1)}{c} \\[1.5mm]
\end{cases}
\end{align*}

\noindent In Case 1 and Case 3, the derivative is clearly negative since $\lambda_1, \lambda_2 > 0$. In Case 2, observe that for all $\bar{e} \in \left[\frac{w (1+\lambda_2)}{c}, \frac{w (1+\lambda_1)}{c}\right]$, we have $\bar{e} > \frac{w}{c}$. This implies $\bar{e}^2 > \frac{w^2}{c^2}$, so $- \frac{1}{2} \bar{e}^2 + \frac{w^2}{2c^2} < 0$. Therefore, $\frac{\partial \Delta V(\bar{e})}{\partial c} < 0$ in all cases, and $\Delta V(\bar{e})$ is decreasing in $c$. It follows that $WTP_{exante} = \E_p [\Delta V(\bar{e})]$ is also decreasing in $c$, and hence increasing in $e_{no-info} = \frac{w}{c}$.
\\[5mm]
\noindent (ii) Assume the worker is inequality-averse (i.e., $\lambda_1>0$, $-1<\lambda_2<0$, with $\lambda_1 \geq |\lambda_2|$). Take any $\bar{e} \in \R_+$. To show $WTP_{exante} \leq 0$, we proceed by cases:
\\[5mm]
\underline{Case 1:} $\bar{e} > \frac{w (1+\lambda_1)}{c}$ 
\\[3mm]
Consider $U(e) = we - \frac{c}{2} e^2$. For $e> \frac{w}{c}$, we have $U'(e) = w - ce < 0$, and so $U(e)$ is strictly decreasing on $[ \frac{w}{c}, \infty )$. Since $e_{info, exante} = \frac{w (1+\lambda_1)}{c} > \frac{w}{c} = e_{no-info}$, we have $U(e_{info, exante}) < U(e_{no-info})$. It follows that 
\begin{align*}
    V_{info, exante}(\bar{e}) \equiv U(e_{info, exante}) \underbrace{- \lambda_1 w \left( \bar{e} - \frac{w (1+\lambda_1)}{c}  \right)}_{<0} 
    &< U(e_{info, exante}) \\
    &< U(e_{no-info}) \equiv V_{no-info}
\end{align*}
\underline{Case 2:} $\bar{e} < \frac{w (1+\lambda_2)}{c}$ 
\\[3mm]
The argument is analogous to Case 1, and we again have $V_{info, exante}(\bar{e}) <  V_{no-info}$.
\\[5mm]
\underline{Case 3:} $\bar{e} \in \left[ \frac{w(1+\lambda_2)}{c}, \frac{w(1+\lambda_1)}{c}\right] $ 
\\[3mm]
In this case, the worker bunches at $\bar{e}$, and thus receives utility of $U(\bar{e})$. Since $e_{no-info}$ is optimal for the utility function $U(\cdot)$, we have
\begin{align*}
    U(\bar{e}) &\leq U(e_{no-info}) \\
\iff V_{info,exante}(\bar{e}) &\leq  V_{no-info}
\end{align*}
\bigskip
\noindent Combining all three cases, for any $\bar{e} \in \mathbb{R}_+$, we have \( V_{info,exante}(\bar{e}) \leq V_{no\text{-}info} \). Taking expectations over any belief distribution $p(\bar{e})$ yields:
\[
WTP_{exante} = \mathbb{E}_p[V_{info,exante}(\bar{e})] - V_{no\text{-}info} \leq 0.
\]

\hfill $\blacksquare$

\bigskip
\subsection*{C.2 Proof of Hypothesis 2.2}

(i) Assume the worker is competitive (i.e., $\lambda_1 \geq \lambda_2 > 0$). With peer information ex ante, the marginal benefit of effort becomes:
\begin{align*}
    MB(e) &= \begin{cases}
    w (1+\lambda_1)  &\text{if } e < \bar{e} \\[1mm]
    w (1+\lambda_2)  &\text{if } e > \bar{e} \\[1mm]
    \end{cases} \; ,
\end{align*}
which is strictly higher than the marginal benefit in the no-information case, that is simply $w$. With the marginal cost of effort $MC(e) = ce$ unchanged, it immediately follows that $e_{info, exante} > e_{no-info}$. 
\\[5mm]
\noindent (ii) Assume the worker is inequality-averse (i.e., $\lambda_1>0$, $-1<\lambda_2<0$, with $\lambda_1 \geq |\lambda_2|$).
We proceed by cases:
\\[5mm]
\underline{Case 1:} $\bar{e} > \frac{w(1+\lambda_1)}{c}$ 
\\[3mm]
With peer information ex ante, the worker chooses $e_{info, exante} = \frac{w (1+\lambda_1)}{c} $, while without information, they choose $e_{no-info} = \frac{w}{c} $. Since $\lambda_1>0$, it follows that $
e_{no-info} \leq e_{info, exante} \leq \bar{e} $.
\\[5mm]
\underline{Case 2:} $\bar{e} < \frac{w(1+\lambda_2)}{c}$ 
\\[3mm]
The argument is analogous to Case 1. Since $\lambda_2<0$, we have  
$\bar{e} \geq e_{info, exante} \geq e_{no-info}$.
\\[5mm]
\underline{Case 3:} $\bar{e} \in \left[ \frac{w(1+\lambda_2)}{c}, \frac{w(1+\lambda_1)}{c} \right] $ 
\\[3mm]
In this case, the worker bunches at $e_{info, exante}^{exante} = \bar{e} $. 
\\[5mm]
Therefore, in all cases, the worker's chosen effort with peer information is weakly closer to the average $\bar{e}$ than the case without information, i.e., $|e_{info,exante} - \bar{e}| \leq |e_{no-info} - \bar{e}|$.  \hfill $\blacksquare$

\bigskip
\subsection*{C.3 Proof of Hypothesis 2.3}

Assume $\delta=1$. Take any belief distribution $p(\bar{e})$ over possible average effort levels $\bar{e} \in \mathbb{R}_+$. It suffices to show that $\E_p[V_{info, exante}(\bar{e})] \geq V_{info,expost}$. 
\\

\noindent For every realization $\bar{e}$, define $e_{info,exante}(\bar{e})$ as the worker's optimal effort choice conditional on observing $\bar{e}$:
$e_{info, exante}(\bar{e}) = \argmax_{e \in \R_{+}} \widetilde{U}(e; \bar{e})$. Let $e_{info,expost}$ denote the effort chosen when the worker must choose effort before knowing $\bar{e}$: $ e_{info,expost} = \argmax_{e \in \R_{+}}  \mathbb{E}_p[\widetilde{U}(e; \bar{e})]$.
\\

\noindent By definition, $e_{info, exante}(\bar{e})$ is optimal for the utility function $\Tilde{U}(\cdot; \bar{e})$ for every realization $\bar{e}$, so we have 
\begin{align*} \Tilde{U}(e_{info, exante}(\bar{e}); \bar{e}) &\geq \Tilde{U}(e_{info, expost}; \bar{e}) 
\end{align*}

\noindent Taking expectations with respect to the belief distribution $p(\bar{e})$ yields:
$$
\mathbb{E}_p[V_{info, exante}(\bar{e})] = \mathbb{E}_p[\widetilde{U}(e_{info, exante}(\bar{e}); \bar{e})] \geq \mathbb{E}_p[\widetilde{U}(e_{info, expost}; \bar{e})] = V_{info, expost},
$$
as desired. \hfill $\blacksquare$

\bigskip
\subsection*{C.4 Proof of Hypothesis 2.4}

(i) The proof is the same as that of Hypothesis 2.2.
\\[5mm]
\noindent (ii) Assume the worker is inequality-averse (i.e., $\lambda_1>0$, $-1<\lambda_2<0$, with $\lambda_1 \geq |\lambda_2|$).
\\

\noindent
In the ex ante case, the marginal benefit of effort is discontinuous at the average $\bar{e}$:
\begin{align*}
    MB_{exante}(e) &= \begin{cases}
    w (1+\lambda_1)  &\text{if } e < \bar{e} \\[1mm]
    w (1+\lambda_2)  &\text{if } e > \bar{e} \\[1mm]
    \end{cases} \; 
\end{align*}
\\
\noindent
In contrast, in the ex post case, the worker does not observe $\bar{e}$ and instead maximizes expected utility over their belief distribution $p(\bar{e})$. The marginal benefit of effort is therefore:
\begin{align*}
    MB_{expost}(e) = w + \lambda_1 w \left( \int_e^{\infty} p(\bar{e}) d\bar{e} \right) + \lambda_2 w \left( \int_0^{e} p(\bar{e}) d\bar{e} \right) 
\end{align*}
\\
Now consider two regions. If $e < \bar{e}$, we have $MB_{exante}(e) \geq MB_{expost}(e)$, so for the same marginal cost, the worker's optimal effort choices must satisfy $e_{info, expost} \leq e_{info, exante} \leq \bar{e}$. If $e > \bar{e}$, we have $MB_{exante}(e) \leq MB_{expost}(e)$, so for the same marginal cost, so the worker's optimal effort choices must instead satisfy $\bar{e} \leq e_{info, exante} \leq e_{info, expost} $. Thus, in both cases, the worker's ex post effort choice lies farther from $\bar{e}$ than in the ex ante case, i.e. $|e_{info,expost} - \bar{e}| \geq |e_{info,exante} - \bar{e}|$.  \hfill $\blacksquare$

\bigskip
\subsection*{C.5 Proof of Hypothesis 3.1}

Take any belief distribution $p(\bar{e})$ over possible average effort levels $\bar{e} \in \mathbb{R}_+$. The WTP for peer information ex ante is given by:
\begin{align*}
WTP_{exante} &\equiv \E_p[V_{info, exante}(\bar{e})] - V_{no-info} \\[2mm]
&= -\E_p[\Theta(\bar{e})] \leq 0,
\end{align*}
since $e_{info, exante}(\bar{e}) = e_{no-info}$ and $\Theta(\bar{e}) \geq 0, \forall \bar{e} \in \R_{+}$. \hfill $\blacksquare$ 

\bigskip
\subsection*{C.6 Proof of Hypothesis 4.1}

We show that $WTP_{exante}$ is decreasing in $\alpha_s$, which implies it is decreasing in $e_{no-info} = \frac{w \alpha_s}{c}$. First, we compute the following derivative using Leibniz's rule:
\begin{align*}
    \frac{d}{d \alpha_s} V_{search}(\bar{y}; \alpha_s) &= \frac{d}{d \alpha_s} \left( \int_{\alpha_s}^{\bar{\alpha}} \frac{w^2 \alpha^2}{2c} dF(\alpha | \bar{y}) + F(\alpha_s | \bar{y}) \cdot \frac{w^2 \alpha_s^2}{2c} - K \right) \\[2mm]
    &= -\frac{w^2 \alpha_s^2}{2c} \cdot f(\alpha_s | \bar{y}) + \frac{w^2 \alpha_s^2}{2c} \cdot f(\alpha_s | \bar{y}) + F(\alpha_s | \bar{y}) \cdot \frac{w^2 \alpha_s}{c} \\[2mm] 
    &= F(\alpha_s | \bar{y}) \cdot \frac{w^2 \alpha_s}{c}
\end{align*}

\noindent We also have:
\begin{align*}
    \frac{d}{d \alpha_s} V_{no-info}(\alpha_s) &= \frac{w^2 \alpha_s}{c}
\end{align*}

\noindent  Since $V_{info}(\bar{y}) = \max\left\{ V_{search}(\bar{y}), V_{no-info} \right\}$, it follows that:
\begin{align*}
    \frac{d}{d \alpha_s} V_{info}(\bar{y}; \alpha_S) \leq \frac{w^2 \alpha_s}{c} 
\end{align*}

\noindent Putting the pieces together, we have
\begin{align*}
    \frac{d}{d \alpha_s} WTP_{exante} (\alpha_s) &= \frac{d}{d \alpha_s} \left( \E_{p(\bar{y})} \left[V_{info}(\bar{y}; \alpha_s)\right] \right) - \frac{d}{d \alpha_s} V_{no-info}(\alpha_s) \\[2mm]
    &= \E_{p(\bar{y})} \left[ \frac{d}{d \alpha_s} V_{info}(\bar{y}; \alpha_s)\right] - \frac{w^2 \alpha_s}{c} \\
    &\leq 0
\end{align*}
\hfill $\blacksquare$

\bigskip
\subsection*{C.7 Proof of Hypothesis 4.2}

With peer information, if the worker chooses not to search for a new strategy, they retain the baseline strategy $s$ and chooses $e_{info, exante} = e_{no-info}$. If the worker searches for a new strategy $s'$, there are two cases. If $\alpha_{s'} > \alpha_{s} $, they adopt the new strategy and choose $e_{info, exante} = \frac{w\alpha_{s'}}{c} > \frac{w\alpha_{s}}{c} = e_{no-info}$. Otherwise, if $\alpha_{s'} \leq \alpha_{s} $, they revert to baseline strategy $s$, and chooses the same effort level as before, $e_{info, exante} = e_{no-info}$. 
\hfill $\blacksquare$

\newpage
\subsection*{C.8 Model Extension: Allowing Beliefs to Affect Utility Without Peer Information}

In the models presented in the main text, we assume that beliefs about peers' performance, $p(\bar{e})$, are acted upon only when peer information is provided (whether ex ante or ex post). We now relax this assumption to allow beliefs to play a role even without peer information. 

Let $\sigma_p^2$ denote the variance of the worker's belief distribution $p(\bar{e})$, and let $\theta(\sigma_p^2) \in [0,1]$ represent the weight attached to the peer-related utility component. In the no-information case, the worker's utility is now given by:

\begin{align*}
    U_{no-info}(e; \bar{e}) = m(we) - c(e) + \theta(\sigma_p^2) \cdot N(\cdot), 
\end{align*}
where $N(\cdot)$ captures the relevant nonstandard utility component: $N(\cdot) = f(we - w\bar{e})$ for social preferences, and $N(\cdot) = -\Theta(\bar{e})$ for stress. 

We assume that $\theta(\sigma_p^2)$ is decreasing in $\sigma_p^2$, meaning that a worker with more diffuse beliefs (higher variance) places less weight on the peer-related utility component $N(\cdot)$. This captures the idea that when workers are less certain about their relative performance, peer-related considerations are less salient in their decision-making process. We impose the boundary conditions $\theta(0)=1$ and $\theta(\infty)=0$. When $\theta = 0$, the model collapses to the baseline formulation in the main text, where beliefs play no role absent peer information. 

When peer information is provided (whether ex ante or ex post), peer-related considerations become fully salient to the worker, so we set $\theta=1$. The worker's utility in each case then coincides with the main-text specification:
\begin{align*}
    U_{info, exante}(e; \bar{e}) &= m(we) - c(e) + N(\cdot), 
    \\
    U_{info, expost}(e; \bar{e}) &= m(we) - c(e) + \delta \cdot N(\cdot), 
\end{align*}
where $\delta$ allows the intensity of nonstandard preferences to depend on information timing.

Accordingly, the worker's WTP in the two scenarios is defined as:
\begin{align*}
WTP_{exante}(c) &\equiv \E_p[V_{info, exante}(c;\bar{e})] - \E_p[V_{no-info}(c;\bar{e})],
\\
WTP_{expost}(c; \delta) &\equiv V_{info, expost}(c; \delta) - \E_p[V_{no-info}(c;\bar{e})],
\end{align*}
where $V_{no-info}$ is the indirect utility derived from $U_{no-info}$ defined above.

This extension preserves all predictions derived in the main text, but additionally predicts that the magnitude (absolute value) of WTP for information ex ante and ex post increases with belief variance, $\sigma_p^2$, across all worker types. We empirically test this prediction in Table \ref{tab:table-wtp2}, showing that workers with higher belief variance display larger absolute WTP, consistent with the model's prediction.

\medskip 
\noindent\textit{Conceptual Note.}---Unlike the social preferences and stress avoidance models, the learning model already internalizes uncertainty about peers' performance through the workers' (prior) belief distribution over strategy productivity, $F(\alpha)$, even in the absence of peer information. Therefore, we do not introduce the additional weighting term, $\theta(\sigma_p^2)$, in that model.

\newpage
\setcounter{table}{0} 
\setcounter{figure}{0}
\renewcommand{\thetable}{D\arabic{table}}
\renewcommand{\thefigure}{D\arabic{figure}}

\section*{D. Experimental Instructions}

The following set of screenshots demonstrates a demo version of the experiment.
\\[2mm]

\begin{figure}[H]
\caption{Part 1 Instructions (Task Description)}
    \begin{center}
		\includegraphics[width=\textwidth]{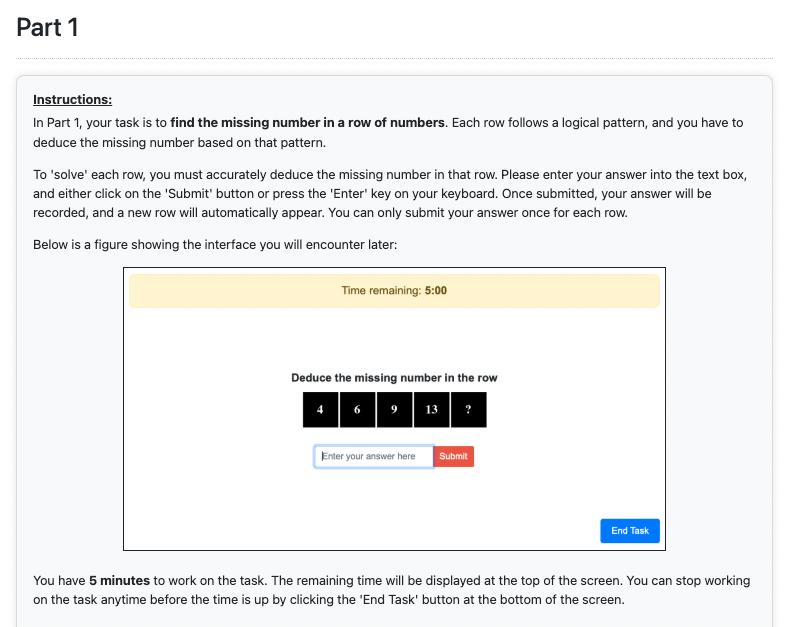}
	\end{center} 
    \begin{minipage}{15cm}
        \singlespacing
		\footnotesize \emph{Notes.} Worker are told that they will earn a bonus of 1 cent for every row solved correctly. If they solve a row incorrectly, they will not earn anything for that row. Before they can advance the page, they need to answer the comprehension question correctly.
	\end{minipage}
\end{figure}

\begin{figure}[H]
\caption{Work Period 1}
    \begin{center}
		\includegraphics[width=\textwidth]{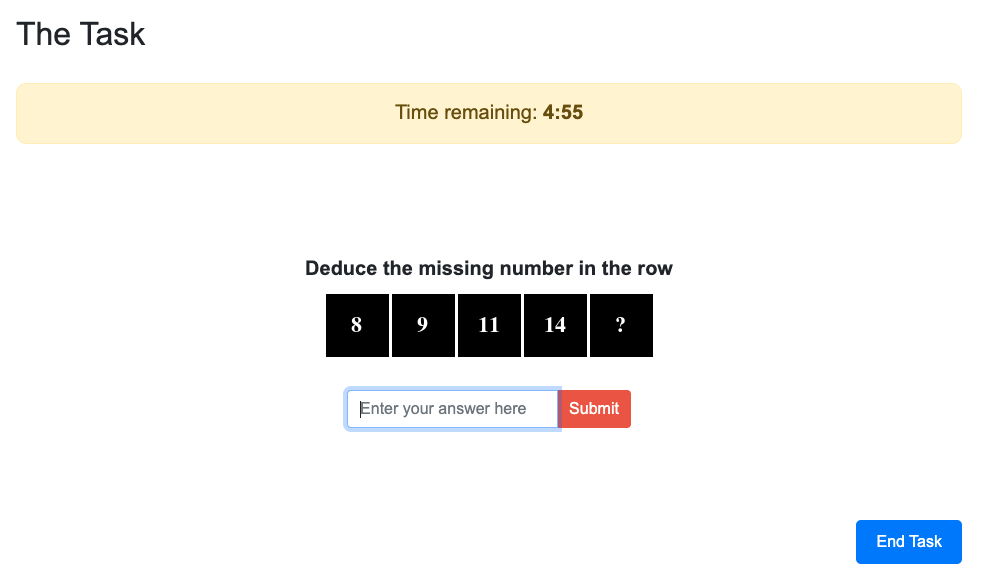}
	\end{center} 
     \begin{minipage}{15cm}
        \singlespacing
		\footnotesize \emph{Notes.} Before the first work period, the worker completes a practice round lasting up to 30 seconds.
	\end{minipage}
\end{figure}

\begin{figure}[H]
\caption{Post-task Assessment (Page 1 of 2)}
\label{fig:post_assessment} 
    \begin{center}
		\includegraphics[width=\textwidth]{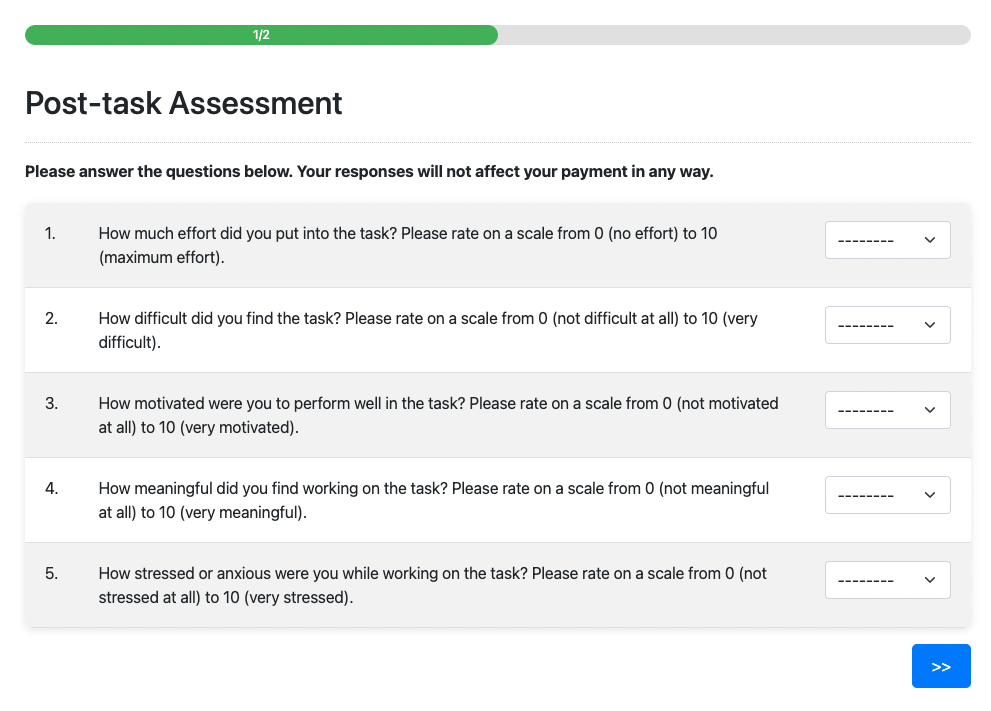}
	\end{center} 
\end{figure}

\begin{figure}[H]
\caption{Post-task Assessment – Belief Elicitation (Page 2 of 2)}
    \begin{center}
		\includegraphics[width=\textwidth]{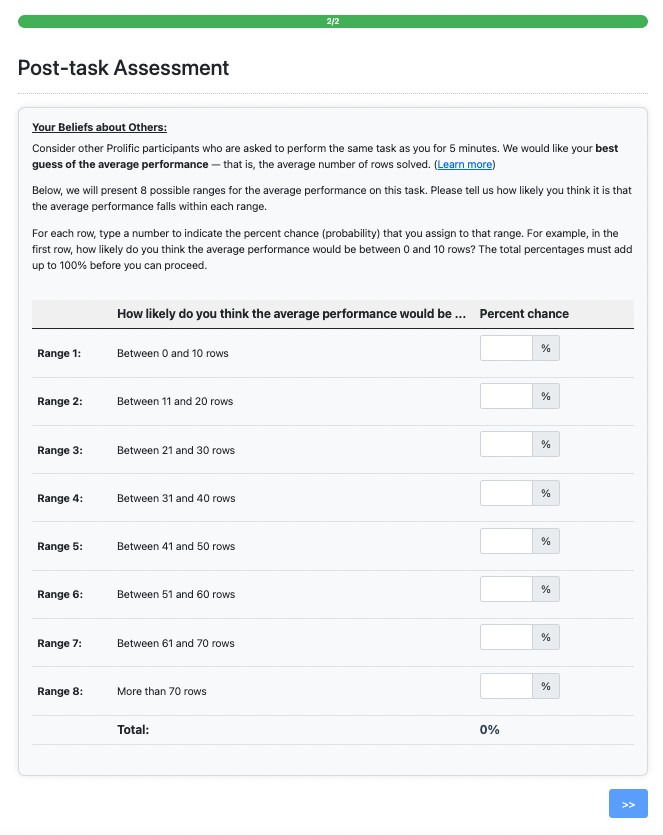}
	\end{center} 
\end{figure}

\begin{figure}[H]
\caption{Part 2 Instructions}
    \begin{center}
		\includegraphics[width=\textwidth]{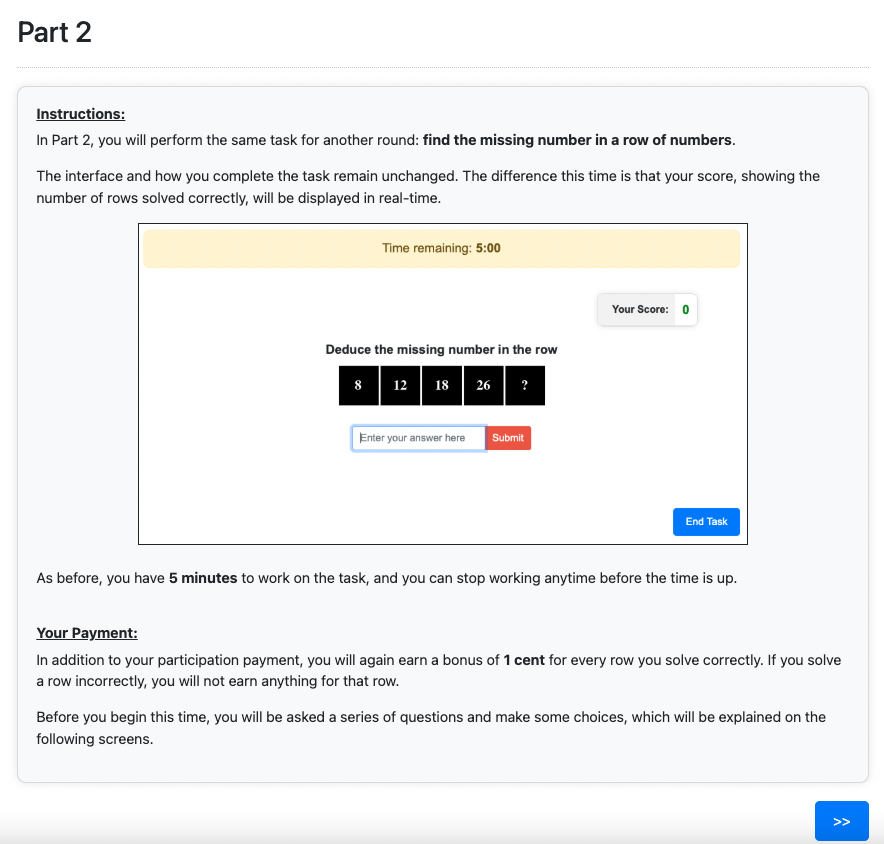}
	\end{center} 
\end{figure}

\begin{figure}[H]
\caption{Peer Information}
    \begin{center}
		\includegraphics[width=\textwidth]{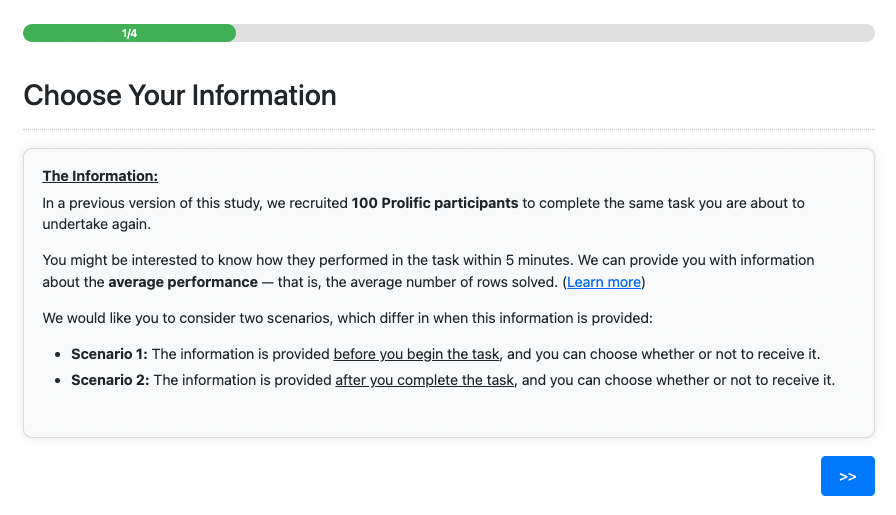}
	\end{center} 
\end{figure}

\begin{figure}[H]
\caption{WTP Elicitation Instructions}
    \begin{center}
		\includegraphics[width=\textwidth]{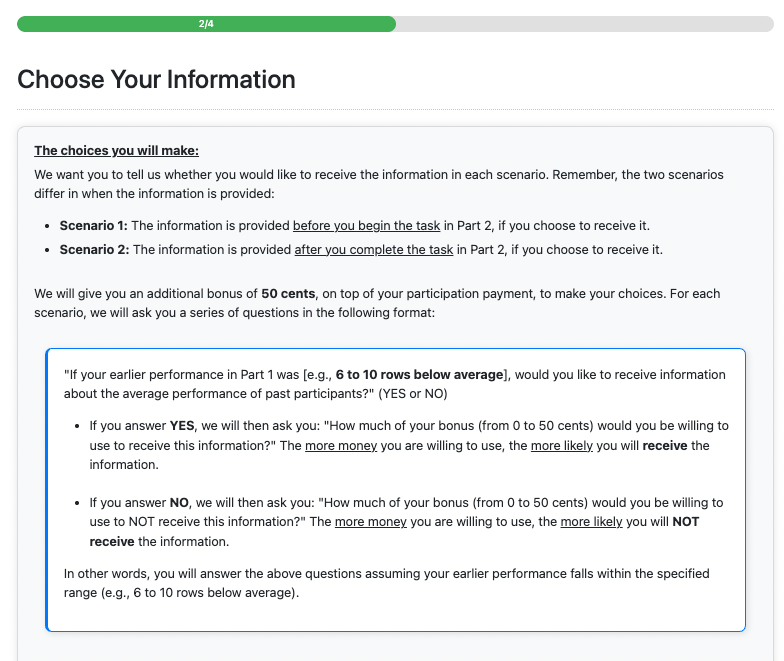}
	\end{center} 
\end{figure}

\begin{figure}[H]
\ContinuedFloat
\caption{WTP Elicitation Instructions (Cont.)}
    \begin{center}
		\includegraphics[width=\textwidth]{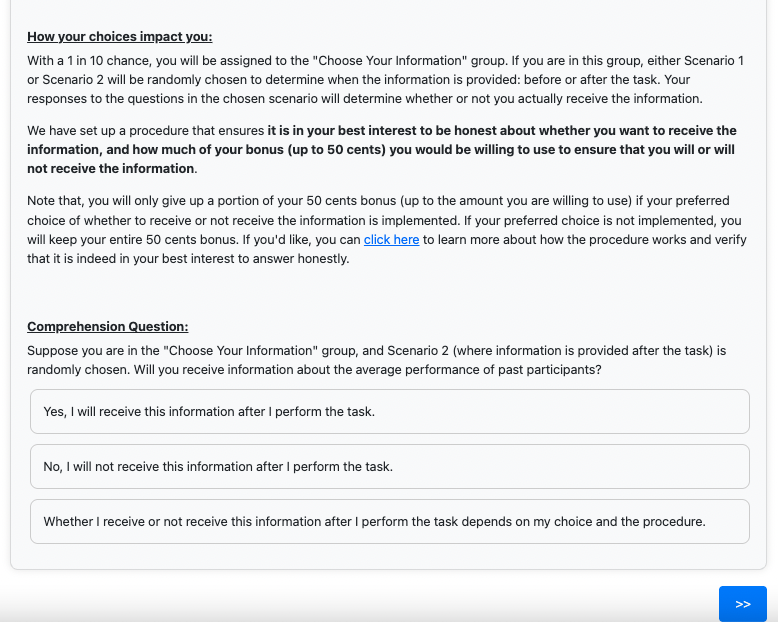}
	\end{center} 
\end{figure}

\begin{figure}[H]
\caption{WTP Elicitation – Step 1 of 2 (Ex ante Scenario)}
    \begin{center}
		\includegraphics[width=\textwidth]{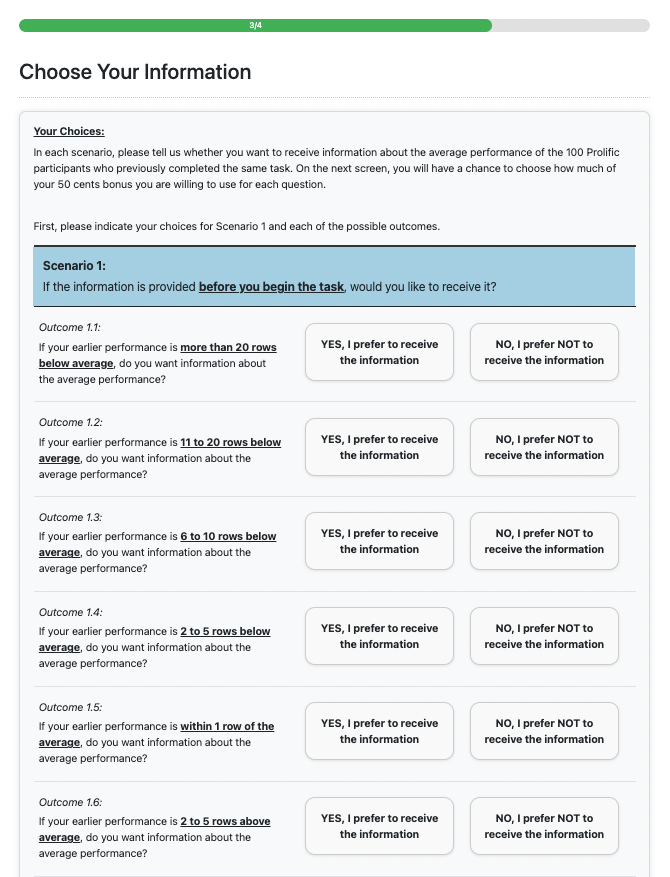}
	\end{center} 
\end{figure}

\begin{figure}[H]
\caption{WTP Elicitation – Step 1 of 2 (Ex post Scenario)}
    \begin{center}
		\includegraphics[width=\textwidth]{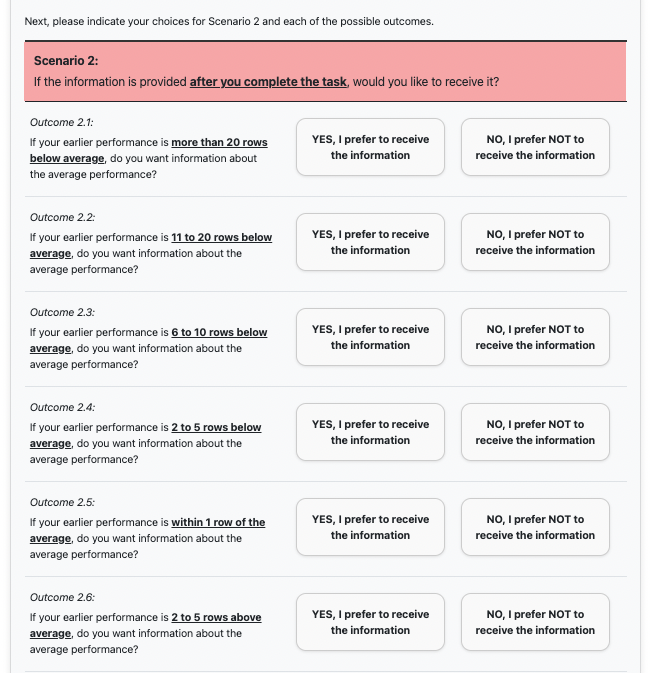}
	\end{center} 
\end{figure}

\begin{figure}[H]
\caption{WTP Elicitation – Step 2 of 2 (Ex ante Scenario)}
    \begin{center}
		\includegraphics[width=\textwidth]{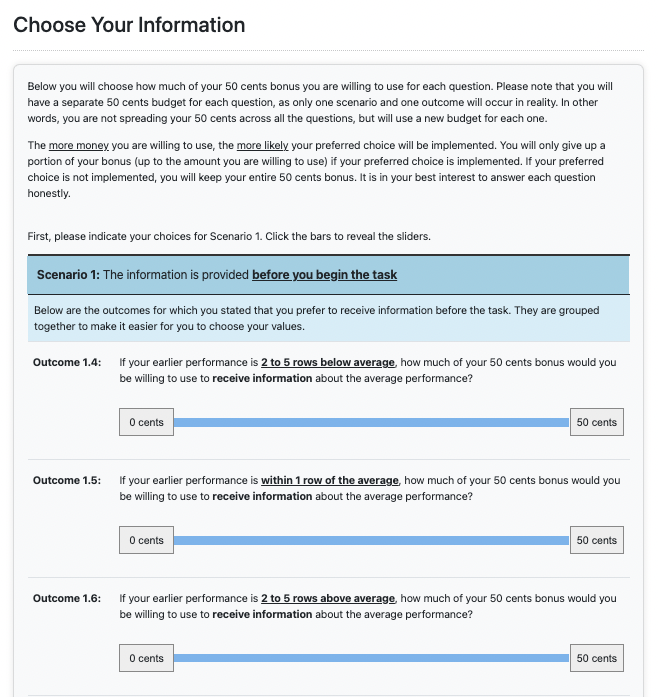}
	\end{center} 
    \begin{minipage}{15cm}
		\scriptsize \singlespacing{\emph{Notes.} The same procedure applies to the ex post scenario, where workers also indicate their choices using sliders.} 
	\end{minipage}
\end{figure}

\newpage
\begin{figure}[H]
\caption{Treatment Assignment (Control Group)}
    \begin{center}
		\includegraphics[width=\textwidth]{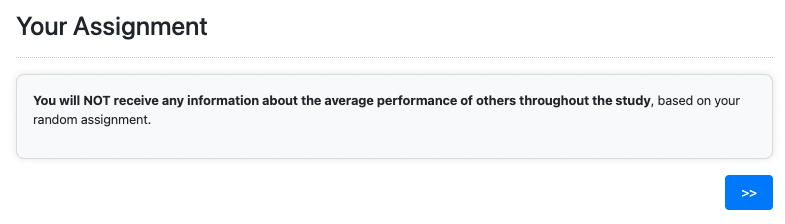}
	\end{center} 
\end{figure}

\begin{figure}[H]
\caption{Treatment Assignment (Ex ante Info Group)}
    \begin{center}
		\includegraphics[width=\textwidth]{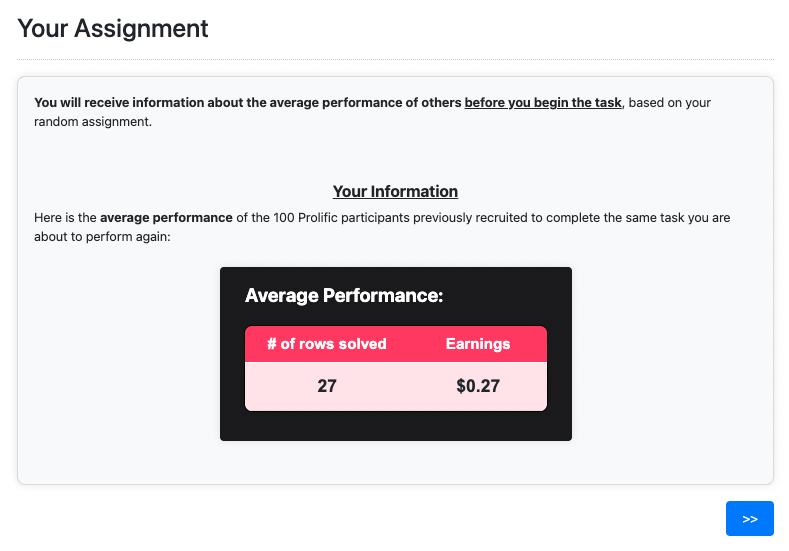}
	\end{center} 
\end{figure}

\begin{figure}[H]
\caption{Treatment Assignment (Ex post Info Group)}
    \begin{center}
		\includegraphics[width=\textwidth]{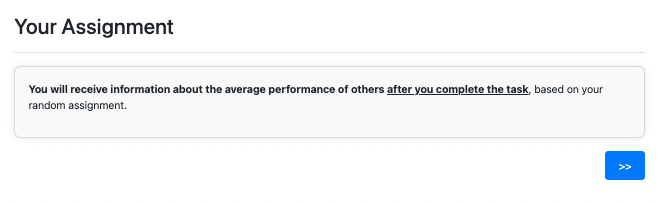}
	\end{center} 
\end{figure}

\begin{figure}[H]
\caption{Treatment Assignment (Choose-Your-Info Group)}
    \begin{center}
		\includegraphics[width=\textwidth]{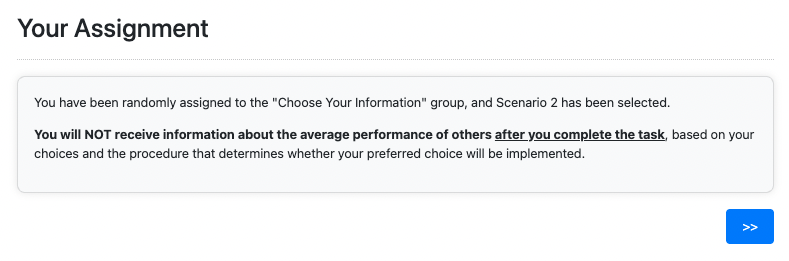}
	\end{center} 
\end{figure}

\begin{figure}[H]
\caption{Work Period 2 (Ex ante Info Group)}
    \begin{center}
		\includegraphics[width=\textwidth]{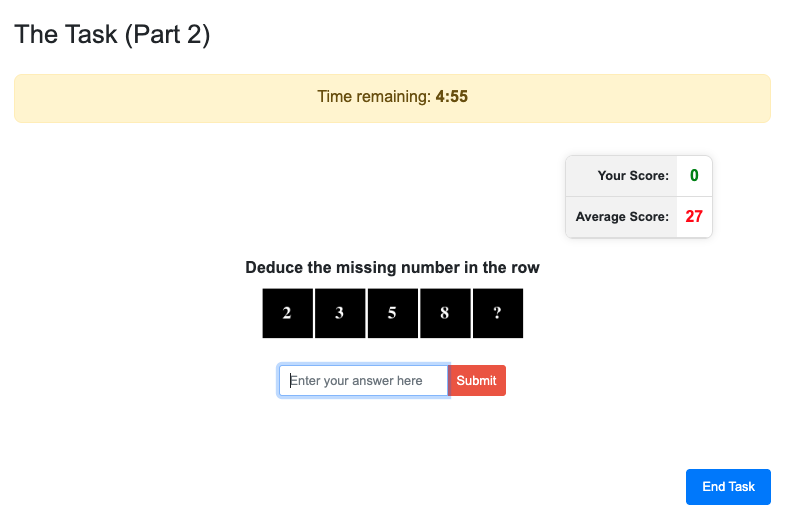}
	\end{center} 
     \begin{minipage}{15cm}
        \singlespacing
		\footnotesize \emph{Notes.} For workers not assigned to receive peer information ex ante, the average score (27 rows) is not shown to them while they perform the task. After the task, we re-elicit the same measures using the same set of questions as in Figure \ref{fig:post_assessment}.  
	\end{minipage}
\end{figure}

\begin{figure}[H]
\caption{Exit Survey}
    \begin{center}
		\includegraphics[width=\textwidth]{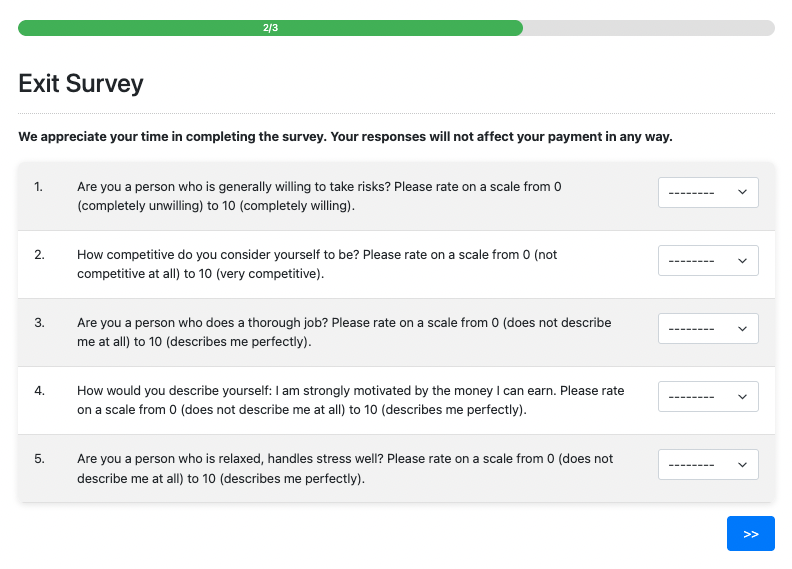}
	\end{center}
    \begin{minipage}{15cm}
        \singlespacing
		\footnotesize \emph{Notes.} Prior to this page, workers provide demographic information on their gender, year of birth, and highest level of education.   
	\end{minipage}
\end{figure}

\begin{figure}[H]
\ContinuedFloat
\caption{Exit Survey (Cont.)}
    \begin{center}
		\includegraphics[width=\textwidth]{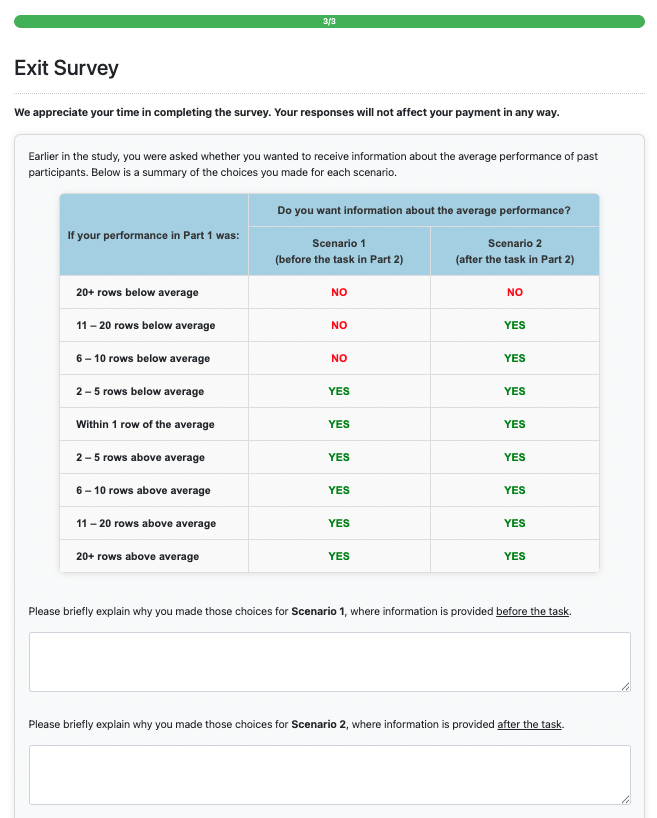}
	\end{center} 
\end{figure}

\end{document}